%% file: main.tex
\documentclass[preprint,
amsmath,amssymb,
 aps, prfluids,
floatfix,
]{revtex4-2}

\usepackage{graphicx}\usepackage{dcolumn}\usepackage{bm}

\input{macros.sty}
\usepackage[hidelinks]{hyperref}

\begin{document}

\title{Generalized Forcing Method: Generation of Diverse Data for Training Linear Transport PDE Closure Models}

\author{Wenyuan Xue}
\author{Ali Mani}\email{Contact author: alimani@stanford.edu}
\affiliation{Department of Mechanical Engineering, Stanford University, California 94305, USA
}

\date{\today}

\begin{abstract}
Data-driven closure modeling for transport partial differential equations requires training data that are accurate, affordable, diverse, and directly tailored to the target closure fields. We develop the Generalized Forcing Method (GFM), a data-generation framework for training linear transport closure models. GFM generates such data by running simulations with a zero initial condition and an extra body force that is constructed compatibly with the reduced dynamics. This framework leads to implicit GFM (iGFM), which prescribes resolved trajectories, and explicit GFM (eGFM), which constructs a basis of admissible forcings.
We apply eGFM to linear transport closure problems involving parallel and spatially inhomogeneous flows, flows with random coefficients, and passive-scalar transport in homogeneous isotropic turbulence. The results show that eGFM can identify accurate and stable reduced models when the reduced variables and model form are consistent with the underlying closure relation.

\end{abstract}

\maketitle

 \input{sections/1.introduction}

 \input{sections/2.GFM}

 \input{sections/3.parallel_flow}

 \input{sections/4.extension}

 \input{sections/5.turbulence}

 \input{sections/6.conclusion}

\begin{acknowledgments}

This work is supported by Boeing and the Office of Naval Research.

\end{acknowledgments}

\section*{Data Availability}

The test datasets, trained models, and scripts used for testing and generating plots are available on Zenodo at \url{https://doi.org/10.5281/zenodo.20424794} \cite{xue_2026_20424794}. Other data and code are not publicly available.

\bibliography{main}

\end{document}

%% file: sections/1.introduction.tex
\section{\label{sec:intro}Introduction}

Closure modeling arises naturally in coarse-graining of multiscale physical systems, particularly when the influence of unresolved small-scale dynamics is not negligible. The closure problem appears in various fields, including fluid dynamics, statistical physics, climate modeling and porous media flow. A canonical example lies in the Reynolds-averaged Navier--Stokes (RANS) and large eddy simulation (LES) approaches to turbulence, where the Reynolds stresses or subgrid-scale stresses require closure~\cite{smagorinsky1963general,pope2001turbulent}.

Traditionally, closure models are derived from asymptotic approximations and physical intuition, with a small number of parameters calibrated from data. More recently, scientific machine learning has emerged as a complementary approach in which data play a crucial role in building models~\cite{sanderse2024scientific}.

A central practical bottleneck in data-driven closure modeling is the generation of reference data.
Ideally, the data-generation framework should satisfy the following ``\textbf{TAAD}'' criteria:
(i). Tailored. It provides data directly probing the target closure fields of interest.
(ii). Accurate. The datasets are physically consistent with the physical system.
(iii). Affordable. The method is computationally efficient to enable systematic data collection.
(iv). Diverse. The datasets should sufficiently span the state space to ensure robust model identification.

In the context of multiscale PDE modeling, specific difficulties arise for these requirements.
Typical approaches such as varying domain geometry, boundary conditions and initial conditions are often not affordable ways to generate diverse datasets. The multiscale nature of many PDE systems requires large numbers of spatial modes to be resolved, which is prohibitively expensive for complex geometries.
Besides, many PDEs of interest are dissipative: datasets generated by naive sampling of initial conditions tend to quickly contract toward low-dimensional attractors, leading to insufficient data for model identification.

To enrich the data, forcing methods are commonly used to sustain statistically stationary dynamics or to probe the closure response. In the context of turbulence, this concept appeared in nonlocal closure modeling of turbulent scalar and momentum transport, where Hamba~\cite{hamba1995analysis,hamba2004nonlocal,hamba2005nonlocal,hamba2022analysis,hamba2025analysis,hamba2025nonlocal} used a Green's function approach to measure the response under an impulse forcing, thereby forming the nonlocal kernel of the closure operator. The Macroscopic Forcing Method (MFM)~\cite{mani2021macroscopic,liu2023systematic,park2024direct,lavacot2024non} generalized this concept by utilizing forcings more general than Dirac delta functions, which enabled a direct measurement of the moments of closure operators without explicitly computing their Green's functions.

Analogous strategies have also appeared in atomistic modeling~\cite{van2023hyperactive}, where adaptive bias forces are added to drive simulations toward high-uncertainty regions of configuration space, thereby enriching the reference data.

While a primary motivation of this research is turbulence modeling, the nonlinearity of the Navier--Stokes equations presents a major barrier to initial method development and testing. Consequently, we focus here on closure modeling for passive-scalar transport. For a prescribed velocity field, the scalar equation is linear and serves as an ideal prototype: it captures the fundamental closure challenge while allowing for formal analysis and detailed validation.

For formal validation, we first revisit the classical Taylor--Aris dispersion problem~\cite{taylor1953dispersion,aris1956dispersion} and derive a set of analytical closure models. These analytical results serve as references for assessing the GFM-based models. We then demonstrate the method's broader capability by applying it to inhomogeneous and random parallel flows. Finally, we consider scalar transport in homogeneous isotropic turbulence.

The remainder of the paper is organized as follows. Secs.~\ref{sec:intro-revisiting}--\ref{sec:intro-forcing} revisit the parallel-flow dispersion problem. Sec.~\ref{sec:gfm} details the theoretical foundations of GFM. Numerical demonstrations and extensions are provided in Secs.~\ref{sec:parallel} and~\ref{sec:extension}. Sec.~\ref{sec:turbulence} applies eGFM to passive-scalar transport in homogeneous isotropic turbulence. Conclusions and perspectives are provided in Sec.~\ref{sec:conclusion}.

\subsection{\label{sec:intro-revisiting}Revisiting the parallel flow dispersion problem}

We revisit a classical prototype of closure modeling: dispersion in a parallel shear flow. It is a simple linear PDE with multiscale structure, and is a special case of the well-known Taylor--Aris dispersion. It has a long history of analytical theory~\cite{taylor1953dispersion,taylor1954conditions,aris1956dispersion,gill1970exact,barton1983method,frankel1989foundations} and has recently been revisited using data-driven closure approaches~\cite{mani2021macroscopic,liu2023systematic}.

Consider an advection-diffusion system under a simple shear flow in the domain $\Omega = \left(-\infty, \infty \right) \times \left[-\pi, \pi \right]$ which is periodic in $y$. Following the rescaling in Ref.~\cite{mani2021macroscopic}, the governing equation can be simplified as
\begin{equation}
    \frac{\partial c}{\partial t} + u \frac{\partial c}{\partial x} = \frac{\partial^2 c}{\partial y^2}, \qquad u(x,y,t)=\cos y,
    \label{eq:intro:ADE-HDM-simple}
\end{equation}
with initial condition $c(x,y,0)=\bar c_0(x)$ (uniform in $y$).

For any field $h(x,y,t)$, we define the cross-sectional average and fluctuation by
\begin{equation}
    \bar{h}(x,t) = \frac{1}{2\pi} \int_{-\pi}^{\pi} h(x,y,t)\,dy, \quad
    h'(x,y,t) = h(x,y,t) - \bar{h}(x,t). \label{eq:intro:averaging}
\end{equation}

\begin{figure}[tbp]
    \centering
        \centering
        \includegraphics{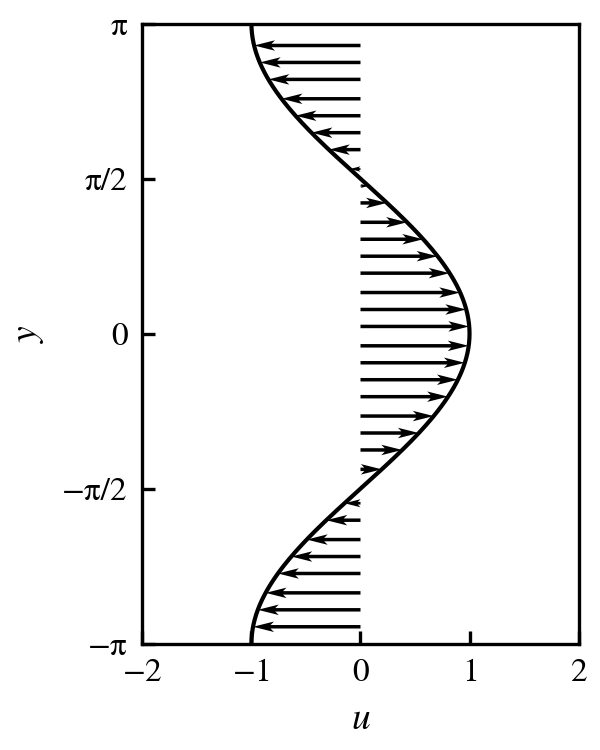}
    \hfill
        \centering
        \includegraphics{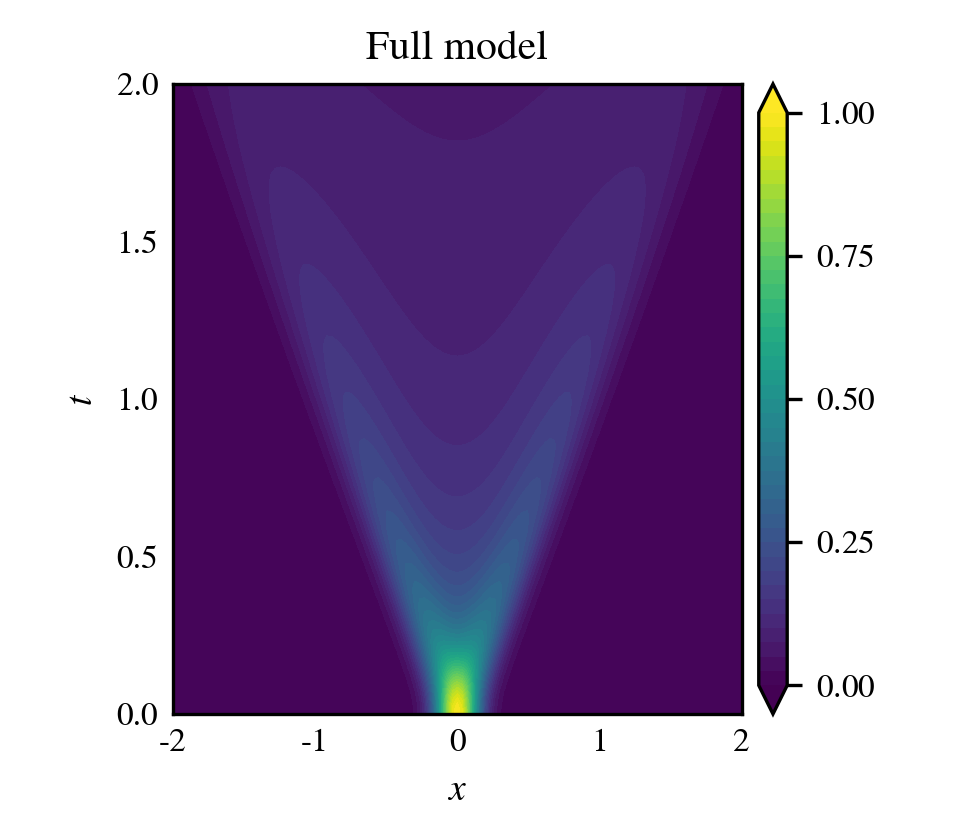}

    \caption{
        Dispersion of passive scalar in parallel flow with $\bar{c}_0(x) = \exp(-40x^2)$.
        Left: velocity profile $u(y)$; right: time history of the cross-sectional mean $\bar c(x,t)$.
    }
    \label{fig:intro:dispersion}
\end{figure}

\begin{figure}[tbp]
        \centering
        \includegraphics{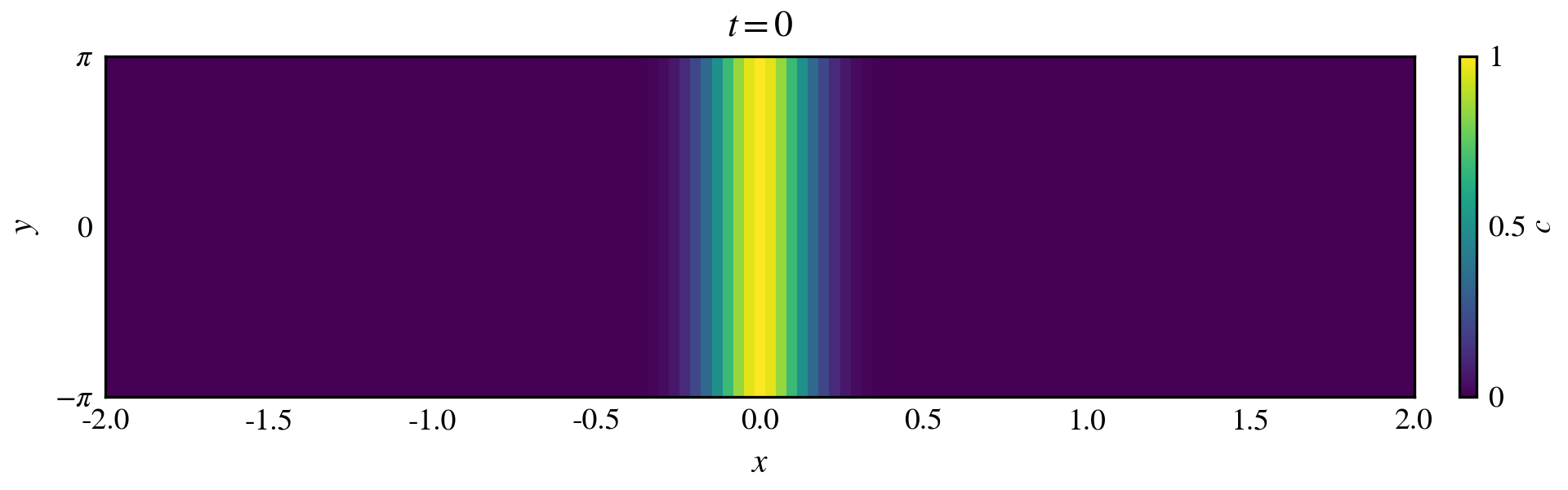}

        \centering
        \includegraphics{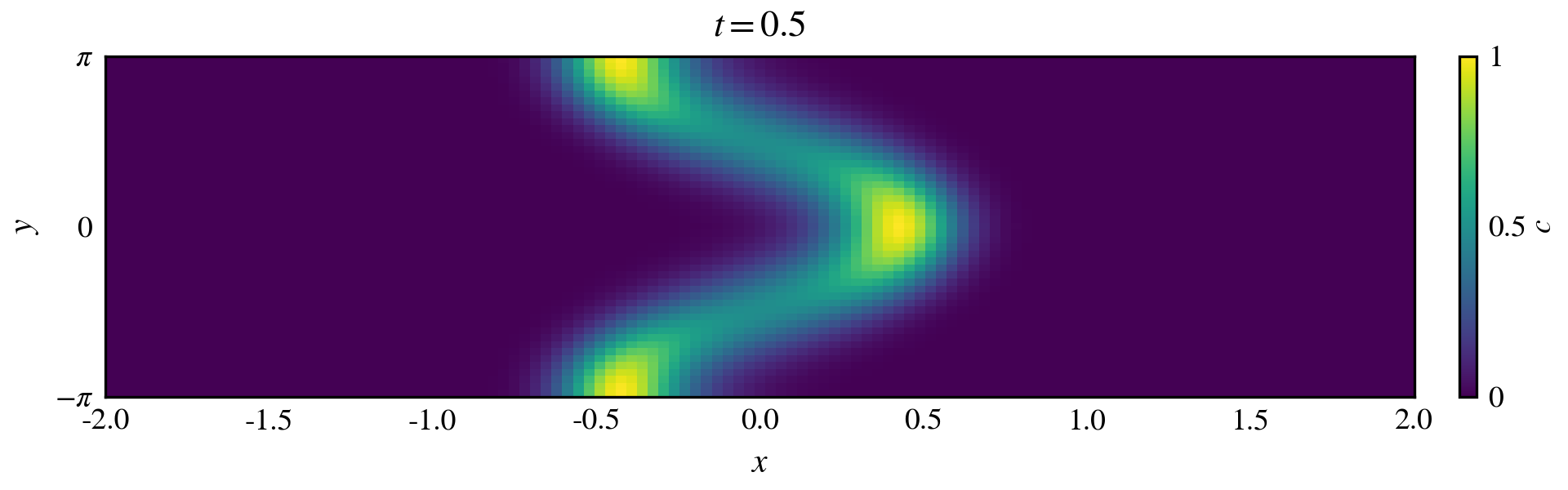}

        \centering
        \includegraphics{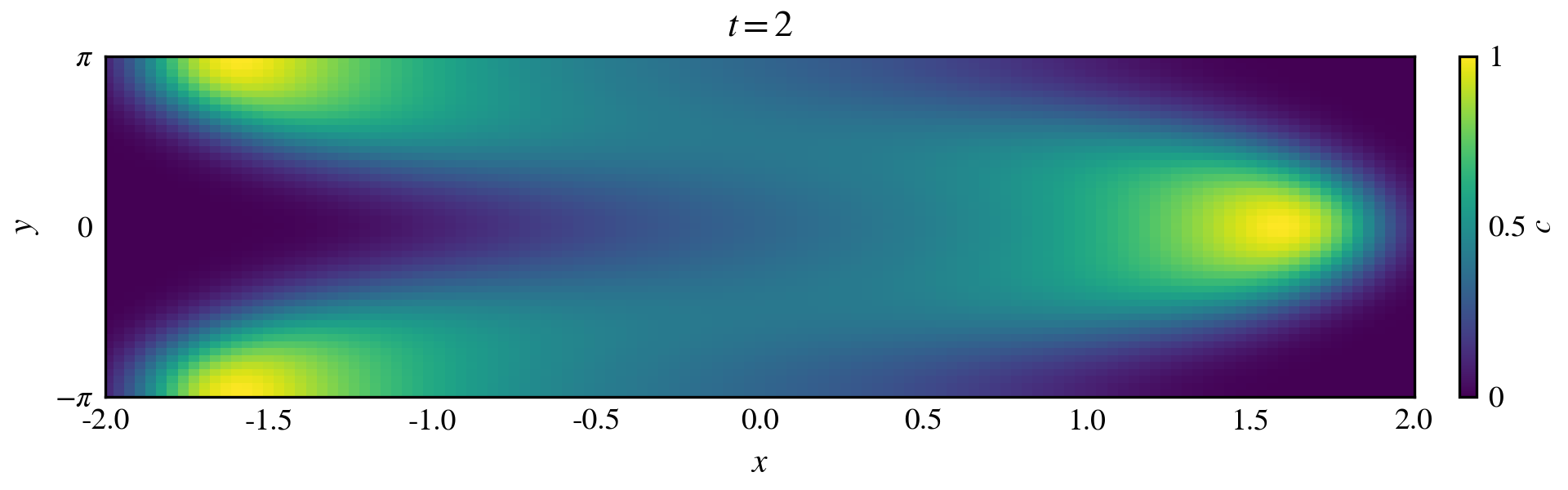}

        \centering
        \includegraphics{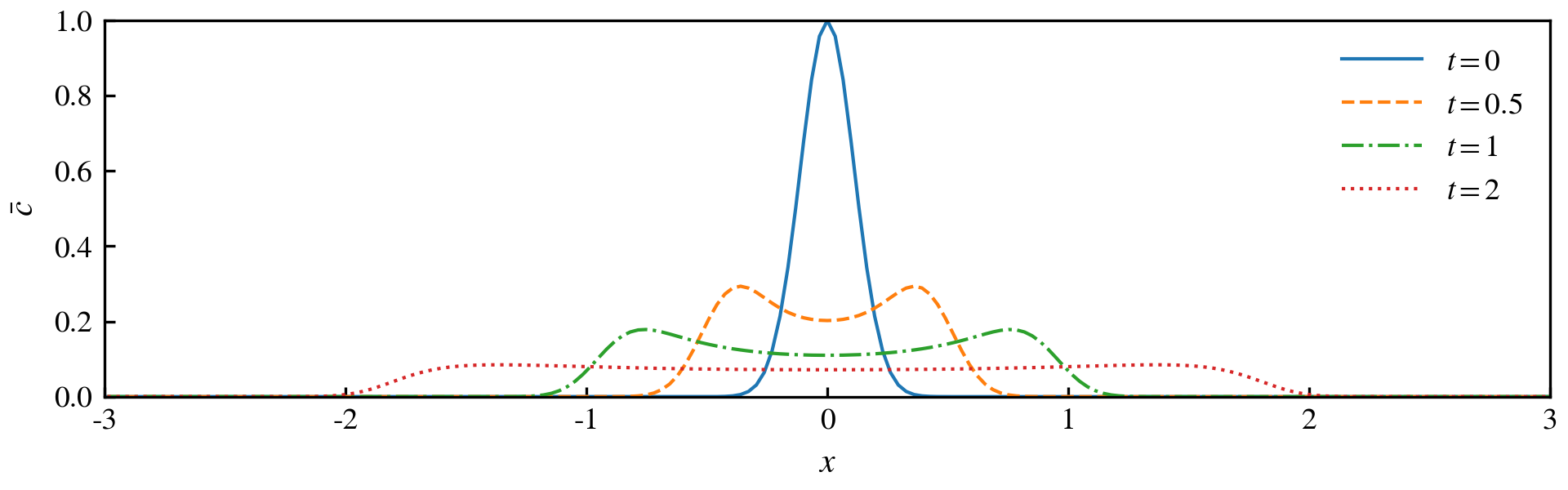}
    \caption{
        Contours of $c(x, y, t)$ and slices of $\bar{c}(x, t)$ with $\bar{c}_0(x) = \exp(-40x^2)$.
    }
    \label{fig:intro:contour}
\end{figure}

The dispersion process of Eq.~\eqref{eq:intro:ADE-HDM-simple} can be solved numerically using the spectral method. Fig.~\ref{fig:intro:dispersion} shows the evolution process of the mean scalar field $\bar c(x,t)$ of a Gaussian initial condition.
Fig.~\ref{fig:intro:contour} shows snapshots of $c(x,y,t)$ and slices of $\bar c(x,t)$.

A question of interest is how to obtain a reduced model for the mean field $\bar c(x,t)$. Averaging the governing equation, Eq.~\eqref{eq:intro:ADE-HDM-simple}, over $y$ yields
\begin{equation}
    \papa{\bar{c}}{t} = - \papa{\overline{u'c'}}{x},
    \label{eq:intro:ADE-HDM-unclosed}
\end{equation}
where the unclosed term $\overline{u'c'}$ requires closure modeling.

In the long-time, large-scale limit, Taylor--Aris theory shows the system behaves as diffusion~\cite{taylor1953dispersion,aris1956dispersion}. Applying the same idea to this flow, the leading-order model can be derived as in Ref.~\cite{mani2021macroscopic}
\begin{equation}
    \papa{\bar{c}}{t} = \frac{1}{2} \papat{\bar{c}}{x}.
    \label{eq:intro:Taylor-model}
\end{equation}

Further improvement over Taylor's model is challenging, as adding higher-order spatial derivatives often leads to worse accuracy~\cite{mani2021macroscopic}. A remarkable improvement is found in~\cite{liu2023systematic}, where the matched moment inverse (MMI) approach is used.

For simplicity, consider a steady problem. In the MMI approach, the nonlocal scalar flux is first approximated via a Taylor series expansion of the nonlocal eddy diffusivity kernel, referred to as the Kramers--Moyal expansion~\cite{moyal1949stochastic}:
\begin{equation} \label{eq:explicit_expansion}
    -\overline{u'c'}(x) = \left[ D^0(x) + D^{1}(x) \frac{d}{dx} + D^{2}(x) \frac{d^2}{dx^2} + \dots \right] \frac{d \bar{c}}{dx},
\end{equation}
where $D^n(x)$ represents the $n$-th spatial moment of the closure kernel~\cite{mani2021macroscopic}. This operator is not guaranteed to converge with adding higher-order terms. To address this issue, MMI approximates the relation using an inverse operator:
\begin{equation} \label{eq:implicit_model}
    \left[ 1 + a_1(x) \frac{d}{dx} + a_2(x) \frac{d^2}{dx^2} \right] (-\overline{u'c'}) = a_0(x) \frac{d \bar{c}}{dx}.
\end{equation}

The coefficients $a_i(x)$ are determined pointwise by a Padé approximation to $D^n(x)$ in spectral space, or equivalently, by matching the Kramers--Moyal expansion of the inverse operator to Eq.~\eqref{eq:explicit_expansion}.

Based on MMI, Liu~\cite{liu2023systematic} derived the following closure model for Eq.~\eqref{eq:intro:ADE-HDM-unclosed}:
\begin{equation}
    \papa{\overline{u'c'}}{t} = - \frac{1}{2}\papa{\bar{c}}{x} - \overline{u'c'} + \frac{1}{16}\papat{\overline{u'c'}}{x}.
    \label{eq:intro:Liu-model}
\end{equation}

Despite the improved accuracy over Taylor's model, the MMI approach has several limitations: (i) extending the moment matching framework to a high-moment model is challenging; (ii) MMI fits the model directly from the kernel moments without exploiting the structure of the governing PDE; and (iii) for inhomogeneous systems, the pointwise fitting of coefficients requires local matrix inversions, which are prone to ill-conditioning.

The proposed GFM framework is developed to address these limitations. It is naturally designed for high-moment models. The resolved variables and the model form are chosen based on physical considerations, and the modeling task is ultimately formulated as a regularized linear regression problem.

To motivate the advantages and benefits of higher-moment closures, we first derive a hierarchy of analytical solutions for Eq.~\eqref{eq:intro:ADE-HDM-simple}.

\subsection{\label{sec:intro-analytical}An analytical closure family}

The dominant balance in Eq.~\eqref{eq:intro:ADE-HDM-simple} is between advection and diffusion. From a Fourier-in-$y$ perspective, the advection term $u \papa{c}{x}$ introduces mixing across the modes, while the diffusion term $\papat{c}{y}$ damps the $k$-th mode at rate $k^2$. This motivates taking the low-$k$ modes as resolved variables and approximating the high-$k$ modes from the quasi-steady assumption.

Define the cosine and sine moments of $c(x,y,t)$ as
\begin{align}
    h_k(x,t) = \overline{\cos(ky)c(x,y,t)}, \quad
    g_k(x,t) = \overline{\sin(ky) c(x,y,t)}. \label{eq:intro:hn-gn-basis}
\end{align}

In particular, $h_0=\bar c$. With these definitions, $c$ can be expanded as
\begin{equation}
    c(x,y,t) = h_0(x,t) + 2 \sum_{n=1}^{\infty} \left( h_n(x,t) \cos(ny) + g_n(x,t) \sin(n y) \right).
\end{equation}

For $u = \cos y$ and a $y$-independent initial condition, $c$ has no sine modes (i.e., $g_k\equiv 0$ for all $k$). Thus the resolved variables are chosen as the first $n$ cosine modes $\{h_k\}_{k=0}^{n-1}$.

Projecting Eq.~\eqref{eq:intro:ADE-HDM-simple} onto $\cos(ky)$ yields the exact evolution equations
\begin{align}
    \papa{h_0}{t} &=  - \papa{}{x}h_{1},\\
    \papa{h_k}{t} &=  - \frac12 \papa{}{x}(h_{k-1} + h_{k+1}) - k^2 h_k, \quad 1 \leq k \leq n-1.
\end{align}

\begin{figure}[tbp]
    \centering
    \includegraphics[width=\textwidth]{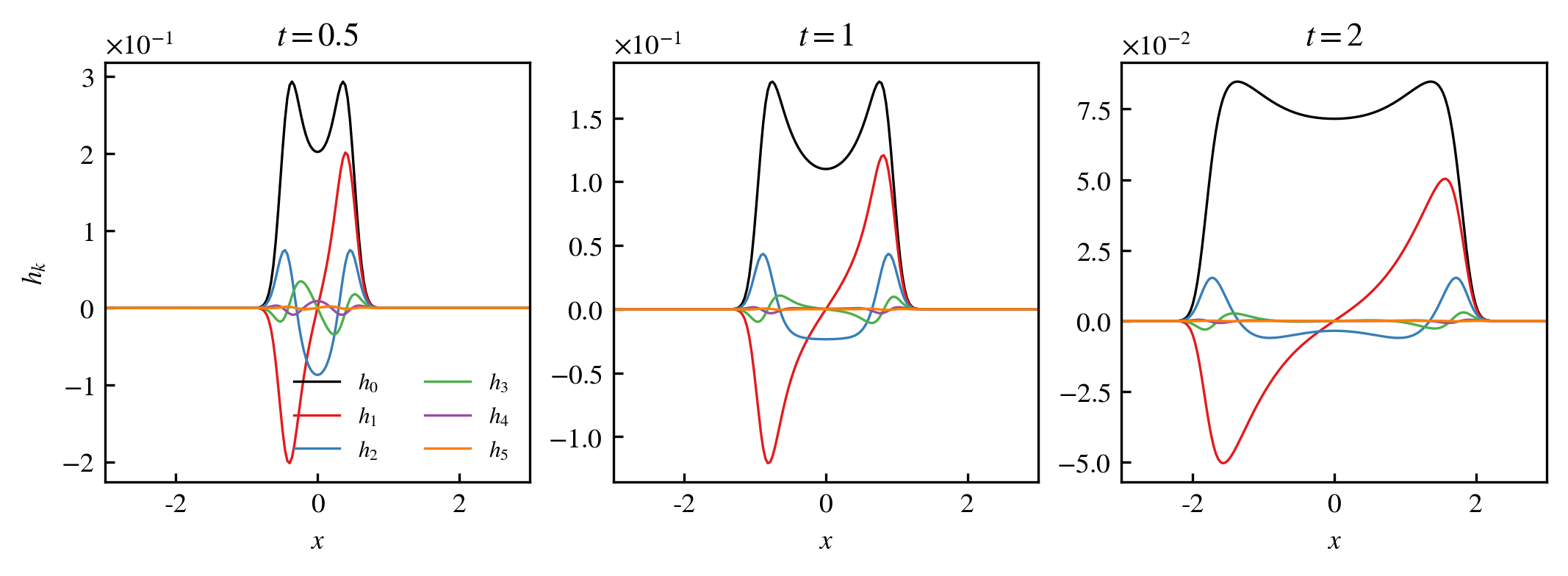}
    \caption{Profiles of the first six cosine modes, $h_0,\ldots,h_5$, for Eq.~\eqref{eq:intro:ADE-HDM-simple} at $t=0.5$, $1$, and $2$.}
    \label{fig:intro:spectral-profiles}
\end{figure}

Figure~\ref{fig:intro:spectral-profiles} shows the first six cosine modes for the same simulation as Fig.~\ref{fig:intro:contour}. The lower modes have larger magnitudes at all three times, while $h_4$ and $h_5$ remain small. This trend supports the use of a low-order spectral truncation.

To close the system, we model $h_n$ by the quasi-steady approximation.
Starting from the exact equation
\begin{equation}
    \papa{h_n}{t}
    = - \frac12\,\papa{}{x}\!\big(h_{n-1} + h_{n+1}\big) - n^2 h_n,
    \label{eq:intro:exact-hn}
\end{equation}

We assume $\partial_t h_n$ is negligible as a fast decaying mode, and the transport term is dominated by the resolved mode $h_{n-1}$.
This gives the algebraic closure
\begin{equation}
    h_n = -\frac{1}{2n^2}\,\papa{}{x} h_{n-1}.
\end{equation}

Since $h_0=\overline{c}, h_1 = \overline{u'c'}$, for $n=1$, we recover the classical Taylor model, Eq.~\eqref{eq:intro:Taylor-model},
\begin{equation}
    \papa{h_0}{t} = \frac{1}{2}\papat{h_0}{x},
\end{equation}
and for $n=2$, we recover Liu's two-equation model, Eq.~\eqref{eq:intro:Liu-model},
\begin{equation}
    \papa{h_0}{t} = - \papa{h_1}{x}, \qquad\papa{h_1}{t} = - \frac{1}{2}\papa{h_0}{x} - h_1 + \frac{1}{16}\papat{h_1}{x}.
\end{equation}

For $n \ge 3$, the closed $n$-equation model is given by
\begin{align}
    \papa{h_0}{t} &=  - \papa{}{x} h_{1},\\
    \papa{h_k}{t} &=  - \frac12 \papa{}{x}(h_{k-1} + h_{k+1}) - k^2 h_k, \quad 1 \leq k \leq n-2,\\
    \papa{h_{n-1}}{t} &=  - \frac12 \papa{}{x} h_{n-2}  - (n-1)^2 h_{n-1} + \frac{1}{4n^2} \papat{h_{n-1}}{x}.
    \label{eq:intro:n-equation-model}
\end{align}

\begin{figure}[tbp]
    \centering
        \includegraphics{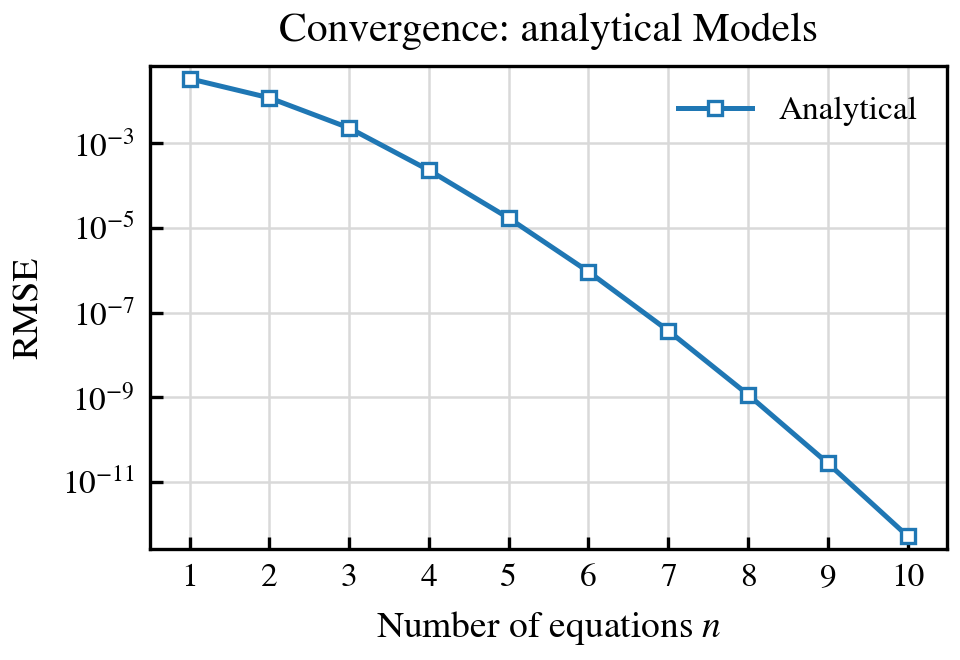}
    \caption{
        Convergence of the analytical model family for the problem in Eq.~\eqref{eq:intro:ADE-HDM-simple} with increasing number of equations, measured by root-mean-square error (RMSE) over $t\in[0,2]$ with initial condition $\bar{c}_0(x)$.
    }
    \label{fig:intro:rmse-convergence}
\end{figure}

Convergence of this analytical model family is shown in Fig.~\ref{fig:intro:rmse-convergence}.

\subsection{\label{sec:intro-forcing}Data-driven modeling framework}

While the analytical family in Eq.~\eqref{eq:intro:n-equation-model} offers theoretical clarity, closed-form derivations are generally intractable for most PDEs. Consequently, our objective is to identify an equivalent reduced model from data.

Constructing a physics-based, explicit, and interpretable closure requires a selection of resolved variables and a model form. Guided by the spectral hierarchy established above, we select the first $n$ resolved modes from Eq.~\eqref{eq:intro:n-equation-model} as our primary basis, collecting them into resolved variables:
\begin{equation}
\bm{f} = [h_0, \ldots, h_{n-1}]^T.
\end{equation}

These variables are linear functions of the full field $c$, which can be denoted compactly as
\begin{equation}
    \bm f = \mathcal M c,
\end{equation}
where $\mathcal M$ is a linear operator.

We propose a linear evolution model for the reduced dynamics:
\begin{equation}
    \frac{\partial \bm{f}}{\partial t} = \bm{A} \bm{f} + \bm{B} \frac{\partial \bm{f}}{\partial x} + \bm{C} \frac{\partial^2 \bm{f}}{\partial x^2} + \dots,
    \label{eq:intro:linear-pde-ansatz}
\end{equation}
where $\bm{A}, \bm{B}, \bm{C} \in \mathbb{R}^{n \times n}$ are constant coefficient matrices to be identified from data.

Guided by the analytical hierarchy in Eq.~\eqref{eq:intro:n-equation-model} and the need for numerical robustness, we truncate the expansion at the second-order spatial derivative. This restricts the closure to a coupled advection-diffusion system, ensuring that the model remains physically interpretable and numerically well-posed.

It is worth noting that the additional resolved variables $h_1,\dots,h_{n-1}$ can also be interpreted as latent variables in machine-learning terminology. From this perspective, the PDE ansatz in Eq.~\eqref{eq:intro:linear-pde-ansatz} is a physics-based reduced model with strong structural constraints and a small number of parameters. In future work, this ansatz could be extended to more flexible neural-network-based models. Such models could be trained using the data-generation strategy introduced in Sec.~\ref{sec:gfm}.

To generate fitting data, we consider a forced system~\cite{mani2021macroscopic}
\begin{equation}
    \papa{c^{(i)}}{t} = \mathcal{L} c^{(i)} + s^{(i)}, \qquad s^{(i)} \in S, \qquad i = 1, \dots, N_s
    \label{eq:intro:forced_HDM}
\end{equation}
where $N_s$ is the number of forcing realizations and the forcing set $S$ is to be designed.

Traditional closure modeling requires deriving the exact evolution of $\bm f$, identifying the unclosed terms, and modeling them term by term. In contrast, we adopt a simpler strategy that fits a complete reduced model in a single regression. From $\bm{f} =\mathcal M c$, its evolution in the forced system can be written abstractly as
\begin{equation}
    \papa{\bm f}{t} = \mathcal M(\mathcal L c) + \left(\papa{\mathcal M}{t}\right) c + \mathcal M s.
\end{equation}

Notice that the forcing contribution $\mathcal M s$ is introduced only to probe the dynamics; it does not need to be modeled since it is absent in the original system. Accordingly, the term is subtracted from the model target, and the quantity to be modeled is
\[
\papa{\bm f}{t} - \mathcal M s.
\]

This combined fitting strategy avoids cumbersome algebraic manipulations, since one does not need to postprocess the right-hand side term-by-term. Under ordinary least squares, it is equivalent to the traditional term-by-term fitting procedure. Since the least-squares fitting is a projection onto the model basis, both approaches will recover the same coefficients. When constraints are considered (e.g. mass conservation of certain flux-like terms), the two approaches may differ slightly. Nevertheless, the combined approach will not impose unphysical behavior as the constraints are still enforced globally (e.g. global conservation instead of term-wise conservation). The combined fitting approach is adopted throughout this paper for its simplicity, and its implementation is discussed in Sec.~\ref{sec:extension-inhomogeneous}.

After running $N_s$ forced simulations, we assemble the data into a least-squares problem based on Eq.~\eqref{eq:intro:linear-pde-ansatz}:
\begin{equation}
    \widehat{\bm A}, \widehat{\bm B}, \widehat{\bm C}
    =
    \arg\min_{\bm A, \bm B, \bm C} \sum_{i=1}^{N_s} \left\| \frac{\partial \bm{f}^{(i)}}{\partial t} - \mathcal M s^{(i)} - \bm{A} \bm{f}^{(i)} - \bm{B} \frac{\partial \bm{f}^{(i)}}{\partial x} - \bm{C} \frac{\partial^2 \bm{f}^{(i)}}{\partial x^2}   \right\|_F^2. \label{eq:intro:LS-problem}
\end{equation}
where $\|\cdot\|_F$ is the Frobenius norm and $\widehat{\bm A}, \widehat{\bm B}, \widehat{\bm C}$ are the approximated model coefficients.

The critical question is how the forcing space $S$ in Eq.~\eqref{eq:intro:forced_HDM} should be designed to ensure that the data-driven model recovers the target $n$-equation closure in Eq.~\eqref{eq:intro:n-equation-model}.
On the one hand, the forcing must be sufficiently diverse so that the model terms (i.e. $\bm{f}, \frac{\partial \bm{f}}{\partial x}, \frac{\partial^2 \bm{f}}{\partial x^2} $) are linearly independent and the regression is well conditioned.
On the other hand, it should not drive the system into regimes that differ from those of the unforced system.
These requirements motivate a systematic analysis of the forced system and lead to the generalized forcing method developed in Sec.~\ref{sec:gfm}.

%% file: sections/2.GFM.tex
\section{\label{sec:gfm}Generalized Forcing Method}

The Generalized Forcing Method (GFM) is a framework for generating datasets for reduced-order modeling of PDE systems. In this work, we specifically consider linear PDEs. The central idea is to introduce a forcing term that actively excites the resolved variables. By exciting and sustaining these dominant modes, this method ensures that the resulting dynamics span the model input space to form a well-conditioned regression problem.

We present the general formulation for an abstract linear evolution equation. To make the derivation intuitive, we use the scalar advection-diffusion problem from Sec.~\ref{sec:intro-analytical} as an example for illustration.

\subsection{\label{sec:gfm-fundamentals}Fundamentals}

We adopt a formal engineering perspective, assuming all fields and operators are sufficiently regular to ensure well-defined manipulations, without addressing functional-analytic subtleties. For simplicity, we consider a periodic 2D domain, though the framework can be extended to more general geometries.

\paragraph{Definition (PDE system).} Let $X=[0,L_x)$ and $Y=[0,L_y)$ be periodic spatial intervals, $T = [0,\infty)$ be the time axis. Let $V$ denote the space of spatial periodic scalar fields on $X\times Y \times T$. Consider a linear evolution equation for a field $c$ with periodic boundary conditions:
\begin{align}
\frac{\partial c}{\partial t} &= \mathcal{L}c + s, \qquad c(x,y,0)=c_0(x,y), \qquad c, s \in V, \label{eq:gfm:ADE-HDM}
\end{align}
where $\mathcal{L}$ is a linear spatial operator (possibly time-dependent), and $s$ is the forcing.

We use the notation $c(t)$ to represent the spatial field at time $t$, and the solution can be written in the Duhamel form:
\begin{equation}
c(t) = (\mathcal{A}s)(t) + (\mathcal{B}c_0)(t), \label{eq:gfm:Duhamel-representation}
\end{equation}
where
\begin{align}
(\mathcal{A}s)(t) &:= \int_{0}^{t} \mathcal{U}(t,\tau)\, s(\tau)\, d\tau,\\
(\mathcal{B}c_0)(t) &:= \mathcal{U}(t,0)\, c_0,
\end{align}
and $\mathcal{U}(t,\tau)$ is the evolution family satisfying
\begin{equation}
\papa{\mathcal{U}(t,\tau)}{t}=\mathcal{L}(t)\,\mathcal{U}(t,\tau),\qquad \mathcal{U}(\tau,\tau)= \mathcal{I},
\end{equation}
where $\mathcal{I}$ is the identity operator. For time-independent $\mathcal L$, $\mathcal U(t,\tau)=e^{(t-\tau)\mathcal L}$.

From the engineering viewpoint, access to the full solution $c(x,y,t)$ is often not needed; instead projecting $c$ to the mean field provides quantities of interest. In this example, consider the cross-sectional average
\[
\bar{c}(x,t)=\frac{1}{L_y}\int_Y c(x,y,t)\,dy,
\]
as the quantity of interest.

The exact evolution equation for $\bar c$ ($\partial_t\bar c = \overline{\mathcal{L}c}$) is unclosed and needs closure modeling; however, a closure model expressed solely in terms of $\bar{c}$ is typically insufficient to characterize the dynamics. Often, higher moments of $c$ are needed as additional degrees of freedom. We introduce the following $Y$-inner product to define these variables by
\begin{equation}
    \langle a,\,b\rangle_Y(x,t) := \frac{1}{L_y}\int_Y a(x,y,t)\,b(x,y,t)\,dy, \qquad a, b \in V. \label{eq:gfm:inner-product}
\end{equation}
In particular, $\bar c=\langle 1,c\rangle_Y$. As another example, $\overline{u'c'}$ introduced in Eq.~\eqref{eq:intro:ADE-HDM-unclosed} can be written as $\langle u',c'\rangle_Y$.

While in PDEs involving advective transport, it is common to define resolved variables of the reduced model based on velocity moments, in what follows, we keep the definition of resolved variables more general without committing to specific moments of $c$.

\paragraph{Definition (Resolved variables).} Let $W$ denote the space of spatial periodic $n$-dimensional vector fields on $X \times T$. Given weight functions $\{\phi_i(x,y,t)\}_{i=0}^{n-1}$ that are linearly independent in the $Y$-inner product, we define the resolved variables:
    \begin{equation}
        \bm{f} = [f_0, \dots, f_{n-1}]^T, \qquad f_i = \langle \phi_i, c \rangle_Y, \qquad \bm{f} \in W.
    \end{equation}
    Equivalently, we introduce the operator $\mathcal M:V\to W$ by
    \begin{equation}
    \bm f = \mathcal M c.
    \end{equation}

For further analysis on the forcing, since $s$ belongs to the full state space $V$, we relate the resolved variables $\bm{f}$ back to $V$.
In doing so, it is convenient to decompose the full state space $V$ into two complementary subspaces, one preserved in the reduced space, while the other eliminated.
To obtain this decomposition, we first construct a right-inverse operator of $\mathcal{M}$ using a biorthogonal basis. Let
$G_{ij}(x,t):=\langle \phi_i,\phi_j\rangle_Y (x,t)$
be the Gram matrix. From linear independence of $\{\phi_i\}_{i=0}^{n-1}$, $G$ is invertible as an $n\times n$ matrix at each $(x,t)$. We define the dual basis functions
\begin{equation}
    \psi_i(x,y,t) := \sum_{j=0}^{n-1} (G^{-1})_{ij}(x,t)\,\phi_j(x,y,t),
\end{equation}
which satisfy the biorthogonality relation $\langle \phi_i,\,\psi_j\rangle_Y(x,t) = \delta_{ij}$.

\paragraph{Definition (Right-inverse operator and projectors).}
    We define $\mathcal{M}^{+}: W \to V$ such that
    \begin{equation}
        (\mathcal{M}^{+}\bm{f})(x,y,t) := \sum_{i=0}^{n-1} f_i(x,t)\, \psi_i(x,y,t).
    \end{equation}
    This operator satisfies $\mathcal{M} \mathcal{M}^{+} = \mathcal{I}_W$. By construction, the composition $\mathcal{M}^+\mathcal{M}$ is a projector on $V$. We define the resolved/unresolved projectors as
    \begin{equation}
    \mathcal P_r:=\mathcal M^+\mathcal M,\qquad
    \mathcal P_u:=\mathcal I_V-\mathcal P_r.
    \end{equation}
    where $\mathcal I_V$ is the identity operator on $V$.
    They induce the decomposition of the full state space into resolved and unresolved subspaces
    \begin{equation}
    V=V_r\oplus V_u,\qquad
    V_r:=\mathrm{Range}(\mathcal P_r),\qquad
    V_u:=\mathrm{Range}(\mathcal P_u)=\mathrm{Ker}(\mathcal M).
    \end{equation}

Accordingly, any field $v \in V$ admits a unique decomposition:
\begin{equation}
    v = v_r + v_u, \qquad v_r = \mathcal P_r v, \qquad v_u = \mathcal P_u v.
\end{equation}

By construction, the resolved subspace $V_r$ is the span of the dual basis functions $\{\psi_i\}_{i=0}^{n-1}$ with varying coefficients on $X \times T$. Since $\{\psi_i\}_{i=0}^{n-1}$ are linear combinations of $\{\phi_i\}_{i=0}^{n-1}$, the subspace $V_r$ can equivalently be characterized as:
\begin{align}
    V_r = \left\{ \sum_{i=0}^{n-1} \alpha_i(x,t)\,\phi_i(x,y,t) \ : \ \alpha_i \text{ are arbitrary scalar fields on $X \times T$}  \right\}. \label{eq:gfm:Vr-basis}
\end{align}

\paragraph{Example (Scalar advection-diffusion with $u=\cos(y)$).}

    In the scalar advection-diffusion problem discussed in Sec.~\ref{sec:intro-forcing}, we select the first $n$ cosine modes as the weight functions, $\phi_k(x,y,t) = \cos(ky)$. Then the operator $\mathcal{M}$ extracts the first $n$ cosine moments:
    \begin{equation}
        \mathcal{M}c = \begin{bmatrix}
            \overline{c} \\
            \overline{c \cos(y)} \\
            \vdots \\
            \overline{c \cos((n-1)y)}
        \end{bmatrix} = \bm{f}.
    \end{equation}

    The Gram matrix is diagonal with $G_{0,0}=1$ and $G_{k,k}=\frac12$ for $k \ge 1$; thus, the dual basis functions are simply rescalings of the weight functions:
    \begin{equation}
        \psi_0(x,y,t) = 1, \qquad \psi_k(x,y,t) = 2\cos(ky) \quad \text{for } k=1,\ldots,n-1.
    \end{equation}

    The operator $\mathcal{M}^{+}$ recovers $\bm{f}$ to the truncated cosine series expansion:
    \begin{equation}
        (\mathcal{M}^{+} \bm{f})(x,y,t) = c_r(x,y,t) = f_0(x,t) + 2\sum_{k=1}^{n-1} f_k(x,t) \cos(ky).
    \end{equation}

    The resolved projector $\mathcal P_r$ extracts the first $n$ cosine modes:
    \begin{equation}
        (\mathcal P_r c)(x,y,t) = \bar{c}(x,t) + 2\sum_{k=1}^{n-1} \overline{c\cos(ky)}(x,t) \cos(ky).
    \end{equation}
    while the unresolved projector $\mathcal P_u$ captures the high-frequency cosine modes and all the sine modes (which are not chosen as weight functions):
    \begin{equation}
        (\mathcal P_u c)(x,y,t) = 2\sum_{k=n}^{\infty} \overline{c\cos(ky)}(x,t) \cos(ky) + 2\sum_{k=1}^{\infty} \overline{c\sin(ky)}(x,t) \sin(ky).
    \end{equation}

    From the above derivation, the resolved subspace $V_r$ can be expressed as:
    \begin{equation}
        V_r = \left\{ \sum_{k=0}^{n-1} \alpha_k(x,t) \cos(ky) \middle| \alpha_k \text{ arbitrary} \right\}.
        \label{eq:gfm:Vr-basis-illustrative}
    \end{equation}

With the above definitions, we proceed to analyze the forced system in Eq.~\eqref{eq:gfm:ADE-HDM}.
We first demonstrate that the closure terms not only depend on the resolved variables $\bm{f}$, but they also depend on the initial condition $c_0$ and the unresolved forcing $s_u$.

Recall the Duhamel form solution $c = \mathcal{A}s + \mathcal{B}c_0$. Applying projectors $\mathcal P_r$ and $\mathcal P_u$ to the variables and operators, and introducing the block operators $\mathcal{A}_{\alpha\beta} := \mathcal{P}_\alpha \mathcal{A} \mathcal{P}_\beta$ and $\mathcal{B}_{\alpha} := \mathcal{P}_\alpha \mathcal{B}$ for $\alpha,\beta \in \{r,u\}$, we obtain the decomposed equations:
\begin{align}
    c_r &= \mathcal{A}_{rr}s_r + \mathcal{A}_{ru}s_u + \mathcal{B}_r c_0, \label{eq:gfm:macroscopic-solution}\\
    c_u &= \mathcal{A}_{ur}s_r + \mathcal{A}_{uu}s_u + \mathcal{B}_u c_0. \label{eq:gfm:fluctuating-solution}
\end{align}

We assume that the resolved forcing produces non-zero production in the resolved subspace, which implies that $\mathcal{A}_{rr}$ admits a right-inverse $\mathcal{A}_{rr}^{+}$ on $V_r$. Solving Eq.~\eqref{eq:gfm:macroscopic-solution} for $s_r$ yields:
\begin{equation}
    s_r = \mathcal{A}_{rr}^{+}(c_r - \mathcal{A}_{ru}s_u - \mathcal{B}_r c_0). \label{eq:macroscopic-forcing}
\end{equation}
Substituting this expression into Eq.~\eqref{eq:gfm:fluctuating-solution} and using $c = c_r + c_u$, we can express $c$ as:
\begin{equation}
    c = \mathcal{A}_1 c_r + \mathcal{A}_2 s_u + \mathcal{A}_3 c_0, \label{eq:gfm:reconstructed-solution}
\end{equation}
where the operators are defined as:
\begin{align}
    \mathcal{A}_1 &:= I + \mathcal{A}_{ur}\mathcal{A}_{rr}^{+}, \\
    \mathcal{A}_2 &:= \mathcal{A}_{uu} - \mathcal{A}_{ur}\mathcal{A}_{rr}^{+}\mathcal{A}_{ru}, \\
    \mathcal{A}_3 &:= (I-\mathcal{A}_1\mathcal P_r)\mathcal{B}.
\end{align}

Let $\bm{g} := \mathcal{N}c$ denote a vector of unclosed terms to be modeled, where $\mathcal{N}$ is a linear operator. Using the relation $c_r = \mathcal{P}_r c =\mathcal{M}^{+} \mathcal{M} c =\mathcal{M}^{+}\bm{f}$ and the reconstruction in Eq.~\eqref{eq:gfm:reconstructed-solution}, the exact closure relation is:
\begin{equation}
    \bm{g} = \mathcal{N}\mathcal{A}_1\mathcal{M}^{+}\bm{f} + \mathcal{N}\mathcal{A}_2 s_u + \mathcal{N}\mathcal{A}_3 c_0.
\end{equation}

This equation reveals that the target closure relation is coupled with terms depending on the unresolved forcing $s_u$ and the initial condition $c_0$. To isolate the closure operator between $\bm{f}$ and $\bm{g}$, we restrict the forcing to the resolved subspace ($s_u \equiv 0$) and initialize from $c_0 \equiv 0$. Then there exists an exact linear map between $\bm{g}$ and $\bm{f}$ which can be approximated from data.

\subsection{\label{sec:gfm-implicit-explicit}Implicit and explicit GFM}

To accurately approximate the closure operator, the datasets must sufficiently sample the mapping between $\bm{f}$ and $\bm{g}$. We propose two strategies to construct forcings.

\paragraph{iGFM: Enforce a prescribed trajectory of the resolved variables.}

A straightforward approach is to prescribe linearly independent trajectories directly to the model input. Let $\bm{f}(x,t)$ be a pre-specified target trajectory in the reduced space $W$. We first reconstruct the corresponding resolved field:
\begin{equation}
    c_r(x,y,t) = (\mathcal{M}^+ \bm{f})(x,y,t), \quad c_r \in V_r.
\end{equation}

From the decomposition $c=c_r+c_u$, using $c_u = \mathcal{P}_u c$ and Eq.~\eqref{eq:gfm:ADE-HDM}, we construct the evolution equation for the full state:
\begin{equation}
    \frac{\partial c}{\partial t} = \frac{\partial c_r}{\partial t} + \mathcal P_u\,\mathcal{L} c + \left(\frac{\partial \mathcal P_u}{\partial t}\right) c. \label{eq:gfm:igfm-evolution}
\end{equation}
where we use the fact that $\mathcal{P}_u s = s_u = 0$.

Once the full state $c$ is obtained, the required forcing is recovered from:
\begin{equation}
    s = \frac{\partial c}{\partial t} - \mathcal{L}c.
\end{equation}

Since the forcing is determined implicitly by $\mathcal{M} c= \bm{f}$, we call this \emph{implicit GFM} (\textbf{iGFM}).
Theoretically, a single iGFM simulation can be sufficient to fit a full model. However, it has two practical drawbacks: (i) if $\mathcal{M}$ is time-dependent, evaluating $\partial_t \mathcal P_u$ can be very expensive; and (ii) relying on a single realization may lead to an inaccurate and overfit model.

\paragraph{Example (Scalar advection-diffusion with $u=\cos(y)$).}
    In the scalar advection-diffusion problem of Sec.~\ref{sec:intro-forcing}, to guarantee the unique solvability of the coefficients in the regression problem in Eq.~\eqref{eq:intro:LS-problem}, we are required to design $\bm{f}$ such that the model input terms $\{\bm{f}, \partial_x\bm{f}, \partial_{xx}\bm{f} \}$ are linearly independent.

    For each model variable $f_i(x,t)$, if $f_i$ contains only one spatial Fourier mode, the $\partial_{xx}f_i$ is proportional to $f_i$. Therefore, each $f_i$ requires at least two spatial modes. A simple time-dependent trajectory can be chosen as
    \begin{equation}
        f_i(x,t) = \left[\cos \left((i+n+1)\frac{2\pi x}{L_x} \right) + \cos \left((i+1) \frac{2\pi x}{L_x} \right) \right] (1 - e^{-t}), \qquad i = 0, \ldots, n-1.
    \end{equation}

\paragraph{eGFM: Explicit forcing basis.}
In contrast, \emph{explicit GFM} (\textbf{eGFM}) adopts a direct approach by choosing a list of forcings $\{s_r^{(k)}\}$ in the resolved subspace $V_r$ and using them to drive independent simulations:
\begin{equation}
    \frac{\partial c^{(k)}}{\partial t} = \mathcal{L}c^{(k)} + s_r^{(k)}.
\end{equation}

eGFM typically requires multiple simulations to ensure linear independence of the model terms. However, these runs are embarrassingly parallel, and the procedure naturally adapts to time-dependent $\mathcal{M}$.

\paragraph{Example (Scalar advection-diffusion with $u=\cos(y)$).}
    Given the structure of $V_r$ in Eq.~\eqref{eq:gfm:Vr-basis-illustrative}, we construct the forcing basis by expanding the coefficients $\alpha_k(x,t)$ using a cosine basis in space and time:
    \begin{equation}
        \begin{aligned}
        S = \Biggl\{& \cos \left(k_1 \frac{2\pi x}{L_x} \right) \cos \left(k_2 \frac{2\pi y}{L_y} \right) \cos (\omega t), \\
        &0\leq k_1 < x_{modes}, \quad 0\leq k_2 < y_{modes}, \quad 0\leq \omega < t_{modes} \Biggr\}.
        \end{aligned}
    \end{equation}

    The condition $S \subset V_r$ strictly requires $y_{modes} \leq n$. For the spatial modes, we select wavenumbers $k_1$ corresponding to the length scales of interest. For the temporal modes, we select frequencies $\omega$ such that the forcing time scale is larger than the system's characteristic time scale. This excites the target dynamics while avoiding unphysical perturbations.

\paragraph{Summary.}
Two different forcing strategies are proposed. iGFM serves as an $n$-equation extension to the inverse MFM used in MMI~\cite{liu2023systematic} for identifying model forms, and eGFM is designed for coefficient calibration after determining the model form.
Both approaches fulfill the GFM requirement (ensuring $s_u = 0, c_0 = 0$), thereby avoiding introducing unphysical behaviors.
In Secs.~\ref{sec:parallel}, \ref{sec:extension}, and \ref{sec:turbulence}, we evaluate eGFM on a range of scalar advection-diffusion problems.

%% file: sections/3.parallel_flow.tex
\section{\label{sec:parallel}General parallel-flow dispersion}

We now apply the GFM framework developed in Sec.~\ref{sec:gfm} to the prototype dispersion problem proposed in Sec.~\ref{sec:intro-forcing}. More generally, we consider the scalar transport problem under a general periodic parallel shear flow. The governing equation is
\begin{equation}
    \frac{\partial c}{\partial t} + u \frac{\partial c}{\partial x} = \frac{\partial^2 c}{\partial y^2}, \qquad u(x,y,t)=u(y),
    \label{eq:parallel:ADE-HDM}
\end{equation}
with the initial condition $c(x,y,0)=\bar c_0(x)$. We refer to this as the full model.

Consider an $n$-equation closure. Given a set of properly defined resolved variables $\bm{f}$, we seek a reduced model of the form
\begin{equation}
    \frac{\partial \bm{f}}{\partial t} = \bm{A} \bm{f} + \bm{B} \frac{\partial \bm{f}}{\partial x} + \bm{C} \frac{\partial^2 \bm{f}}{\partial x^2},
    \label{eq:parallel:ADE-RM}
\end{equation}
where $\bm{A}, \bm{B}, \bm{C} \in \mathbb{R}^{n \times n}$ are constant coefficient matrices. The matrix $\bm A$ describes local coupling and relaxation, $\bm B$ describes transport effects, and $\bm C$ describes diffusion effects. We truncate the model at the second-order spatial derivative to avoid numerical stiffness.

We also enforce mass conservation in the reduced model, which requires the first row of $\bm{A}$ to be zero:
\begin{equation}
\bm A_{1k}=0, \qquad 1 \leq k \leq n.
\label{eq:parallel:mass-constraint}
\end{equation}

For this problem with general $u(y)$, we first extend the methodology in Sec.~\ref{sec:intro-analytical} to derive a family of analytical closure models, which serve as analytical benchmarks. Then, in applying the GFM framework, we consider two choices of resolved variables and fit the corresponding reduced models using datasets generated by eGFM forcings. Finally, for a specific test case, we compare the performance of the eGFM-based reduced models with the full model solution, the analytical reduced models, and reduced models fitted directly from the test data.

\subsection{\label{sec:parallel-analytical}An analytical closure family}

We extend the analytical derivation in Sec.~\ref{sec:intro-analytical} to a general periodic profile $u(y)$ on $[0, 2\pi]$. We assume that $u$ is well approximated by a truncated Fourier expansion,
\begin{equation}
	u(y) \approx \sum_{k=0}^M a_k \cos(ky) + \sum_{k=1}^M b_k \sin(ky).
	\label{eq:velocity-truncation}
\end{equation}

We represent $c$ with the Fourier coefficients $h_k(x,t), g_k(x,t)$ as defined in Eq.~\eqref{eq:intro:hn-gn-basis}. Substituting Eq.~\eqref{eq:velocity-truncation} into Eq.~\eqref{eq:parallel:ADE-HDM} and projecting onto cosine and sine modes yields the coupled equations
\begin{align}
	\papa{h_k}{t} &=
	- \frac12 \papa{}{x}\left( \sum_{j=0}^M a_j (h_{k+j} + h_{k-j}) + \sum_{j=1}^M b_j (g_{k+j} - g_{k-j})  \right) - k^2 h_k, \qquad k \ge 0
	\label{eq:spectral-h-equation}\\
	\papa{g_k}{t} &=
	- \frac12  \papa{}{x}\left( \sum_{j=1}^M b_j (h_{k-j} - h_{k+j}) + \sum_{j=0}^M a_j (g_{k+j} + g_{k-j}) \right) - k^2 g_k, \qquad k \ge 1.
	\label{eq:spectral-g-equation}
\end{align}
Here we use the standard symmetry conventions for negative indices:
$g_0\equiv 0$, $h_{-k}=h_k$, and $g_{-k}=-g_k$.

To construct an \(n\)-equation reduced model, we retain the low Fourier modes as resolved variables. In general, we keep
\begin{equation}
	h_0, \ldots, h_{n_1}, \quad g_1, \ldots, g_{n_2}, \quad n_1 + n_2 = n-1.
\end{equation}

The choice of $n_1$ and $n_2$ depends on the symmetry of $u$. If $u$ is symmetric, i.e.\ $u(y)=u(-y)$, then $b_k=0$ for all $k$. The $y$-independent initial condition implies $g_k(t=0) = 0$, and Eq.~\eqref{eq:spectral-g-equation} implies $g_k(t)\equiv 0$ for symmetric $u$. In this case we shall take $n_1 = n-1, n_2=0$. If \(u\) is not symmetric, we choose the number of cosine modes and sine modes almost equal by setting $n_2=\lfloor n/2\rfloor$, $n_1=n-n_2-1$. We refer to this choice of variables as spectral-based variables, and their corresponding model as spectral-based models.

Following the quasi-steady assumptions used in Sec.~\ref{sec:intro-analytical}, the unclosed terms
\begin{equation}
    h_{n_1+j}, \quad g_{n_2+j}, \quad 1 \leq j \leq M,
\end{equation}
in Eqs.~\eqref{eq:spectral-h-equation} and~\eqref{eq:spectral-g-equation} are approximated from the resolved modes as
\begin{align}
	h_{n_1+j} &= \frac{-1}{2(n_1+j)^2} \papa{}{x} \left(\sum_{\substack{0 \le k \le M \\ |n_1+j-k| \le n_1}}  a_k h_{n_1+j-k} -  \sum_{\substack{1 \le k \le M \\ |n_1+j-k| \le n_2  }}  b_k g_{n_1+j-k} \right),\label{eq:closure-h}\\
	g_{n_2+j} &= \frac{-1}{2(n_2+j)^2} \papa{}{x} \Bigg( \sum_{\substack{1 \le k \le M \\ |n_2+j-k| \le n_1}} b_k h_{n_2+j-k} \notag\\
	&\qquad\qquad + \sum_{\substack{0 \le k \le M \\ |n_2+j-k| \le n_2}} a_k g_{n_2+j-k} \Bigg).\label{eq:closure-g}
\end{align}

Substituting Eqs.~\eqref{eq:closure-h} and~\eqref{eq:closure-g} into Eqs.~\eqref{eq:spectral-h-equation} and~\eqref{eq:spectral-g-equation} yields a closed reduced model, which we refer to as the analytical model.

After deriving the analytical closure family as reference models, we now specify the two types of resolved variables used in the GFM-based modeling procedure.

\subsection{\label{sec:parallel-problem}Resolved variables}

Motivated by the analytical derivation above and the classic ideas in advective-transport modeling, we consider two different definitions of the resolved variables.

\paragraph{Spectral-based variables.}

For an $n$-equation model, after defining the spectral-based variables in Sec.~\ref{sec:parallel-analytical}, they are collected into $\bm{f}^s = [f_0^s, \ldots, f_{n-1}^s]^T$, where the superscript \(s\) denotes spectral-based variables.

Precisely, for symmetric $u$, we define
\begin{equation}
	f_k^s = h_k, \quad k = 0, \ldots, n-1.\label{eq:parallel:spectral-var-cos}
\end{equation}

For asymmetric $u$, we define
\begin{align}
	f_{2k}^s &= h_{k}, \quad 0 \le 2k < n,\label{eq:parallel:spectral-var-h}\\
	f_{2k+1}^s &= g_{k+1}, \quad 0 \le 2k+1 < n. \label{eq:parallel:spectral-var-g}
\end{align}

\paragraph{Velocity-based variables.}

A more classical choice for advection-dominated transport is to define resolved variables from velocity moments. We collect these variables into $\bm{f}^v = [f_0^v, \ldots, f_{n-1}^v]^T$ with
\begin{equation}
    f_0^{v} = \bar{c}, \quad \text{and} \quad f_k^{v} := \overline{u'^k c'}, \quad k = 1, \ldots, n-1,
    \label{eq:parallel:velocity-vars}
\end{equation}
where the superscript $v$ denotes velocity-based variables.

This choice is reasonable because \(\overline{u'^n c'}\) appears as an unclosed term in the exact governing equation for $\overline{u'^{n-1}c'}$. For a general velocity profile, however, transverse diffusion also generates terms such as $\overline{(\partial_y^2u)c'}$, which may not lie in the span of the selected velocity moments. Therefore, velocity-based variables are less general than the spectral-based variables for higher-moment reduced models.

In the following, reduced models based on both \(\bm f^s\) and \(\bm f^v\) are constructed and compared.

\subsection{\label{sec:parallel-numerical}Numerical discovery of the reduced model}

\subsubsection{\label{sec:parallel-simulation-settings}Simulation settings}

We briefly summarize the numerical solvers used for the full model in Eq.~\eqref{eq:parallel:ADE-HDM} for training and test data generation, and for the reduced model in Eq.~\eqref{eq:parallel:ADE-RM} for model evaluation. We truncate the domain in $x$ to a finite periodic domain $[-L_x/2, L_x/2]$.

\paragraph{Full model.}
We discretize Eq.~\eqref{eq:parallel:ADE-HDM} using a Fourier pseudo-spectral method in $(x,y)$ and advance in time with the DOP853 scheme implemented in SciPy~\cite{virtanen2020scipy}. FFTs are computed with FFTW~\cite{frigo2005design} through the pyFFTW Python interface. A standard $3/2$-dealiasing procedure is applied to the advection term~\cite{canuto2006spectral}.

\paragraph{Reduced model.}
After calibrating the model coefficient matrices $\bm A,\bm B,\bm C$ in Eq.~\eqref{eq:parallel:ADE-RM}, the reduced system can be solved accurately after Fourier transforming in \(x\). Writing
\[
\bm f(x,t)=\sum_{m}\widehat{\bm f}_m(t)e^{i\kappa_m x},
\qquad
\kappa_m=\frac{2\pi m}{L_x},
\]
we obtain, for each Fourier mode \(m\),
\begin{equation}
    \frac{d\widehat{\bm f}_m}{dt}
    =
    \left(
    \bm A + i\kappa_m \bm B - \kappa_m^2 \bm C
    \right)\widehat{\bm f}_m.
\end{equation}

This is a linear ODE system, which we solve using the matrix exponential computed by \texttt{expm} in SciPy~\cite{virtanen2020scipy}.

\subsubsection{\label{sec:parallel-gfm-forcing}GFM forcing construction}

To identify the reduced model in Eq.~\eqref{eq:parallel:ADE-RM}, we generate training data using eGFM forcings. As derived in Sec.~\ref{sec:gfm-implicit-explicit}, the admissible forcings are determined by the weight functions associated with the chosen reduced variables. Thus, we discuss the forcing construction for the two types of resolved variables separately.

\paragraph{Spectral-based variables.}
For the spectral-based variables in Eqs.~\eqref{eq:parallel:spectral-var-cos},~\eqref{eq:parallel:spectral-var-h}, and~\eqref{eq:parallel:spectral-var-g},
we write $f_k^{s}=\langle \phi_k^{s},\,c\rangle_Y$, where $\phi_k^{s}(y)$ are the corresponding Fourier basis functions (either cosine or sine functions, depending on the construction).
An explicit eGFM forcing basis is obtained by spanning the $\alpha_i(x,t)$ in Eq.~\eqref{eq:gfm:Vr-basis} with cosine modes in $x$. Let $n_{x\mathrm{mode}}$ and $n_{y\mathrm{mode}}$ denote the number of designed modes in $x$ and $y$, and let $\kappa_p:=2\pi p/L_x$. The forcing family is
\begin{equation}
    s^{s}_{p,q}(x,y,t) = \cos(\kappa_p x)\,\phi^{s}_{q}(y),
    \quad p = 0,\ldots,n_{x\mathrm{mode}}-1,\quad q = 0,\ldots,n_{y\mathrm{mode}}-1,\quad (p,q)\neq(0,0),
    \label{eq:parallel:eGFM-forcing-spectral}
\end{equation}

\paragraph{Velocity-based variables.}
For the velocity-based variables in Eq.~\eqref{eq:parallel:velocity-vars}, their inner-product representation is
\begin{align}
	f_0^{v} &= \bar c = \langle \phi_0^{v},\,c\rangle_Y, \qquad \phi_0^{v}(y)=1,\\
	f_k^{v} &= \overline{u'(y)^k\,c'} = \langle \phi_k^{v},\,c\rangle_Y,\qquad
	\phi_k^{v}(y)=u'(y)^k-\overline{u'(y)^k},\quad k=1,\ldots,n-1.
\end{align}

The forcing family is constructed similarly:
\begin{equation}
    s^{v}_{p,q}(x,y,t) = \cos(\kappa_p x)\,\phi^{v}_{q}(y),
    \quad p = 0,\ldots,n_{x\mathrm{mode}}-1,\quad q = 0,\ldots,n_{y\mathrm{mode}}-1,\quad (p,q)\neq(0,0),
    \label{eq:parallel:eGFM-forcing-velocity}
\end{equation}

As derived in Sec.~\ref{sec:gfm-implicit-explicit}, the identification of an \(n\)-equation reduced model requires the number of admissible \(y\)-forcing modes to satisfy \(n_{y\mathrm{mode}} \le n\). The effect of the number of forcing modes will be investigated numerically in Sec.~\ref{sec:parallel-results}. The full trajectories from the GFM simulations are used for fitting, thereby capturing the essential physics of the transient behavior.

For reference, Table~\ref{tab:parallel:variables} summarizes the key parameters used throughout this section.

\begin{table}[tbp]
\caption{Key parameters used in Secs.~\ref{sec:parallel} and~\ref{sec:extension}.}
\label{tab:parallel:variables}
\begin{ruledtabular}
\begin{tabular}{cl}
\textbf{Parameter} & \textbf{Description} \\
\hline
$n$              & Number of equations in the reduced model \\
$n_1$            & Highest retained cosine mode index (cosine modes $h_0,\ldots,h_{n_1}$ are resolved) \\
$n_2$            & Highest retained sine mode index (sine modes $g_1,\ldots,g_{n_2}$ are resolved) \\
$n_{x\mathrm{mode}}$ & Number of forcing modes in $x$ \\
$n_{y\mathrm{mode}}$ & Number of forcing modes in $y$ \\
\end{tabular}
\end{ruledtabular}
\end{table}

\subsection{\label{sec:parallel-results}Results and discussion}

We consider two cases of shear flow: a simple cosine profile $u(y) = \cos(y)$ and a complex asymmetric parallel flow, given as
\begin{equation}
	u(y) = 0.45 \cos(y) -0.31 \cos(2y-1) - 0.14 \cos(3y+0.8) + 0.1 \cos(4y+1).
    \label{eq:parallel:complex_u}
\end{equation}

The eGFM training data are generated from a coarse grid with $L_x = 2\pi, N_x = 32, N_y = 32, \Delta t = 8\times 10^{-3}, T_{\max} = 1.6$, and zero initial condition $\bar{c}_0 \equiv 0$.
We consider reduced models with $n = 1, 2, \dots, 10$ equations for both spectral-based and velocity-based variables.
The eGFM datasets are generated using the forcing families in Eqs.~\eqref{eq:parallel:eGFM-forcing-velocity} and~\eqref{eq:parallel:eGFM-forcing-spectral}. The largest forcing family uses $n_{x\mathrm{mode}}=15$ and $n_{y\mathrm{mode}}=10$ and contains $(15\times10)-1=149$ nontrivial forcing cases for each variable type, while the uniform case $(p,q)=(0,0)$ is omitted.

The test solution, which serves as the reference for model evaluation, solves Eq.~\eqref{eq:parallel:ADE-HDM} on a fine grid with $L_x = 4\pi$, $N_x = 384$, $N_y = 48$, $\Delta t = 2\times 10^{-3}$, $T_{\max} = 2$, and initial condition $\bar{c}_0(x) = \exp(-40x^2)$.

In model testing, the reduced system in Eq.~\eqref{eq:parallel:ADE-RM} is solved on the same fine \(x\)-grid as the full model. The only difference is that the reduced model involves \(n\) equations and does not require discretization in \(y\).

In addition to the eGFM-based models, we also consider a reference modeling procedure where the reduced model ansatz in Eq.~\eqref{eq:parallel:ADE-RM} is fitted directly to the test data.
We refer to this approach as \emph{direct-fit}, which provides an approximate optimal performance for the choice of reduced variables.

\subsubsection{Example 1: Simple parallel flow}

We return to the case $u=\cos(y)$. For this flow, the velocity-based variables are linear combinations of the spectral-based variables; thus, the two reduced model families are equivalent. Accordingly, we report only the results from spectral-based models. The evolution of the mean field $\bar{c}$ for this case has been shown in Figs.~\ref{fig:intro:dispersion} and~\ref{fig:intro:contour}. We now compare the reduced models with this reference solution.

\begin{figure}[tbp]
    \centering
    \includegraphics{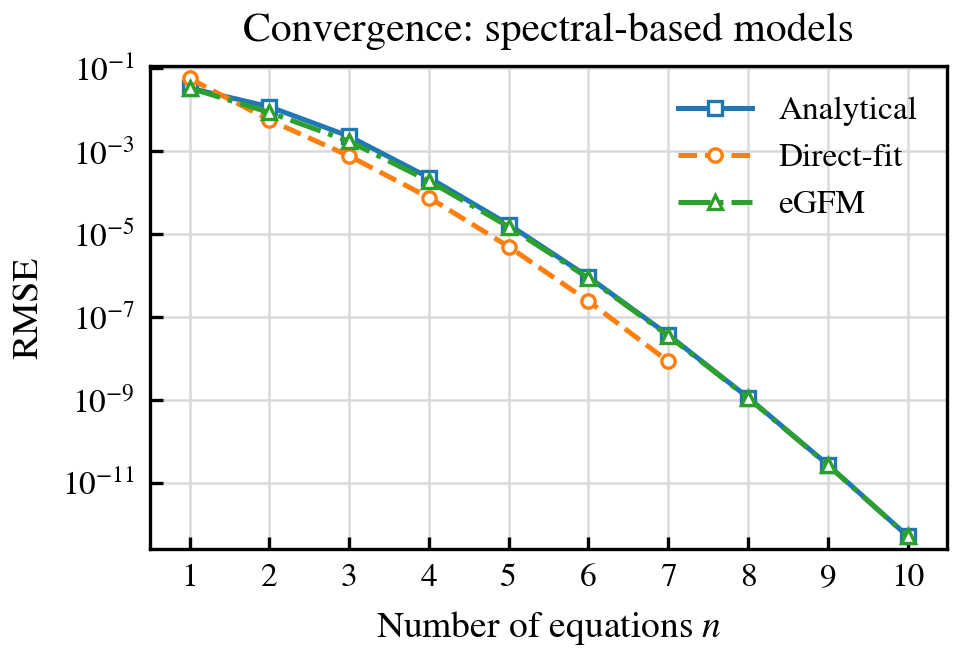}
    \caption{
        Comparison of the RMSE of \(\bar c\) for spectral-based reduced models.
        For the eGFM-based models, we use \(n_{x\mathrm{mode}}=15\) and \(n_{y\mathrm{mode}}=n\).
    }
    \label{fig:parallel:simple-summary}
\end{figure}

An overall comparison of model accuracy is presented in Fig.~\ref{fig:parallel:simple-summary}, which shows the RMSE of \(\bar c\) as a function of the number of equations \(n\). All three approaches exhibit rapid error decay as \(n\) increases, indicating spectral convergence. The direct-fit models perform slightly better than the other two approaches, since they are fitted from the test data itself. Meanwhile, the eGFM-based models have similar performance compared to the analytical models for all \(n\).

\begin{figure}[tbp]
    \centering
    \includegraphics{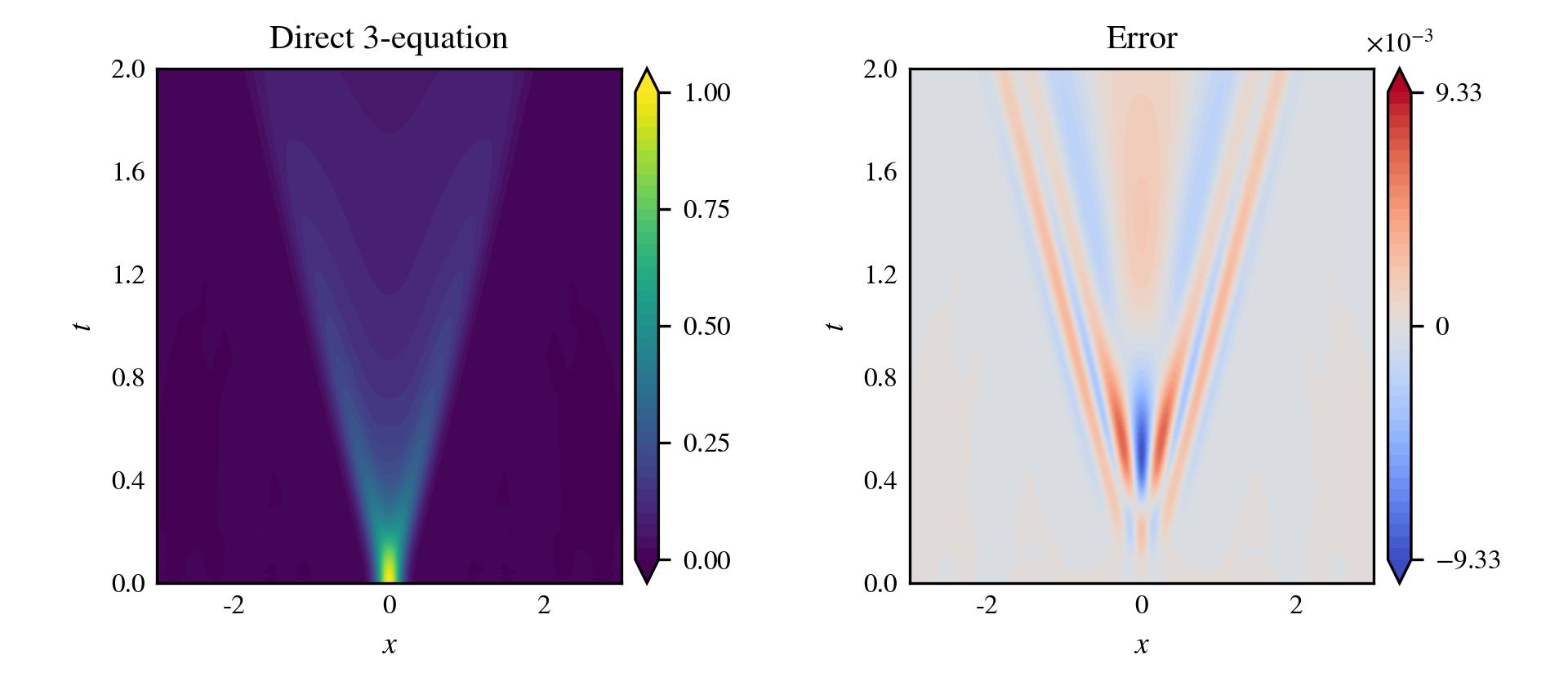}
    \caption{
        Time histories of \(\bar c\) (left) and the error in \(\bar c\) (right) for the 3-equation reduced model obtained from direct-fit. The error is the full-model result minus the reduced-model prediction.
    }
    \label{fig:parallel:simple-err-direct}
\end{figure}

\begin{figure}[tbp]
    \centering
    \includegraphics{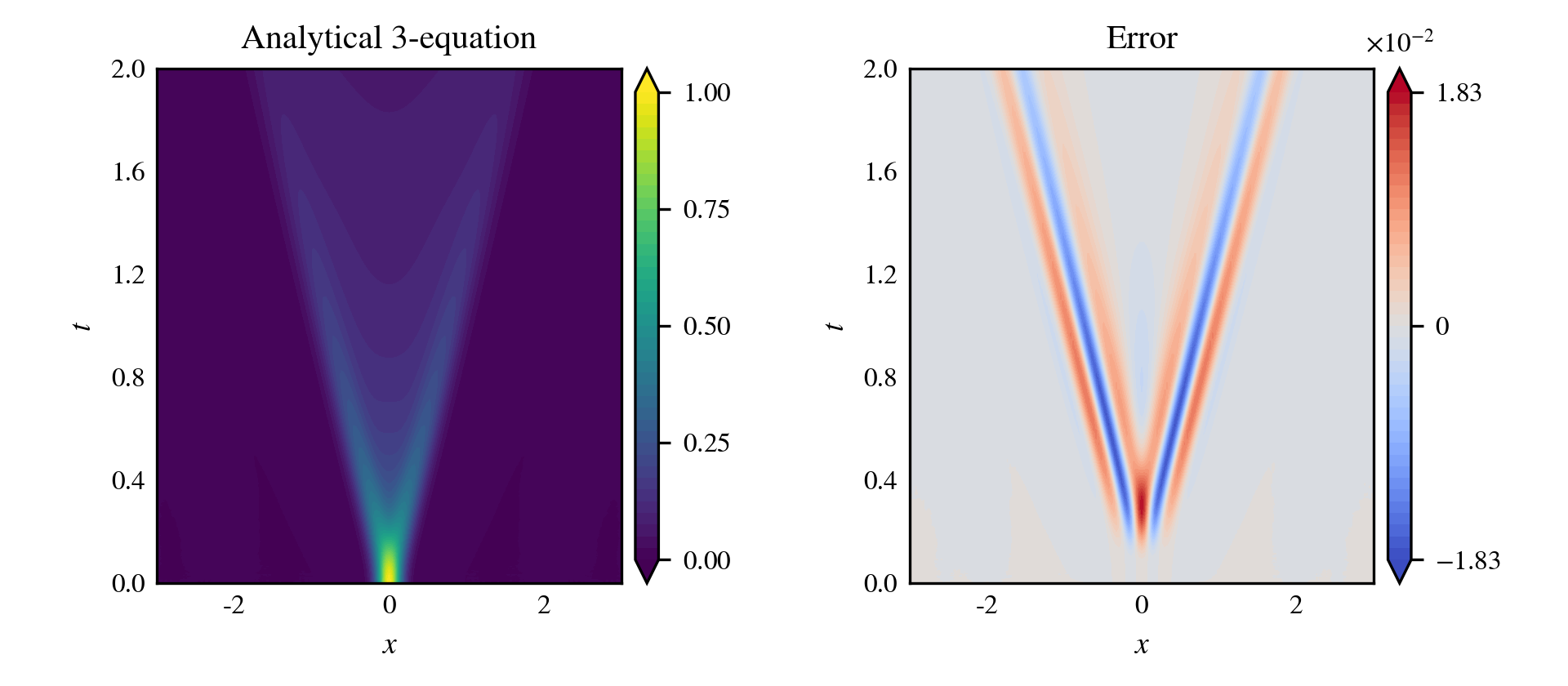}
    \caption{
        Time histories of \(\bar c\) (left) and the error in \(\bar c\) (right) for the 3-equation reduced model obtained from the analytical closure. The error is the full-model result minus the reduced-model prediction.
    }
    \label{fig:parallel:simple-err-anal}
\end{figure}

\begin{figure}[tbp]
    \centering
    \includegraphics{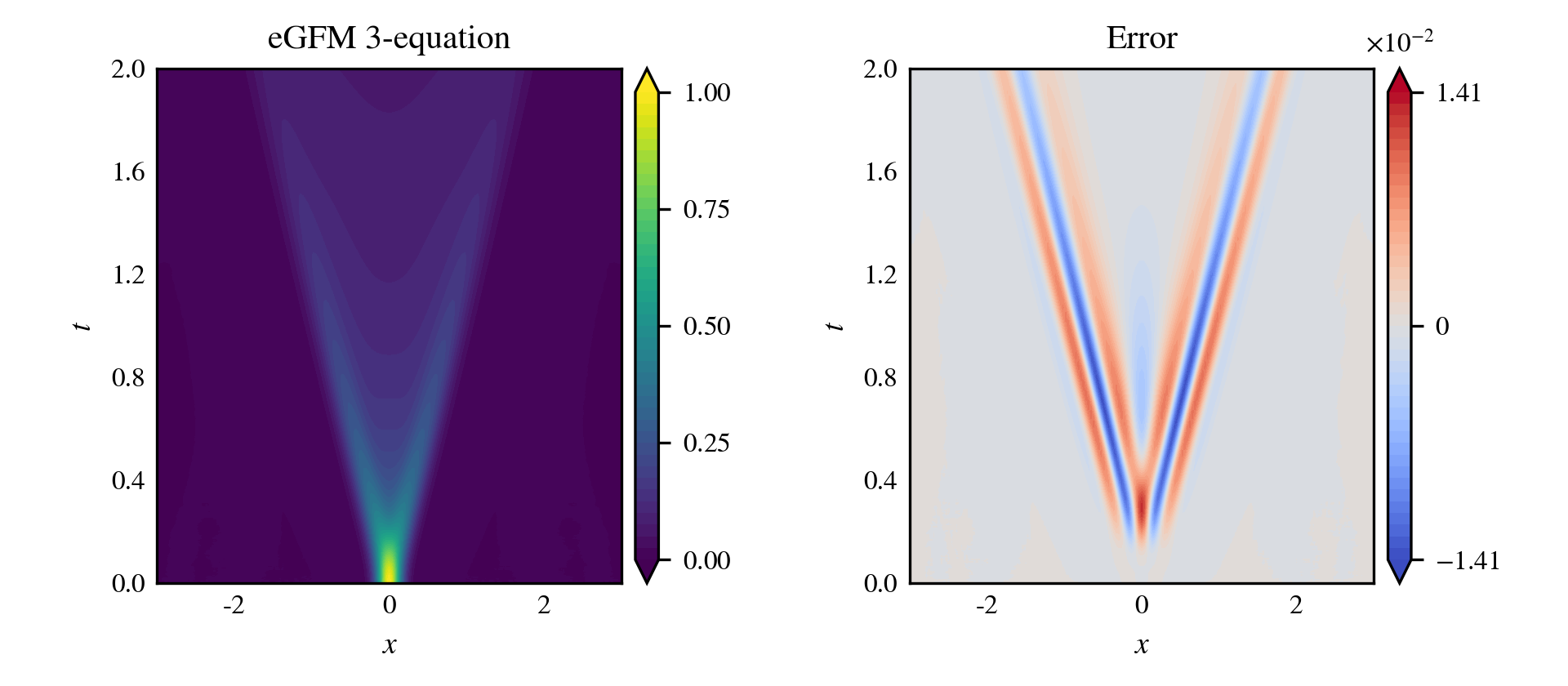}
    \caption{
        Time histories of \(\bar c\) (left) and the error in \(\bar c\) (right) for the 3-equation reduced model obtained from eGFM, with \(n_{x\mathrm{mode}}=15\) and \(n_{y\mathrm{mode}}=3\). The error is the full-model result minus the reduced-model prediction.
    }
    \label{fig:parallel:simple-err-egfm}
\end{figure}

To provide a more detailed comparison, Figs.~\ref{fig:parallel:simple-err-direct},~\ref{fig:parallel:simple-err-anal}, and~\ref{fig:parallel:simple-err-egfm} show the time histories of \(\bar c\) and its error for the three-equation models obtained from the direct-fit, analytical, and eGFM approaches, respectively. Consistent with the RMSE trends, the direct-fit model is the most accurate, while the eGFM-based model remains very close to the analytical model. The alternating error bands in Figs.~\ref{fig:parallel:simple-err-anal} and~\ref{fig:parallel:simple-err-egfm} propagate approximately along \(x=\pm t\), consistent with the advective fronts shown in Fig.~\ref{fig:intro:contour}. They reflect the unresolved higher-mode contributions near these fronts.

\begin{figure}[tbp]
    \centering
    \includegraphics{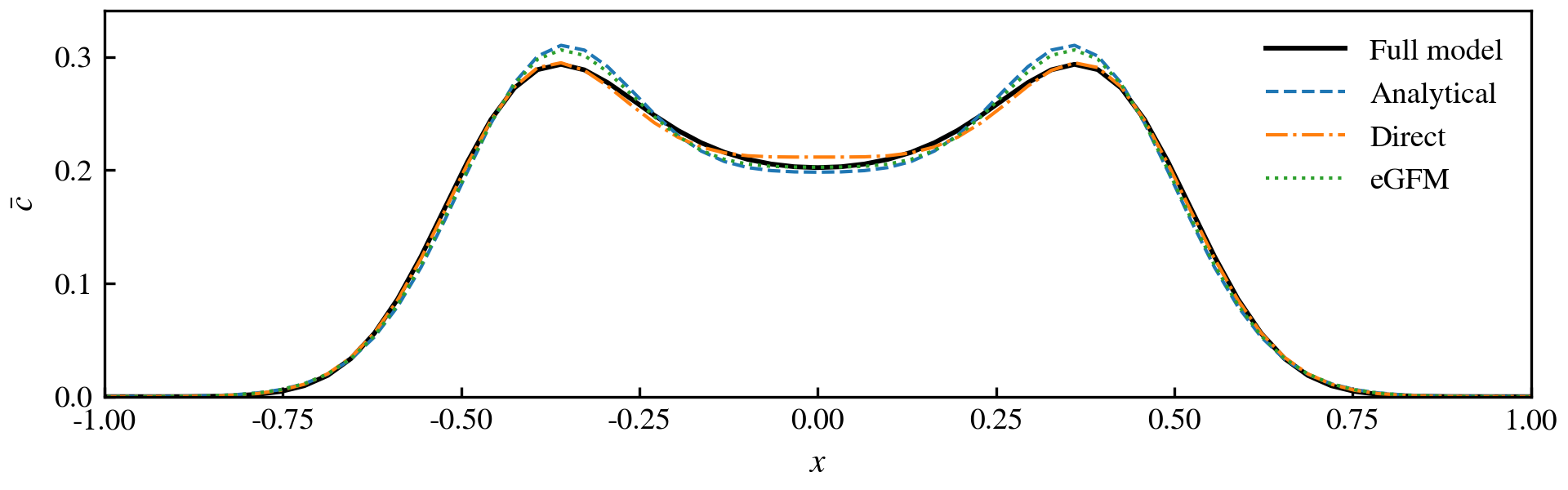}
    \caption{Comparison of \(\bar c\) at \(t=0.5\) from the full model and the 3-equation reduced models.}
    \label{fig:parallel:simple-slice-eq3}
\end{figure}

Figure~\ref{fig:parallel:simple-slice-eq3} provides a closer comparison at \(t=0.5\). All three reduced models closely match the full-model profile. The analytical and eGFM models slightly overpredict the two maxima.

\begin{figure}[tbp]
    \centering
        \centering
        \includegraphics{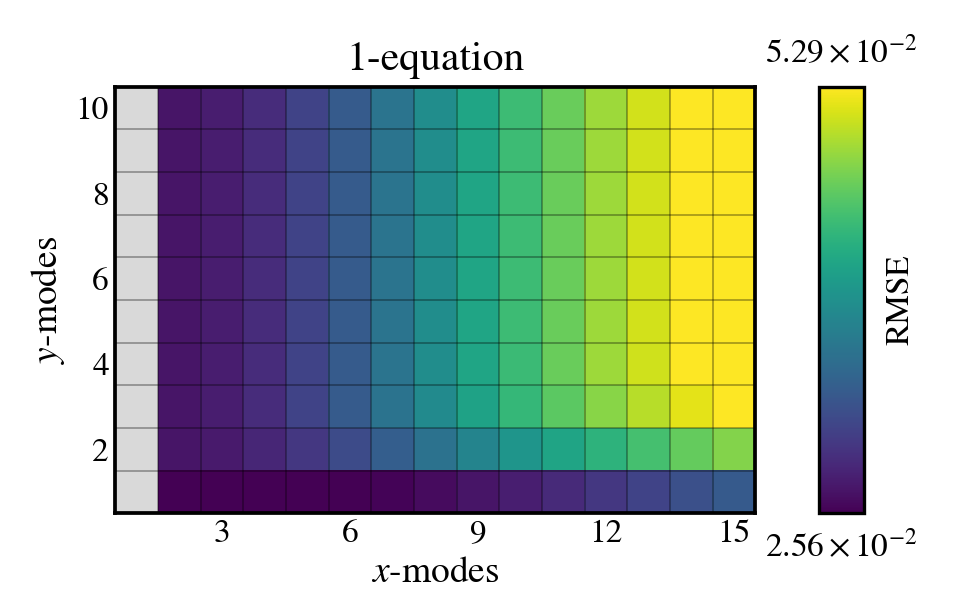}
    \hfill
        \centering
        \includegraphics{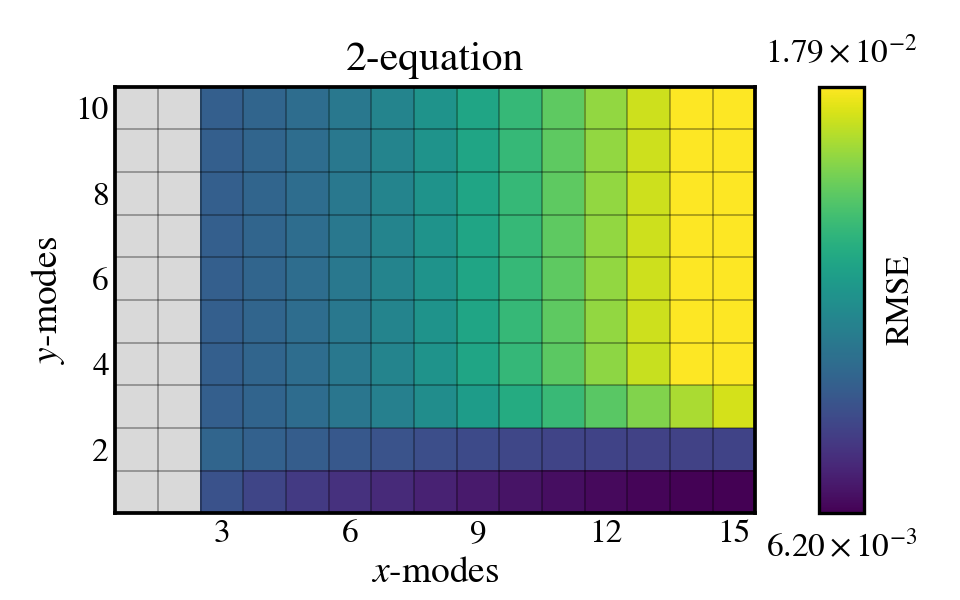}

        \centering
        \includegraphics{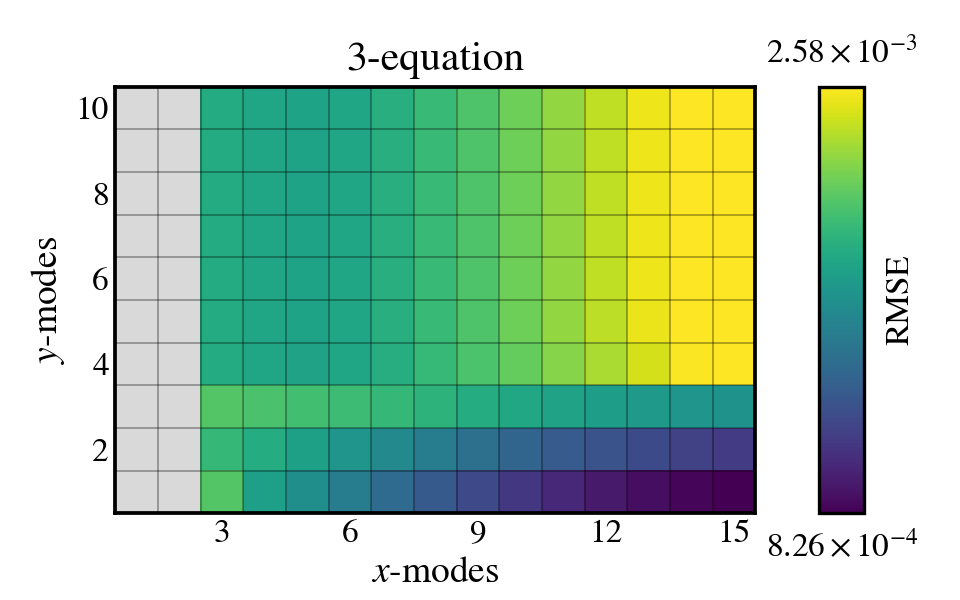}
    \hfill
        \centering
        \includegraphics{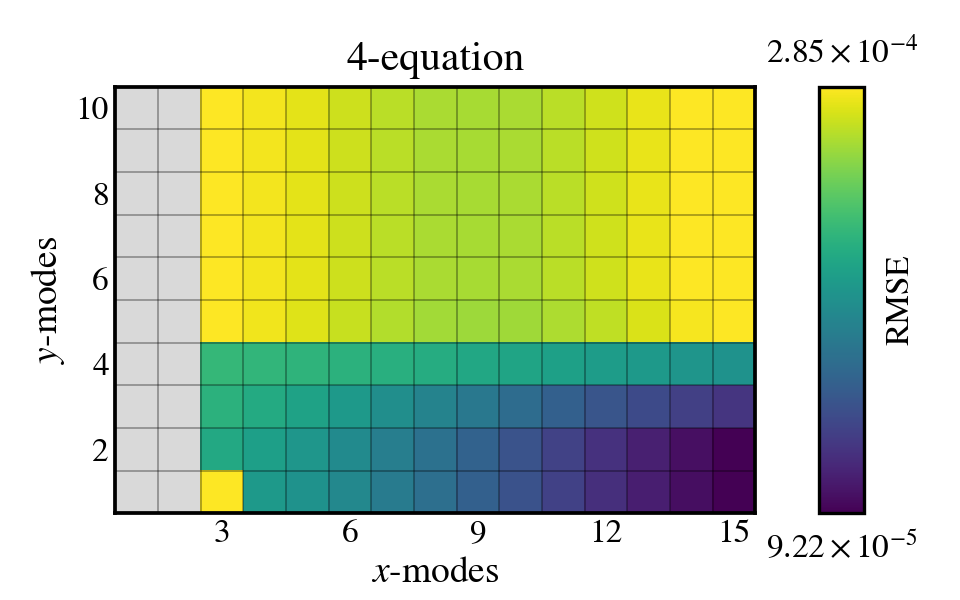}

        \centering
        \includegraphics{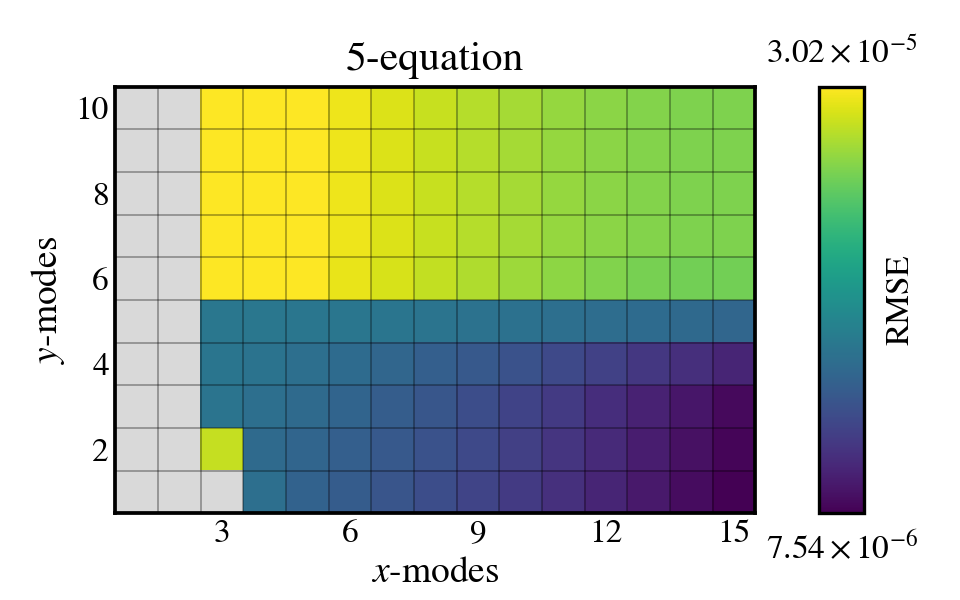}
    \hfill
        \centering
        \includegraphics{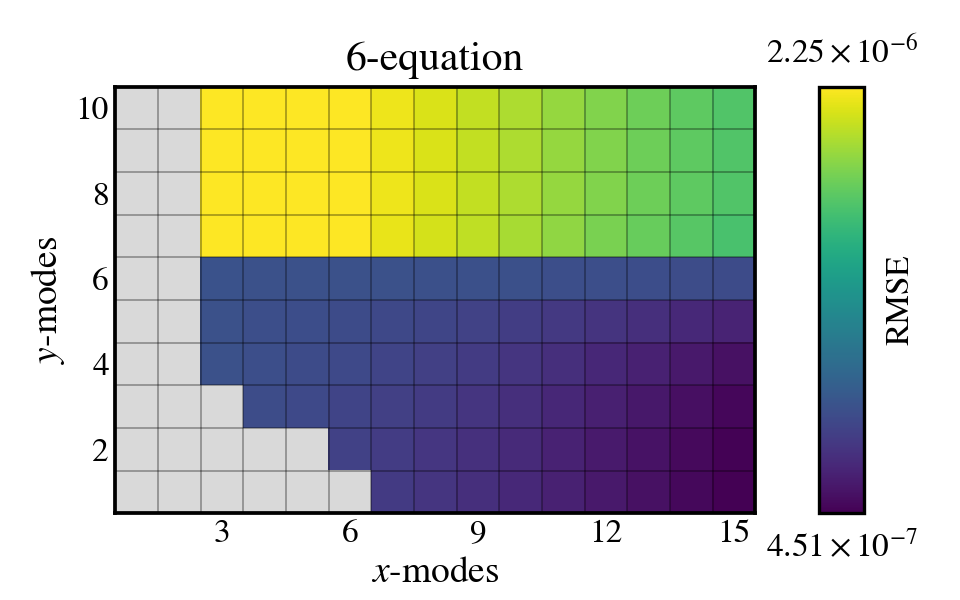}

    \caption{
        RMSE of eGFM spectral-based reduced models over the space of forcing subsets for \(n=1,\ldots,6\). Gray blocks indicate an empty forcing subset, fitting failure, or runtime blowup.
    }
    \label{fig:parallel:simple-dataset-1-6}
\end{figure}

\begin{figure}[tbp]
    \centering
        \centering
        \includegraphics{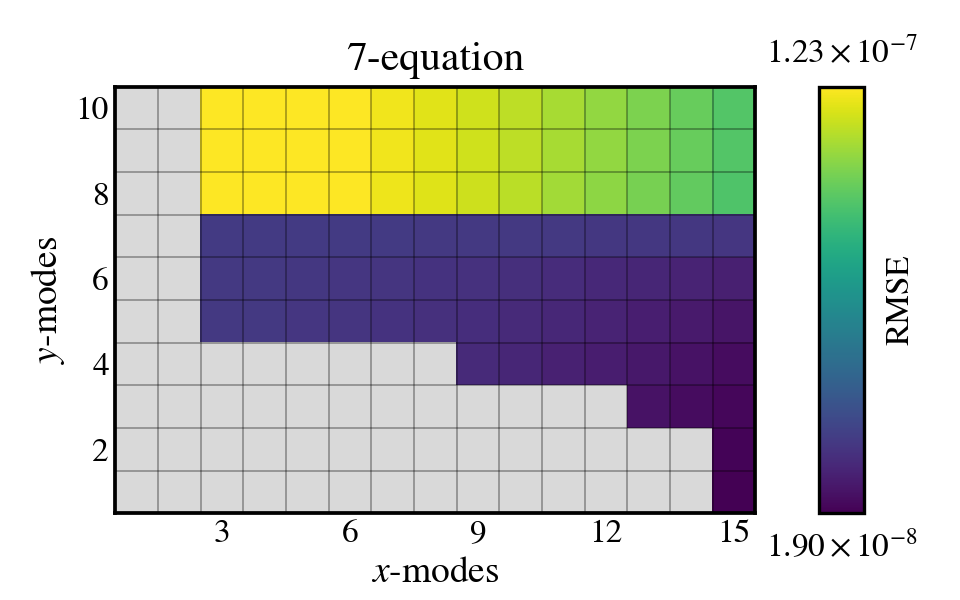}
    \hfill
        \centering
        \includegraphics{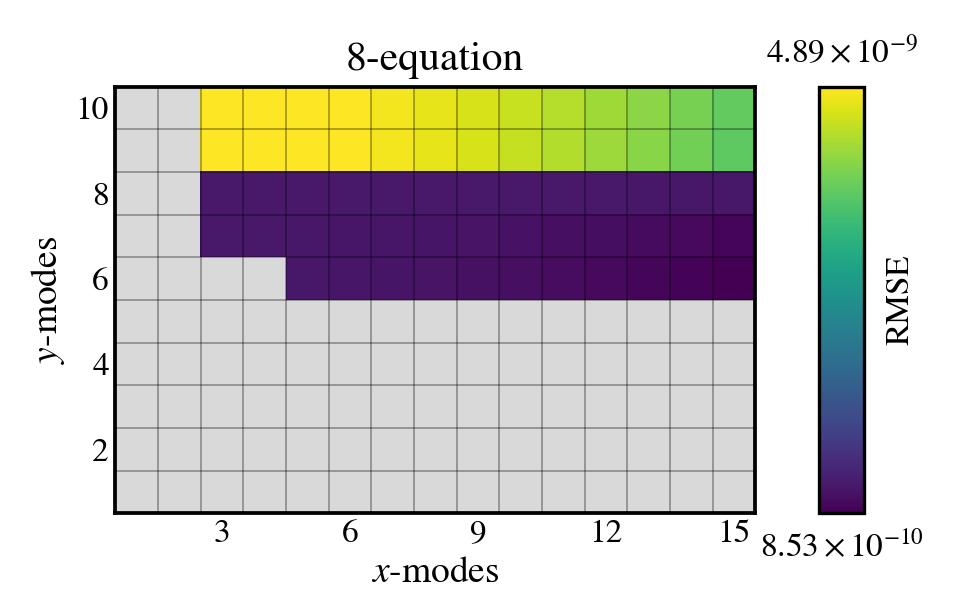}

        \centering
        \includegraphics{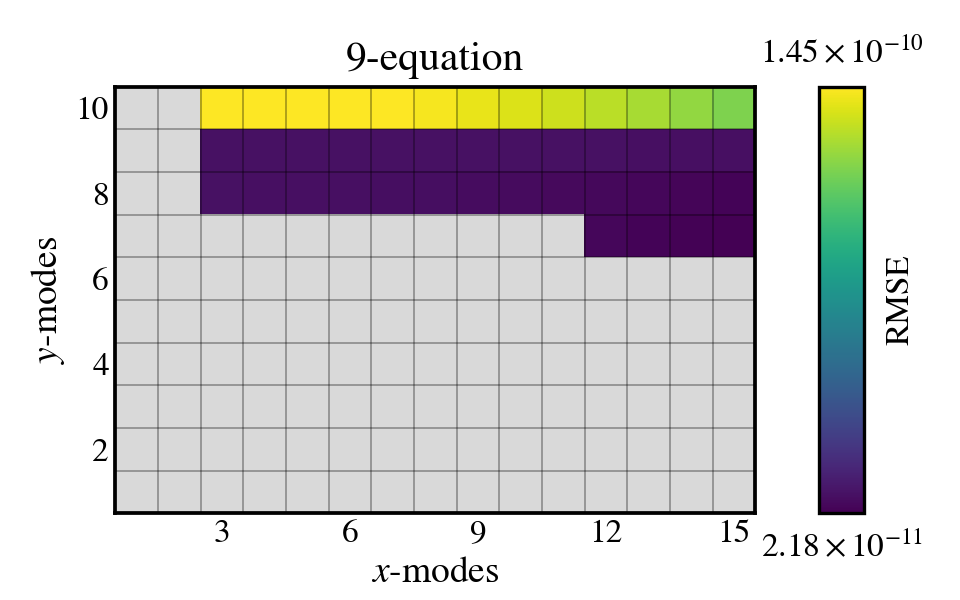}
    \hfill
        \centering
        \includegraphics{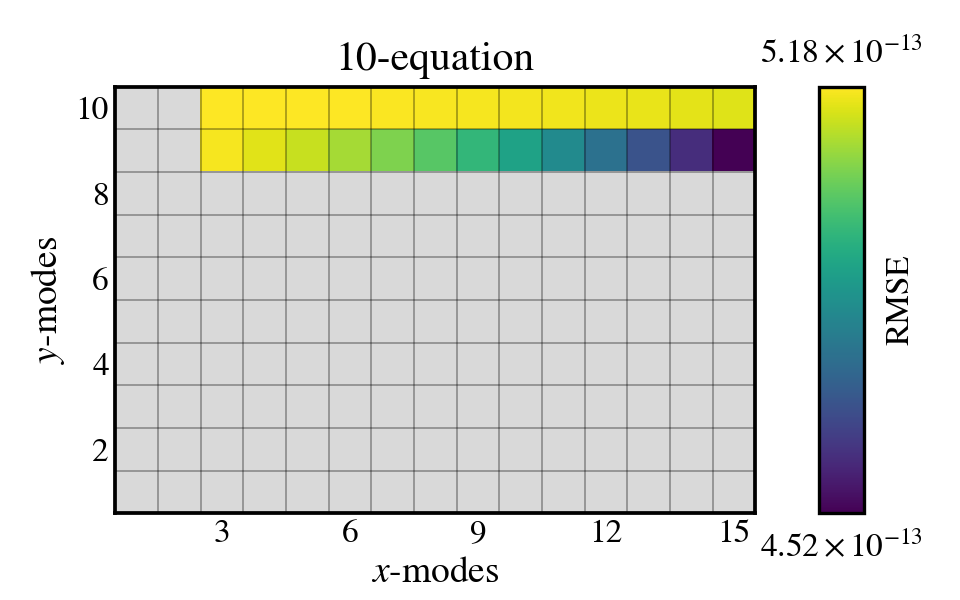}

    \caption{
        RMSE of eGFM spectral-based reduced models over the space of forcing subsets for \(n=7,\ldots,10\). Gray blocks indicate an empty forcing subset, fitting failure, or runtime blowup.
    }
    \label{fig:parallel:simple-dataset-7-10}
\end{figure}

We next investigate the effect of the training dataset on the identified models. For each \(n\), Figs.~\ref{fig:parallel:simple-dataset-1-6} and~\ref{fig:parallel:simple-dataset-7-10} show the RMSE of the reduced model over the space of forcing subsets parameterized by \((n_{x\mathrm{mode}}, n_{y\mathrm{mode}})\). We observe the following patterns, which support the GFM forcing principle.

First, a minimum number of $x$-forcing modes is required for a successful fit. For the one-equation model, at least two \(x\)-modes are needed, while for \(n\ge 2\), at least three are required.

Second, as \(n\) grows, more \(y\)-forcing modes are required. For high-moment models, choosing \(n_{y\mathrm{mode}} \ll n\) can lead to fitting failure or runtime blowup, indicating that the forcing is insufficient to identify the model.

Third, for admissible datasets with \(n_{y\mathrm{mode}}\le n\), which is consistent with the GFM principle, increasing $n_{x\mathrm{mode}}$ tends to reduce the error since the forcing subspace spans the admissible space more completely. For \(n_{y\mathrm{mode}}\ge n+1\), the model error increases substantially. This behavior is consistent with the theoretical expectation that forcing outside the admissible subspace would excite modes that cannot be captured by the corresponding reduced variables.

\subsubsection{Example 2: Complex asymmetric parallel flow}

\begin{figure}[tbp]
    \centering
        \centering
        \includegraphics{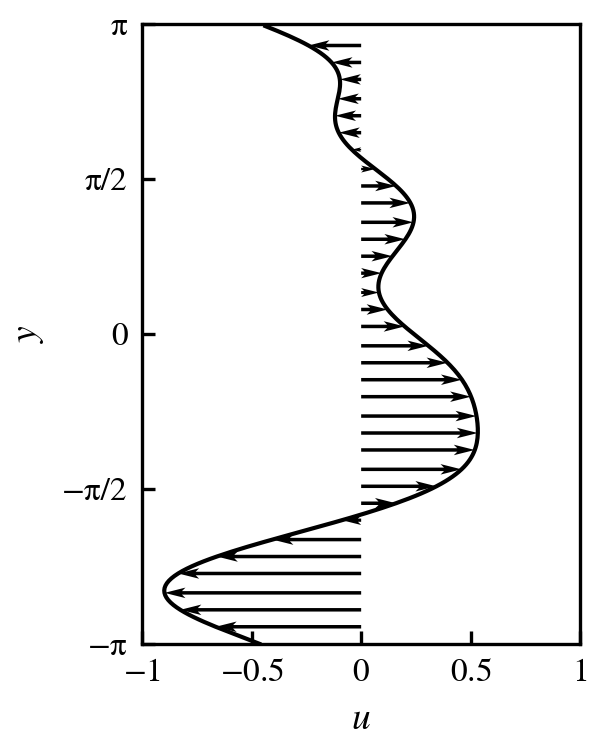}
    \hfill
        \centering
        \includegraphics{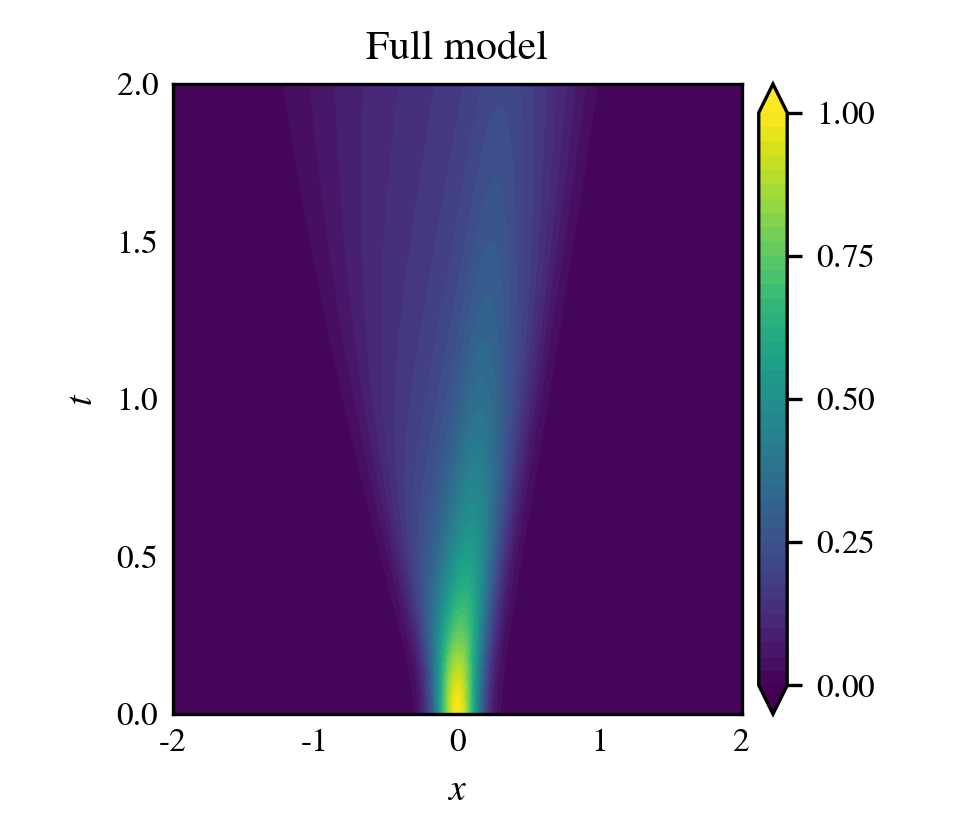}

    \caption{
        Dispersion of passive scalar in the complex asymmetric parallel flow in Eq.~\eqref{eq:parallel:complex_u} for the test case with $\bar{c}_0(x) = \exp(-40x^2)$.
        Left: velocity profile $u(y)$; right: time history of the cross-sectional mean $\bar c(x,t)$.
    }
    \label{fig:parallel:complex-dispersion}
\end{figure}

We next consider the more complex asymmetric shear flow profile in Eq.~\eqref{eq:parallel:complex_u}. The velocity profile and the time history of \(\bar c\) for the test case are shown in Fig.~\ref{fig:parallel:complex-dispersion}.

\begin{figure}[tbp]
        \centering
        \includegraphics{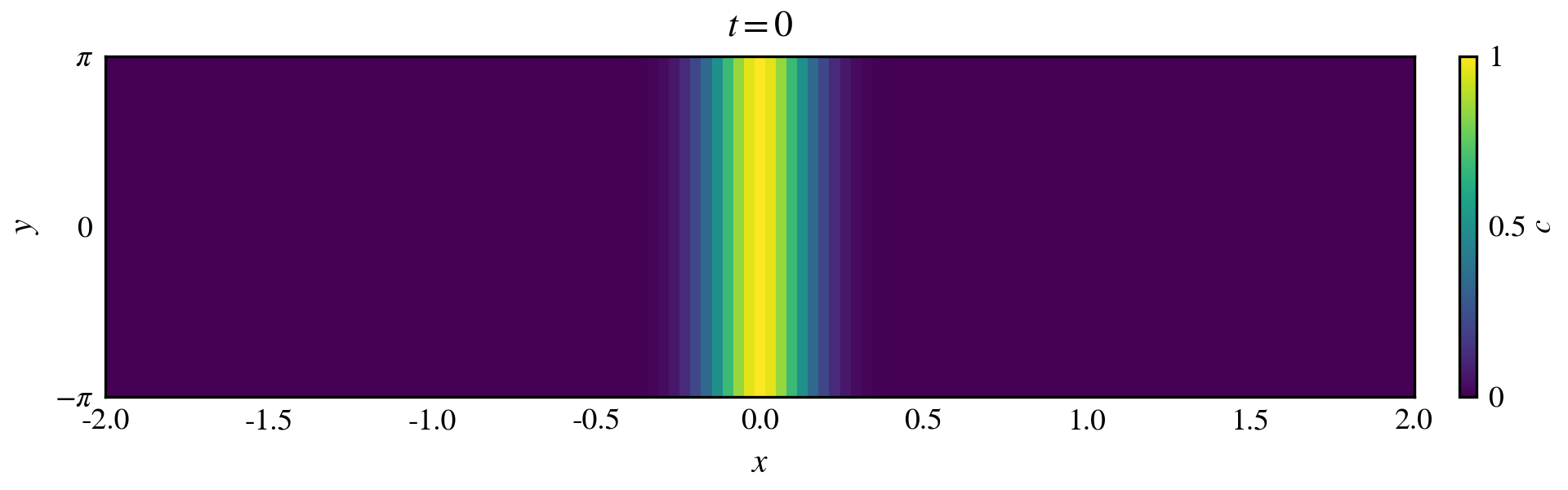}

        \centering
        \includegraphics{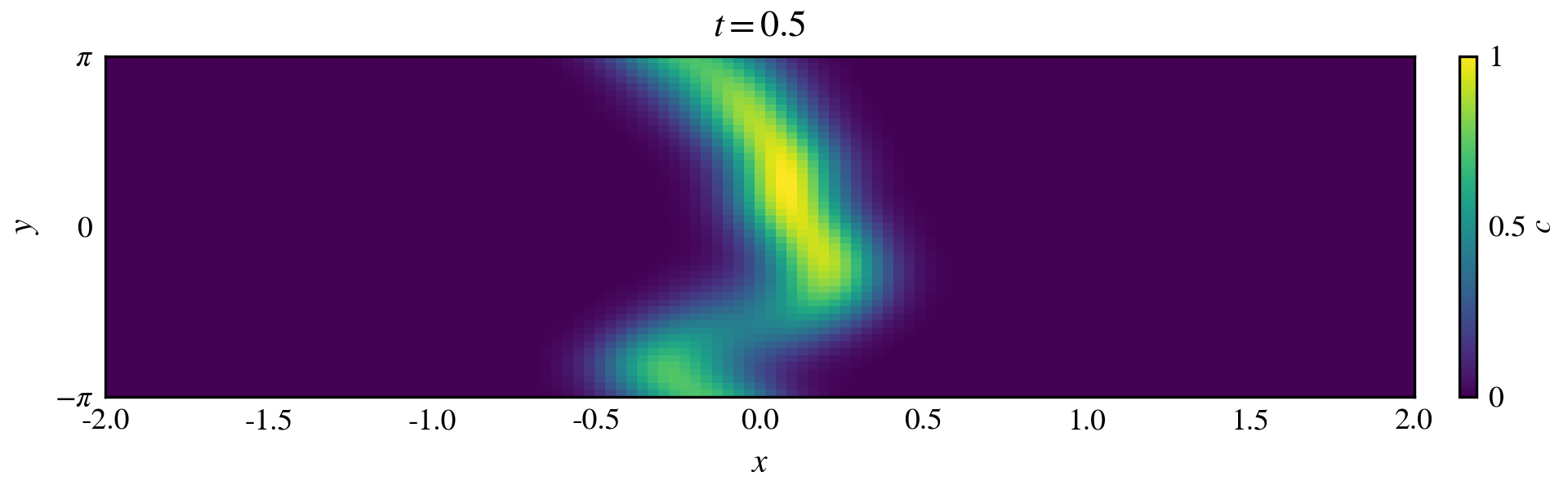}

        \centering
        \includegraphics{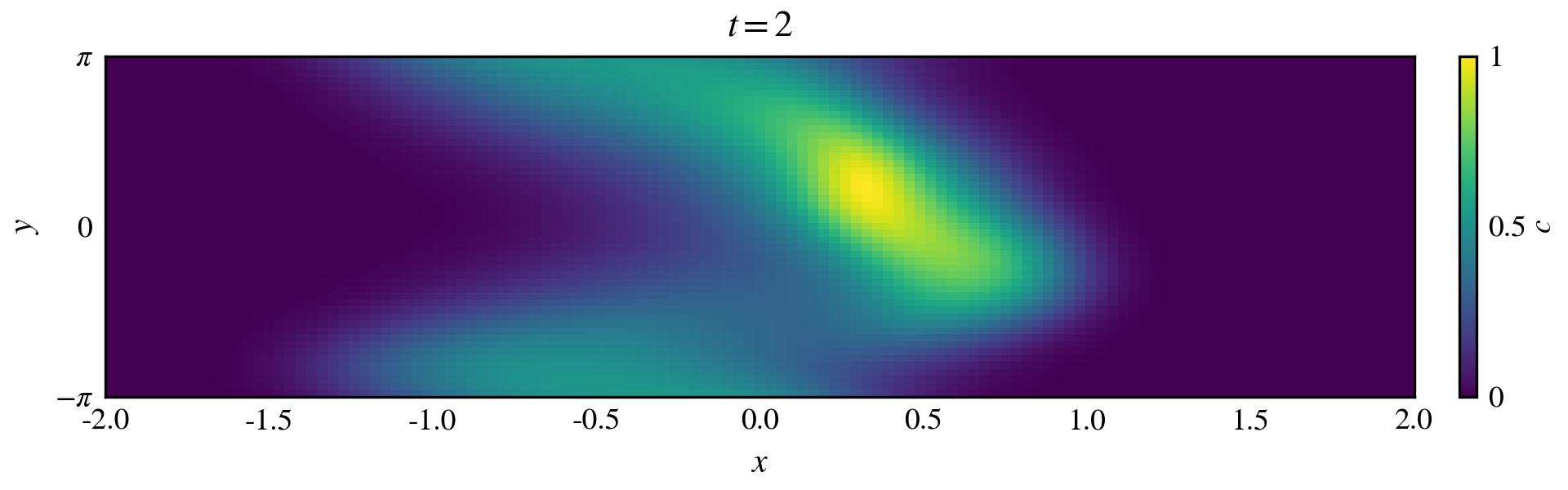}

        \centering
        \includegraphics{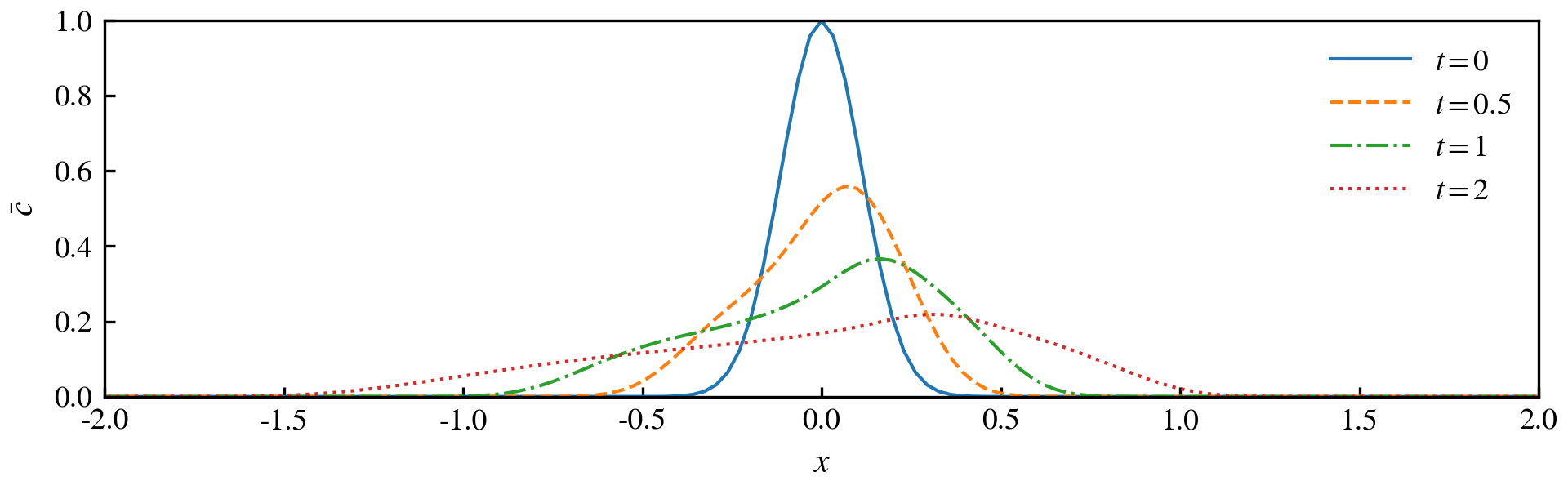}
    \caption{
        Contours of \(c(x,y,t)\) and slices of \(\bar c(x,t)\) for the complex asymmetric parallel flow with initial condition \(\bar c_0(x)=\exp(-40x^2)\).
    }
    \label{fig:parallel:complex-contour}
\end{figure}

Further details of the full model solution are shown in Fig.~\ref{fig:parallel:complex-contour}. In contrast to the simple cosine shear flow case, $\bar{c}$ is no longer symmetric in \(x\) due to the asymmetry of the underlying velocity profile, which makes the modeling problem more challenging.

\begin{figure}[tbp]
    \centering
    \includegraphics[width=\textwidth]{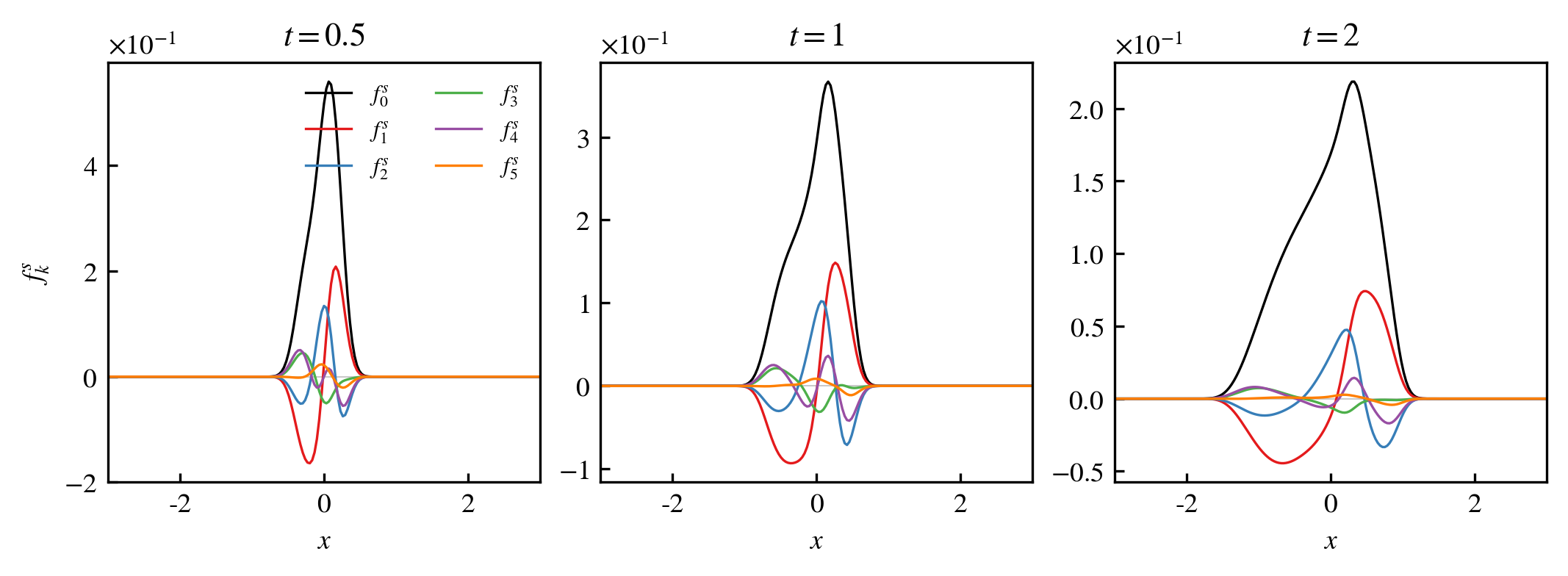}
    \caption{Profiles of the first six spectral-based resolved variables, $f_0^s,\ldots,f_5^s$, for the complex asymmetric parallel flow at $t=0.5$, $1$, and $2$.}
    \label{fig:parallel:complex-spectral-profiles}
\end{figure}

Fig.~\ref{fig:parallel:complex-spectral-profiles} shows the first six spectral-based resolved variables for the test case. Both cosine and sine components are present because the velocity profile is asymmetric. Their magnitudes generally decrease as the variable index increases.

For this case, the spectral-based and velocity-based models exhibit different behavior, and thus we discuss them separately.

\begin{figure}[tbp]
    \centering
    \includegraphics{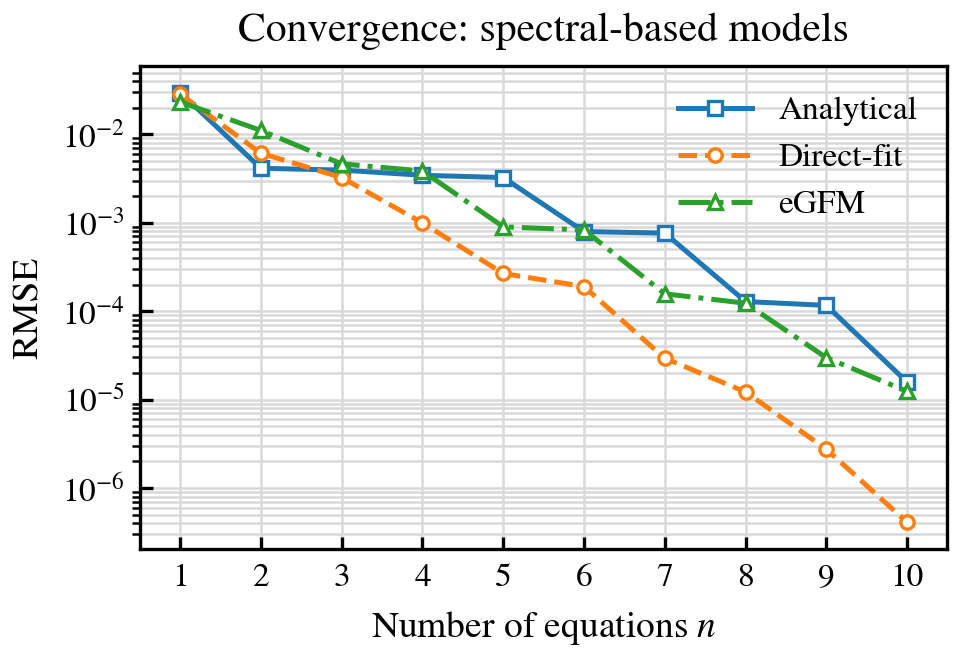}
    \caption{
        Comparison of the RMSE of \(\bar c\) for spectral-based reduced models in the complex asymmetric flow. For the eGFM-based models, we use \(n_{x\mathrm{mode}}=15\) and \(n_{y\mathrm{mode}}=n\).
    }
    \label{fig:parallel:complex-summary-spectral}
\end{figure}

For the spectral-based variables, the RMSE comparison is shown in Fig.~\ref{fig:parallel:complex-summary-spectral}. All three approaches improve as \(n\) increases, although the convergence is slower than the previous example. The direct-fit models are the most accurate among all, while the analytical and eGFM-based models show oscillatory convergence patterns, where noticeable improvement often happens after two additional degrees of freedom are introduced. This is consistent with the spectral convergence rate when both cosine and sine modes are present. Overall, the eGFM-based models perform slightly better than the analytical models.

\begin{figure}[tbp]
    \centering
    \includegraphics{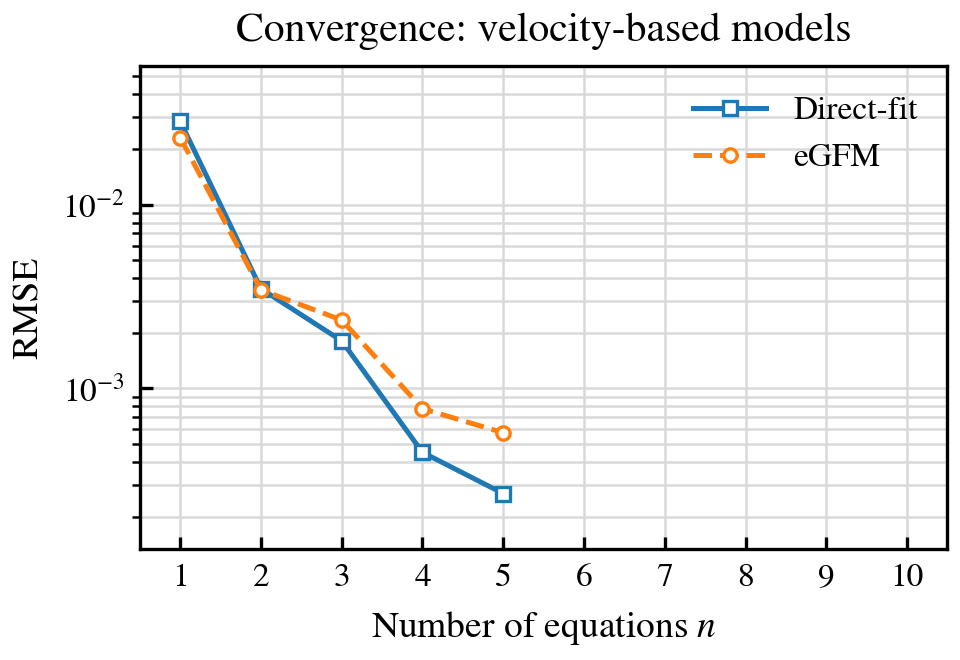}
    \caption{
        Comparison of the RMSE of \(\bar c\) for velocity-based reduced models in the complex asymmetric flow. For the eGFM-based models, we use \(n_{x\mathrm{mode}}=15\) and \(n_{y\mathrm{mode}}=n\). For \(n\ge 6\), no successful models were obtained and these cases are omitted.
    }
    \label{fig:parallel:complex-summary-velocity}
\end{figure}

For the velocity-based variables, the RMSE comparison is shown in Fig.~\ref{fig:parallel:complex-summary-velocity}. In this case, analytical models are not available. Stable reduced models are only found for $n\leq 5$ for both approaches, and the direct-fit models still outperform the eGFM-based models slightly. Both approaches suggest that velocity moments are not an ideal choice of resolved variables for higher-moment closure in this setting.

\begin{figure}[tbp]
    \centering
        \centering
        \includegraphics{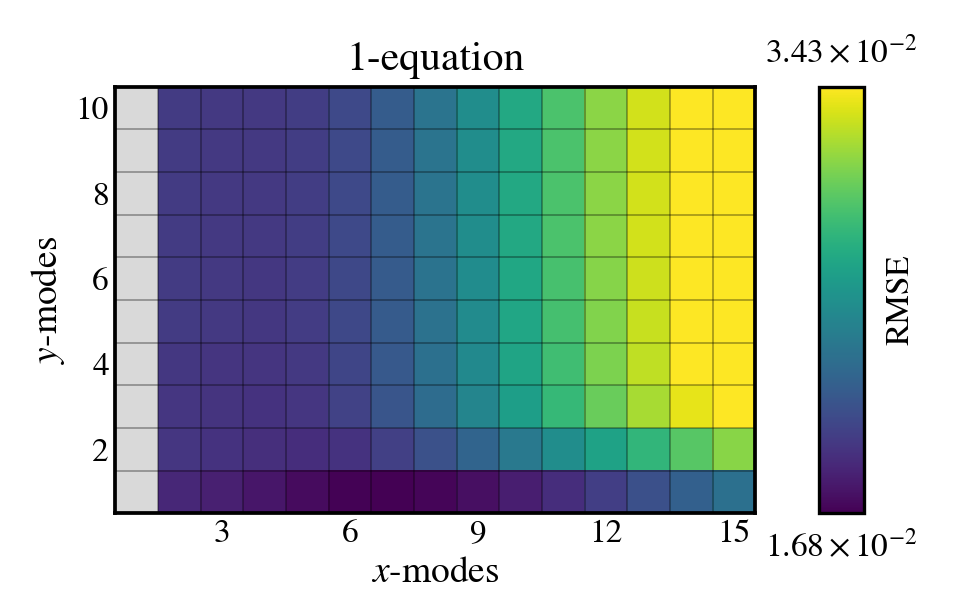}
    \hfill
        \centering
        \includegraphics{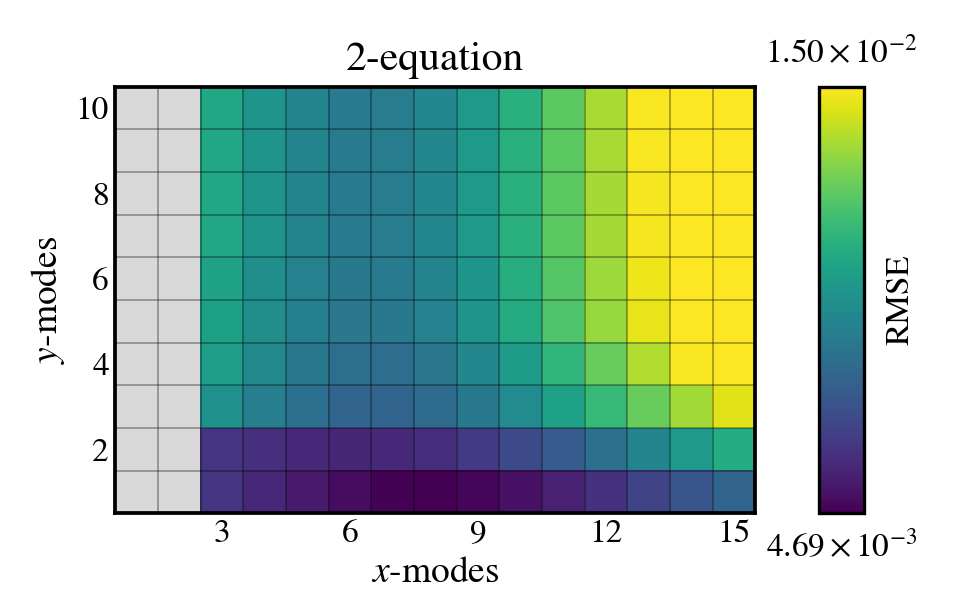}

        \centering
        \includegraphics{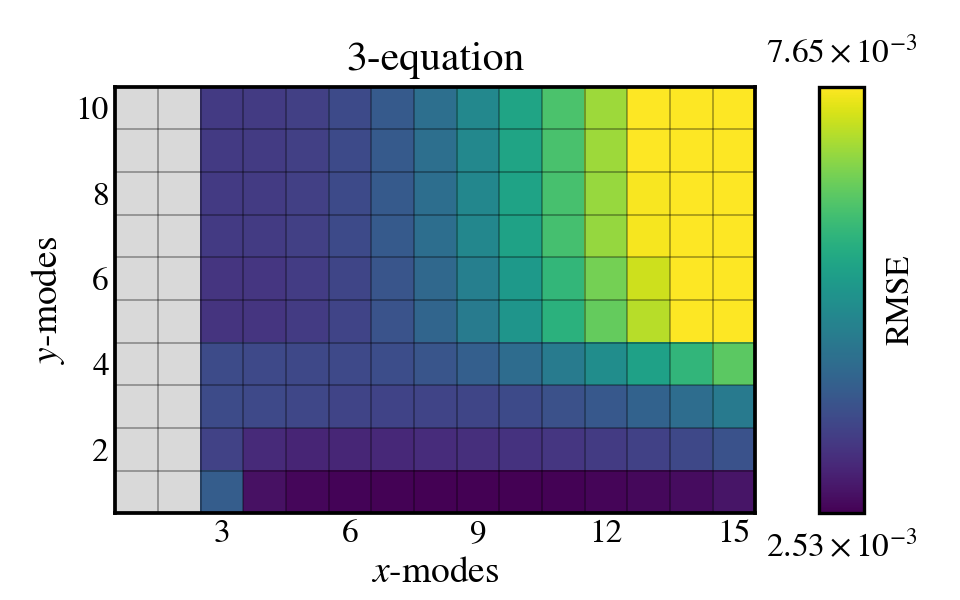}
    \hfill
        \centering
        \includegraphics{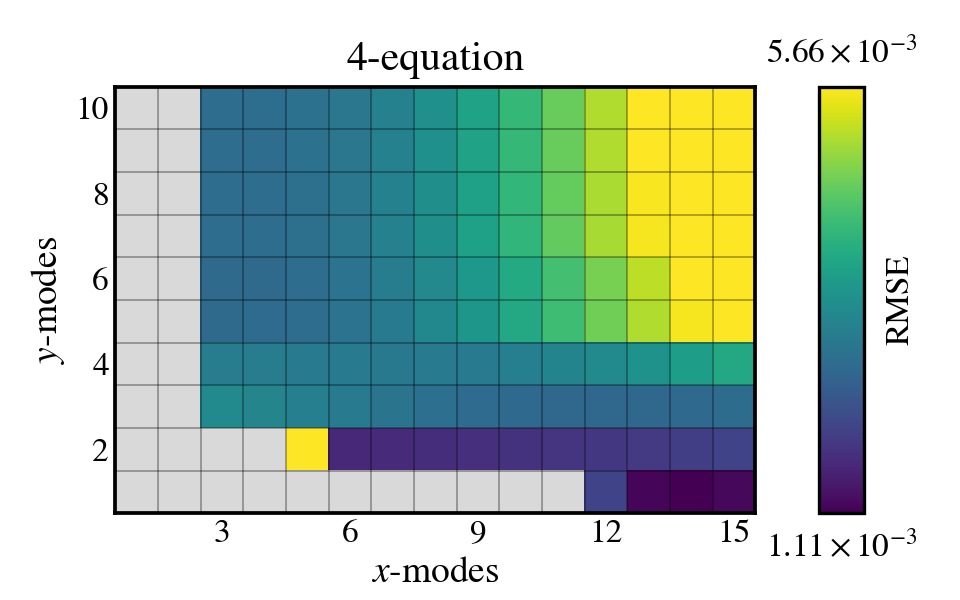}

        \centering
        \includegraphics{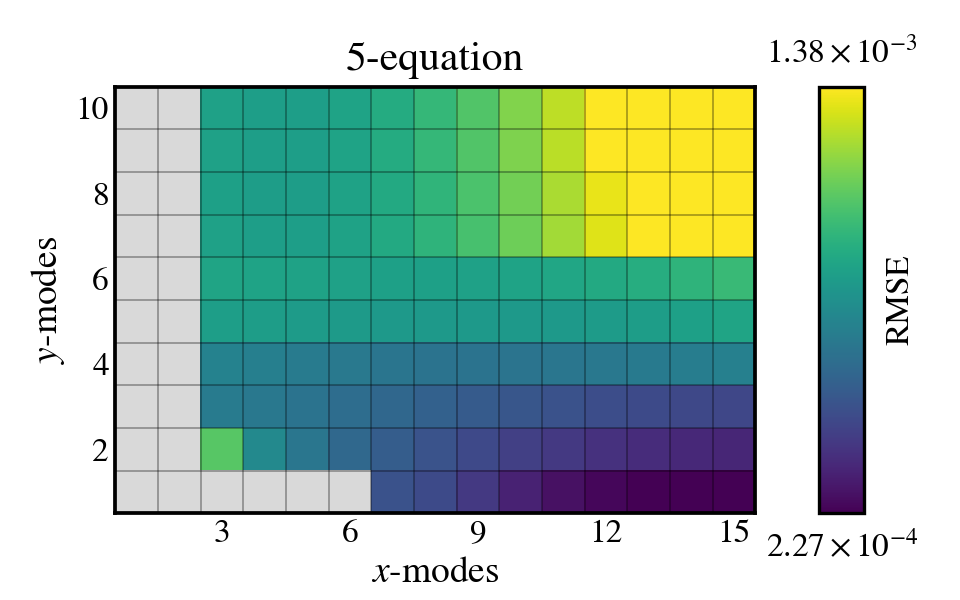}
    \hfill
        \centering
        \includegraphics{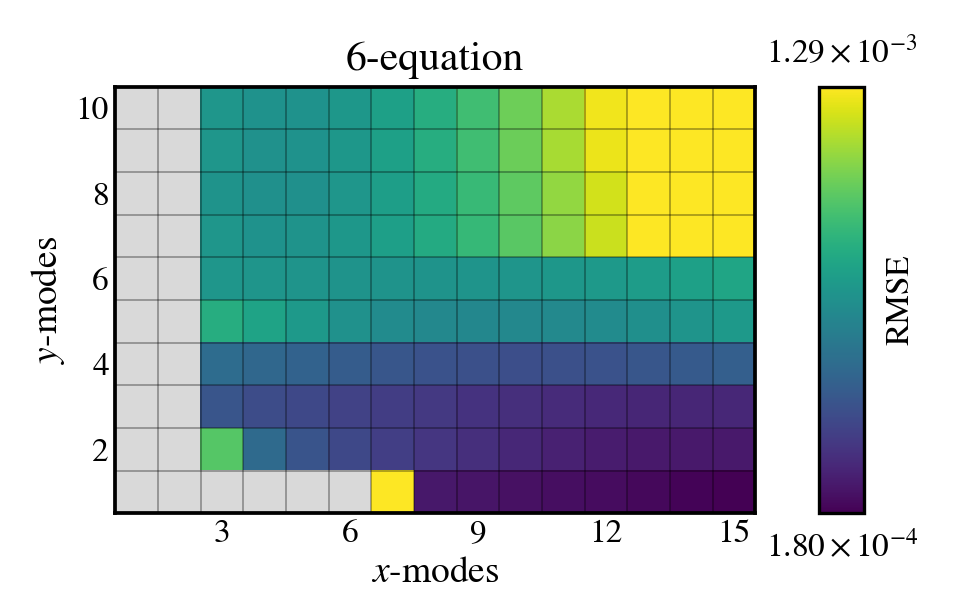}

    \caption{
        RMSE of eGFM spectral-based reduced models over the space of forcing subsets for \(n=1,\ldots,6\) in the complex asymmetric flow. Gray blocks indicate an empty forcing subset, fitting failure, or runtime blowup.
    }
    \label{fig:parallel:complex-spectral-dataset-1-6}
\end{figure}

\begin{figure}[tbp]
    \centering
        \centering
        \includegraphics{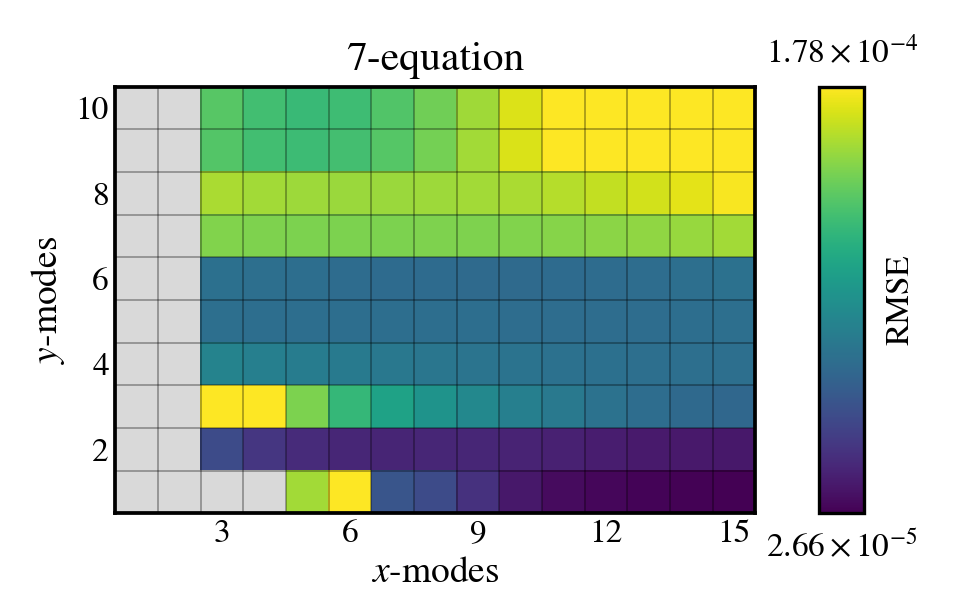}
    \hfill
        \centering
        \includegraphics{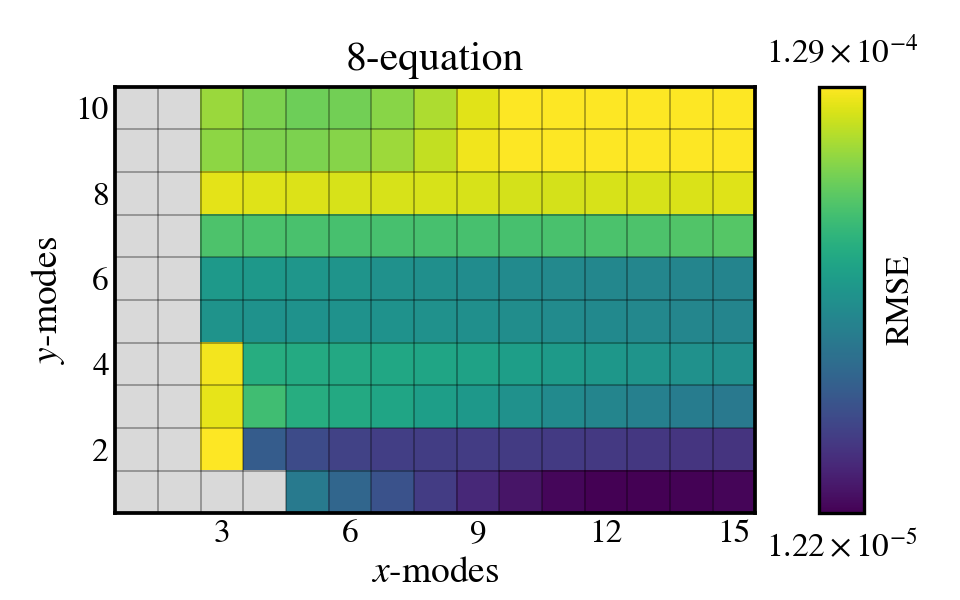}

        \centering
        \includegraphics{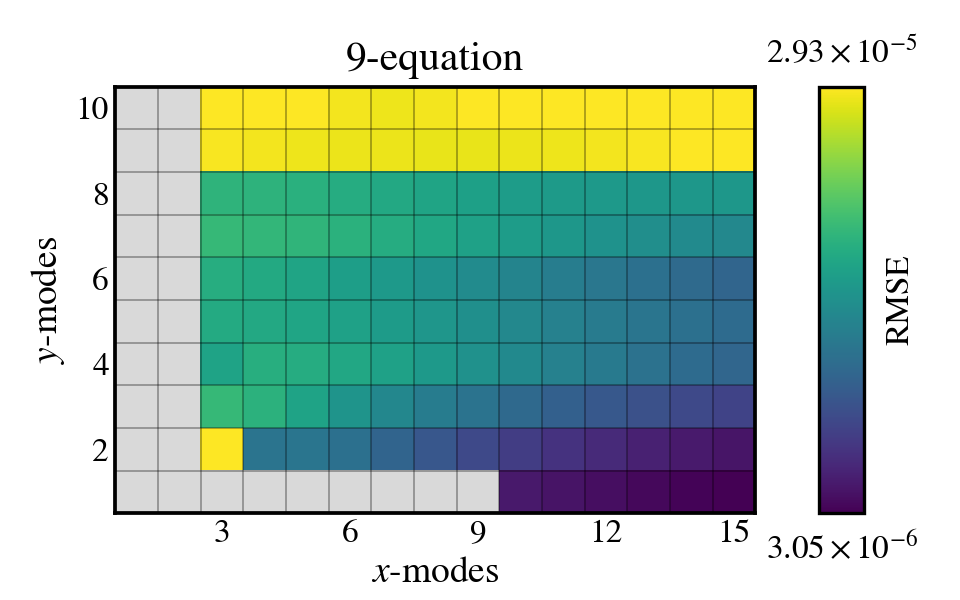}
    \hfill
        \centering
        \includegraphics{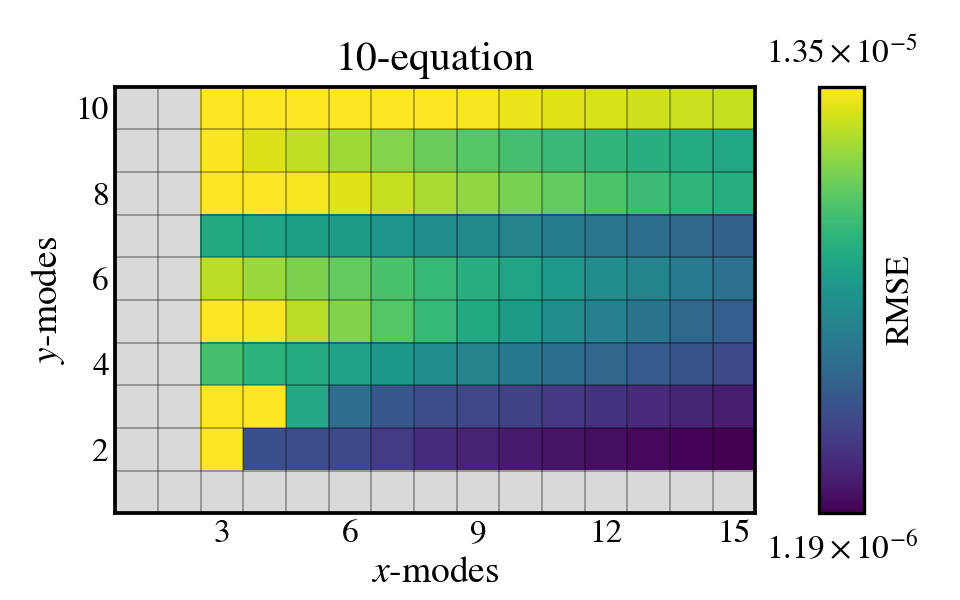}

    \caption{
        RMSE of eGFM spectral-based reduced models over the space of forcing subsets for \(n=7,\ldots,10\) in the complex asymmetric flow. Gray blocks indicate an empty forcing subset, fitting failure, or runtime blowup.
    }
    \label{fig:parallel:complex-spectral-dataset-7-10}
\end{figure}

We next examine the effect of the training dataset on eGFM spectral-based reduced models. Figs.~\ref{fig:parallel:complex-spectral-dataset-1-6} and~\ref{fig:parallel:complex-spectral-dataset-7-10} show the RMSE over the space of forcing subsets. A clear degradation in accuracy is observed when \(n_{y\mathrm{mode}} > n\).
Meanwhile, the optimal models are often obtained with \(1\le n_{y\mathrm{mode}}\le 3\) even for larger \(n\).
A possible explanation is that forcing with large $y$-modes often excites fast decaying modes, which are less representative of the long-time dispersion dynamics in the test case. Since in the fitting procedure all forced simulations are weighted equally, including many high-$y$-mode forcings can potentially increase the test error.

\begin{figure}[tbp]
    \centering
        \centering
        \includegraphics{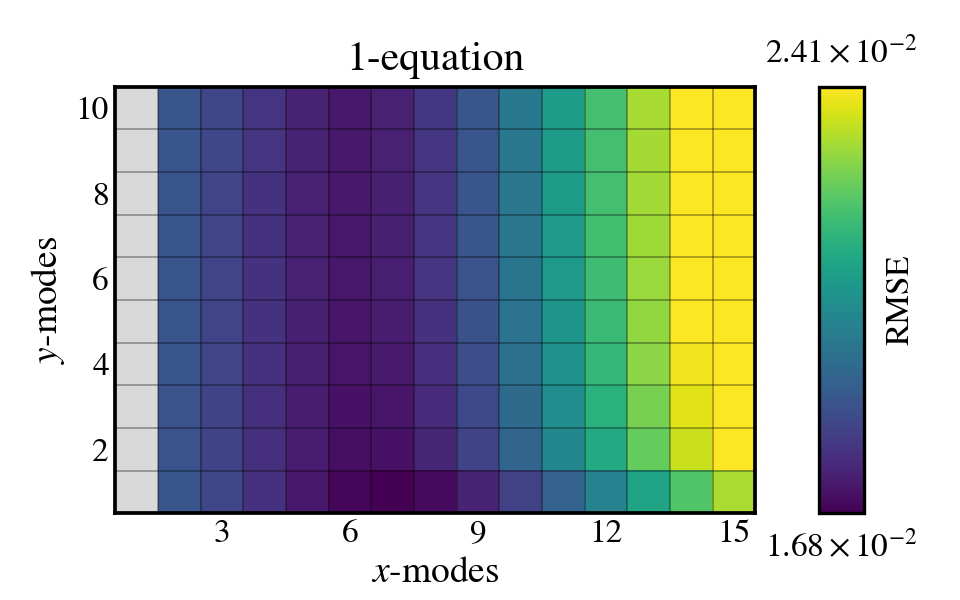}
    \hfill
        \centering
        \includegraphics{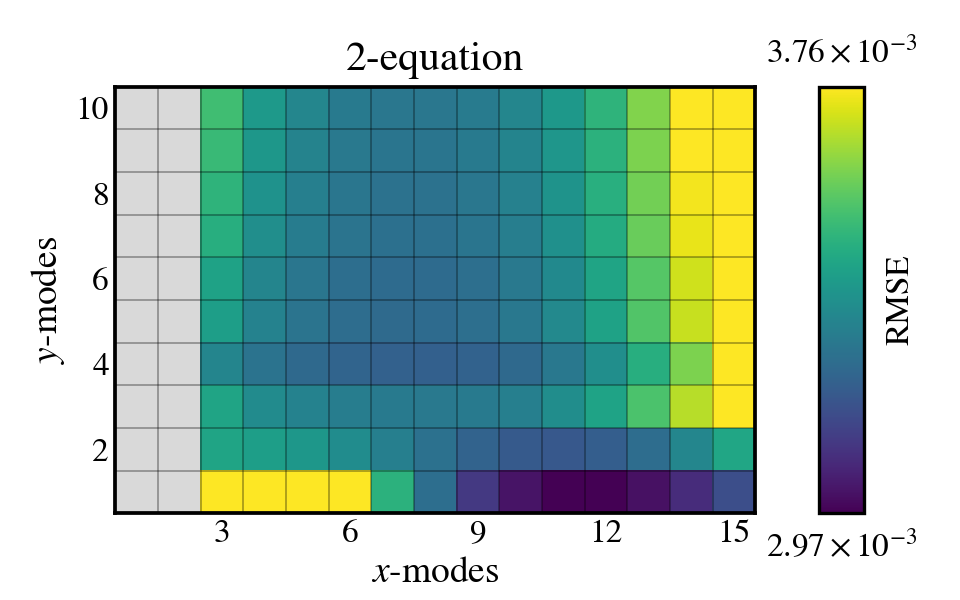}

        \centering
        \includegraphics{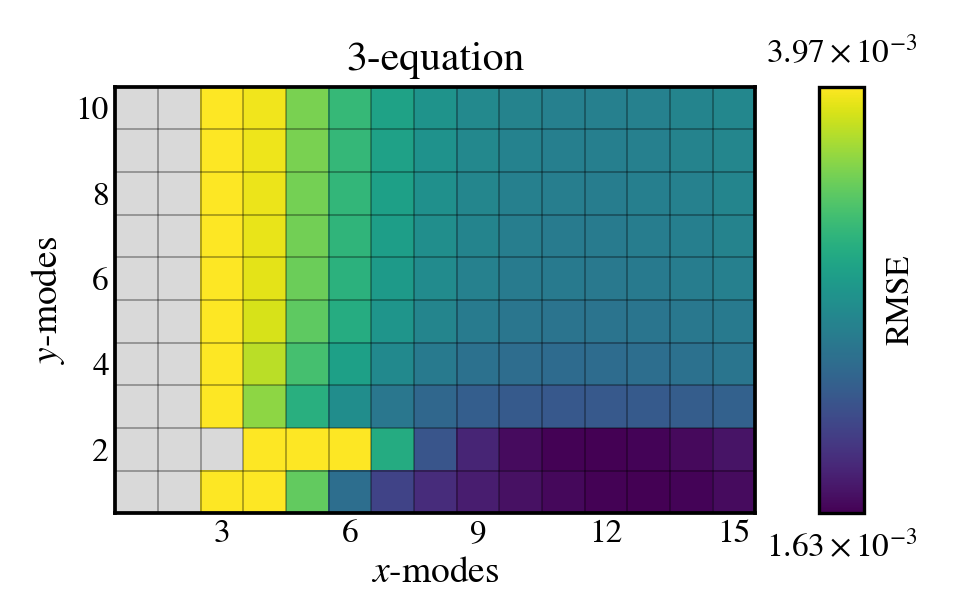}
    \hfill
        \centering
        \includegraphics{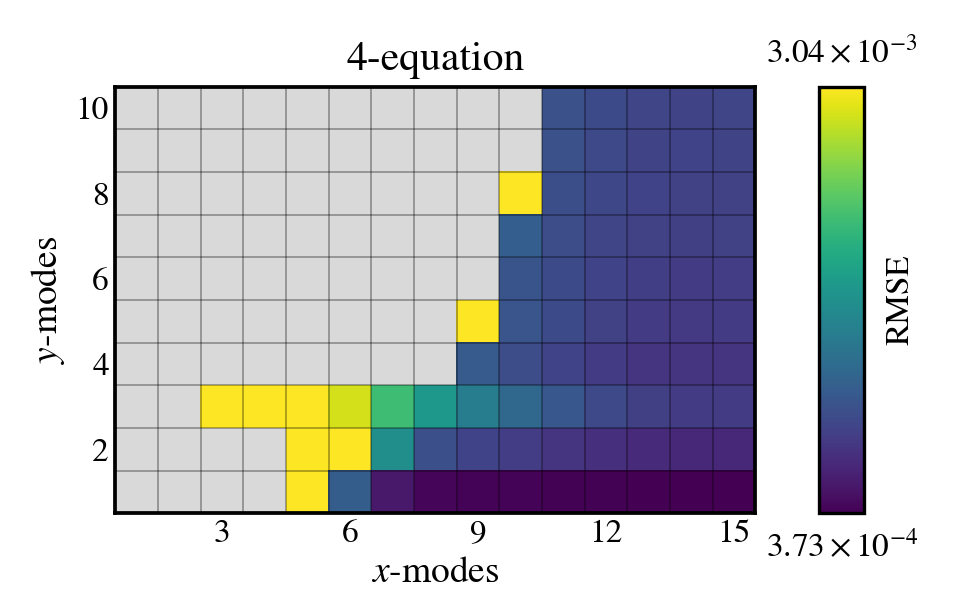}

        \centering
        \includegraphics{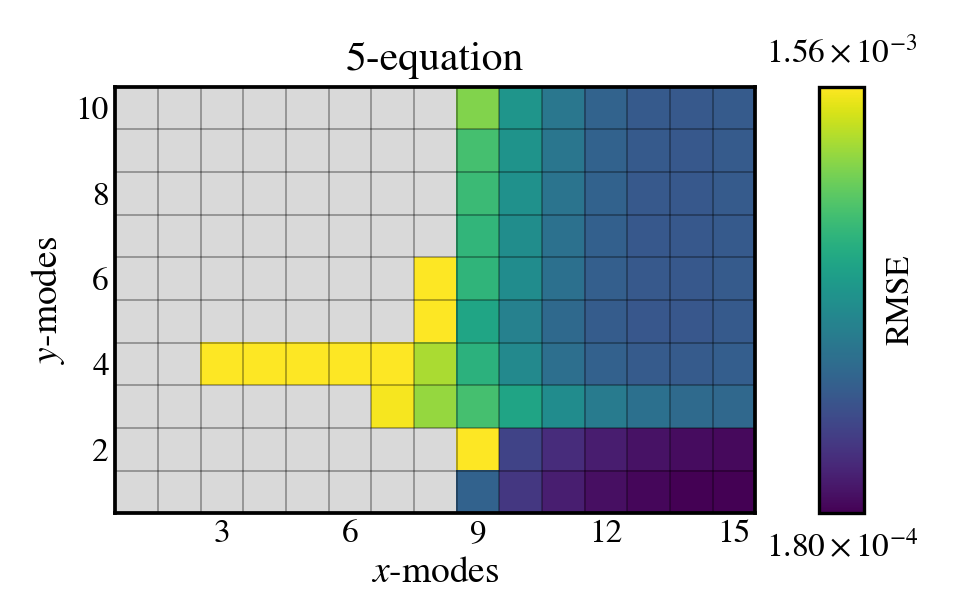}

    \caption{
        RMSE of eGFM velocity-based reduced models over the space of forcing subsets for \(n=1,\ldots,5\) in the complex asymmetric flow. Gray blocks indicate an empty forcing subset, fitting failure, or runtime blowup. For \(n\ge 6\), no successful models were obtained and these cases are omitted.
    }
    \label{fig:parallel:complex-velocity-dataset-1-6}
\end{figure}

For the velocity-based variables, the effect of the training dataset is less systematic, as shown in Fig.~\ref{fig:parallel:complex-velocity-dataset-1-6}. For \(n\ge 3\), increasing \(n_{x\mathrm{mode}}\) generally improves the model, but the dependence on \(n_{y\mathrm{mode}}\) is much less clear. This behavior reflects a limitation of the velocity-based variables, since additional unclosed terms are neglected when $n\ge 3$. This is further supported by the poor convergence of the direct-fit models.

Overall, this example shows that eGFM remains effective for general parallel flows. We also find that the choice of reduced variables strongly affects the model performance. In the next section, we apply the framework further to inhomogeneous flows and parallel shear flows with random coefficients.

%% file: sections/4.extension.tex
\section{\label{sec:extension}Extensions to various flow scenarios}

In this section, we extend the parallel-flow problem to cases with spatial inhomogeneity and temporal randomness, as a step toward configurations relevant to engineering flows such as turbulence.

\subsection{\label{sec:extension-inhomogeneous}Inhomogeneous flows}

\subsubsection{\label{sec:extension-inhomogeneous-problem}Problem statement}

We consider a two-dimensional unsteady inhomogeneous advection-diffusion problem, adapted from Eq.~\eqref{eq:parallel:ADE-HDM} following Ref.~\cite{liu2023systematic}:
\begin{equation}
    \frac{\partial c}{\partial t} + u_1(x,y) \frac{\partial c}{\partial x} + u_2(x,y) \frac{\partial c}{\partial y} = 0.05 \frac{\partial^2 c}{\partial x^2} + \frac{\partial^2 c}{\partial y^2}, \quad \Omega = \left[-\pi, \pi \right] \times \left[-\pi, \pi \right].\label{eq:extension:ADE-HDM-inhom}
\end{equation}
The coefficient \(0.05\) results from the directional nondimensionalization in Ref.~\cite{liu2023systematic}.
The system is periodic in both $x$ and $y$, with initial condition $c(x,y,0) = \bar{c}_0(x)$. The averaging operator $\bar{\bullet}$ denotes the cross-sectional average in $y$, defined in Eq.~\eqref{eq:intro:averaging}.

We consider two inhomogeneous flow cases~\cite{liu2023systematic}. Inhomogeneous flow I is defined as
\begin{equation}
    u_1(x,y) = \frac{1}{2}(2+\cos(x)) \cos(y), \quad u_2(x,y) =  \frac{1}{2}\sin(x) \sin(y),
    \label{eq:extension:inhomogeneous-flow-I}
\end{equation}
and inhomogeneous flow II is defined as
\begin{equation}
    u_1(x,y) = (1+\cos(x)) \cos(y), \quad u_2(x,y) = \sin(x) \sin(y).
    \label{eq:extension:inhomogeneous-flow-II}
\end{equation}

\begin{figure}[tbp]
    \centering
        \centering
        \includegraphics{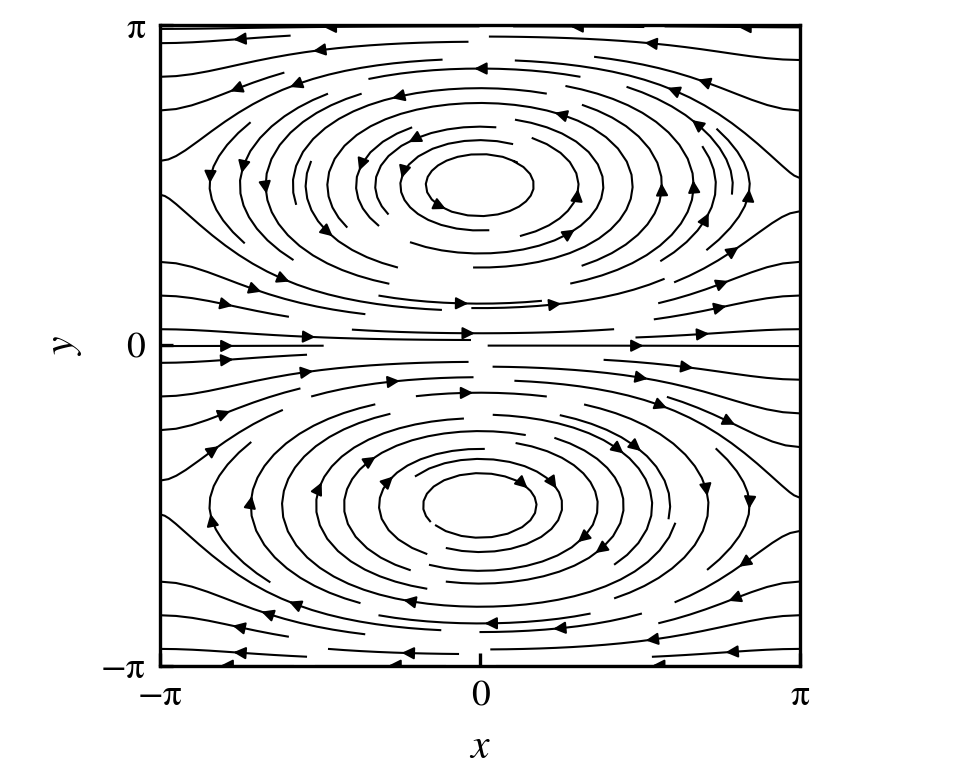}
    \hfill
        \centering
        \includegraphics{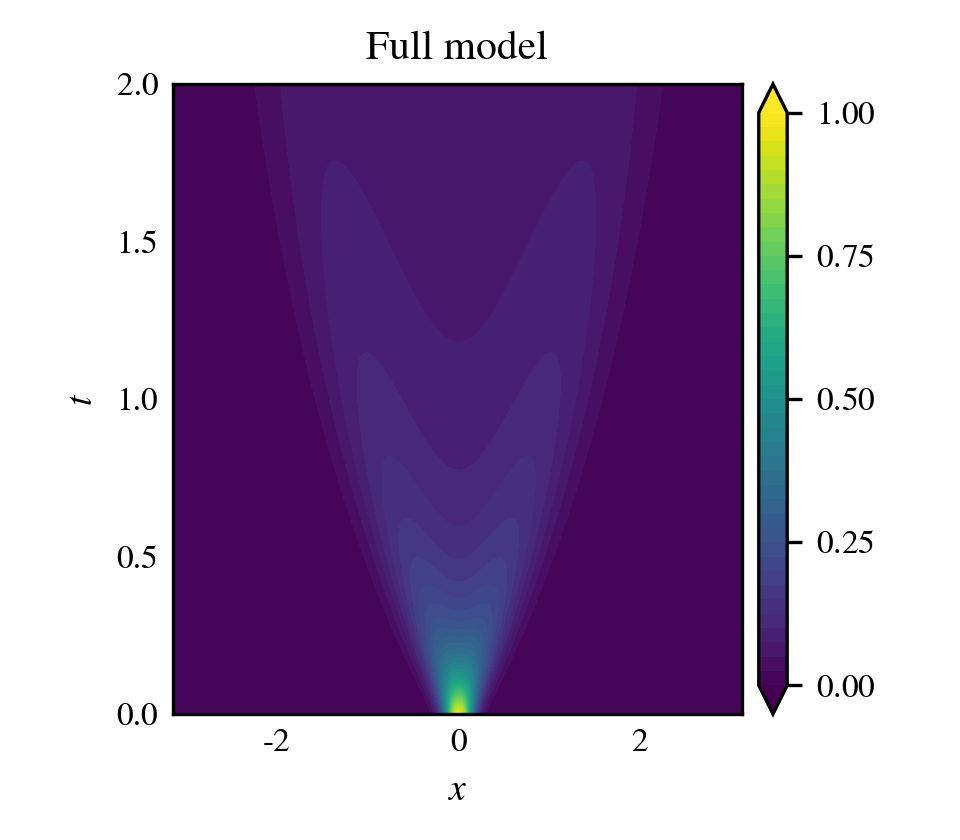}
    \caption{Passive scalar dispersion in inhomogeneous flow I, Eq.~\eqref{eq:extension:inhomogeneous-flow-I}, with initial condition $\bar{c}_0(x) = \exp(-40x^2)$. Left: streamline plot; right: time history of the cross-sectional mean $\bar c(x,t)$.}
    \label{fig:extension:flow-I-dispersion}
\end{figure}

\begin{figure}[tbp]
    \centering
        \centering
        \includegraphics{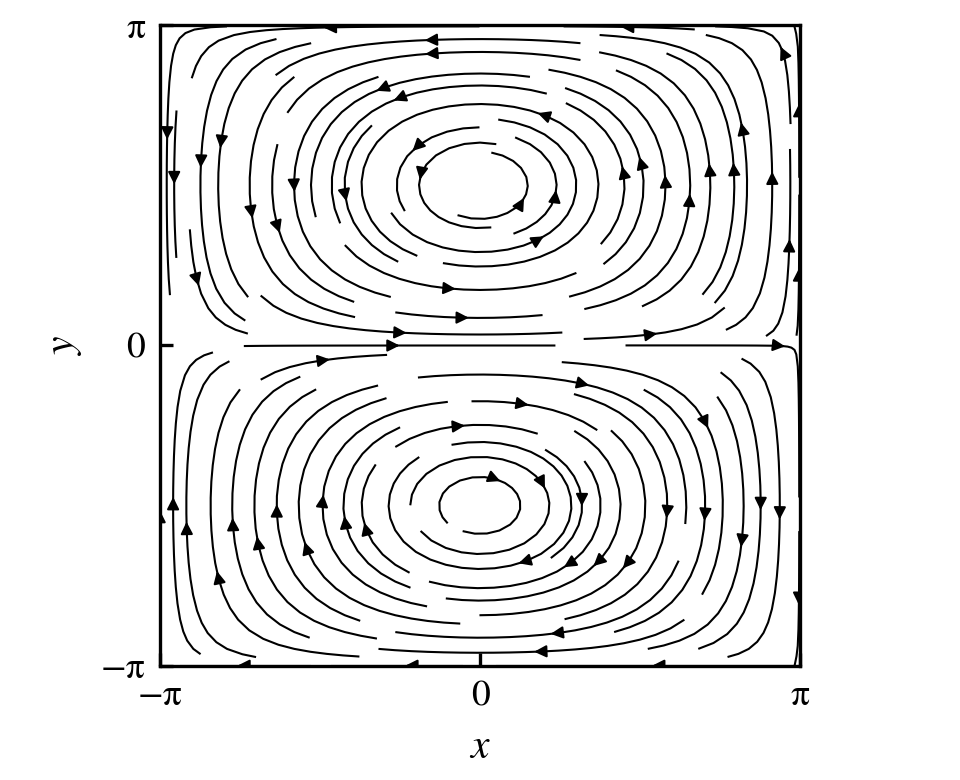}
    \hfill
        \centering
        \includegraphics{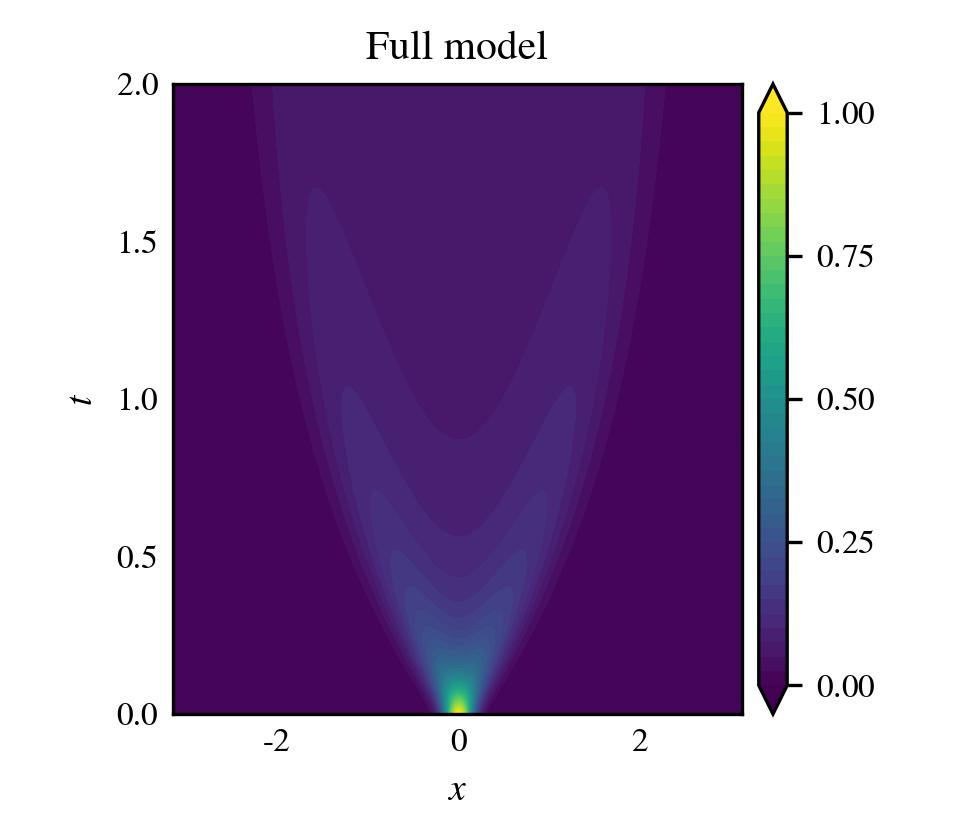}
    \caption{Passive scalar dispersion in inhomogeneous flow II, Eq.~\eqref{eq:extension:inhomogeneous-flow-II}, with initial condition $\bar{c}_0(x) = \exp(-40x^2)$. Left: streamline plot; right: time history of the cross-sectional mean $\bar c(x,t)$.}
    \label{fig:extension:flow-II-dispersion}
\end{figure}

Visualizations of the streamline and time history of the dispersion process for both flow cases are shown in Figs.~\ref{fig:extension:flow-I-dispersion} and~\ref{fig:extension:flow-II-dispersion}.

\subsubsection{\label{sec:extension-modeling}Numerical discovery of the reduced model}

\paragraph{Reduced variables and GFM forcing construction.}

Following the discussion in Sec.~\ref{sec:parallel-problem}, we consider both spectral-based and velocity-based reduced variables. For the spectral-based variables, we adopt the cosine-mode definition in Eq.~\eqref{eq:parallel:spectral-var-cos}, motivated by the symmetry of the velocity field in $y$. For the velocity-based variables, the exact evolution equation of $\bar{c}$ contains one unclosed term $\overline{u_1' c'}$. To avoid the additional complexity of introducing mixed moments, we restrict the construction to the $u_1$ component:
\begin{equation}
    f_0^{v} = \bar{c}, \quad \text{and} \quad f_k^{v} := \overline{u_1'^k c'}, \quad k = 1, \ldots, n-1.
\end{equation}
Although this choice is less physically motivated, it provides insight into how GFM behaves when the reduced variables are defined heuristically.

With this definition of the reduced variables, the eGFM forcings follow Sec.~\ref{sec:parallel-gfm-forcing}, using both sine and cosine functions in $x$.

\paragraph{Model form.}
Due to the spatial inhomogeneity of the physical system, we propose the following reduced model ansatz:
\begin{equation}
    \papa{\bm{f}}{t} = \bm{A}(x) \bm{f} + \bm{B}(x) \papa{\bm{f}}{x} + \bm{C}(x) \papat{\bm{f}}{x},\label{eq:extension:ADE-RM}
\end{equation}
where $\bm{A}(x)$, $\bm{B}(x)$, and $\bm{C}(x)$ are $\mathbb{R}^{n \times n}$-valued, smooth functions of $x$. As in the previous sections, the model is truncated at second-order spatial derivatives.

One important physical constraint on the reduced system is mass conservation. Since $f_0 = \bar{c}$, conservation of total mass is equivalent to
\begin{equation}
    0 = \frac{\partial}{\partial t}\int f_0\,dx
    =
    \int \bm{e}_1^\top \left(\bm{A} - \frac{\partial \bm{B}}{\partial x} + \frac{\partial^2 \bm{C}}{\partial x^2}\right)\bm{f}\,dx,
\end{equation}
where $\bm{e}_1$ is the first standard basis vector, and integration by parts and periodic boundary conditions are used. Therefore, the coefficient functions must satisfy the constraint
\begin{equation}
	\bm{e}_1^\top \left(\bm{A} - \papa{\bm{B}}{x} + \papat{\bm{C}}{x}\right) = 0.\label{eq:extension:LS-constraint}
\end{equation}

\paragraph{Simulation settings.}

For the full model, we use the same numerical schemes as in Sec.~\ref{sec:parallel-simulation-settings}. For the reduced model, because the coefficient functions depend on $x$, the system can no longer be solved exactly in Fourier space. We therefore discretize the reduced model with the same spectral method in $x$, with proper dealiasing.

\paragraph{Optimization problem.}

Following the data generation procedure in Sec.~\ref{sec:intro-forcing}, after generating $N_s$ forced trajectories, we determine the coefficient functions by solving the least-squares problem
\begin{align}
    \widehat{\bm A}, \widehat{\bm B}, \widehat{\bm C}
    =&
    \arg\min_{\bm A, \bm B, \bm C} \frac{1}{N_s} \sum_{i=1}^{N_s} \left\| \frac{\partial \bm{f}^{(i)}}{\partial t} - \mathcal M s^{(i)} - \bm{A} \bm{f}^{(i)} - \bm{B} \frac{\partial \bm{f}^{(i)}}{\partial x} - \bm{C} \frac{\partial^2 \bm{f}^{(i)}}{\partial x^2}   \right\|_F^2 \notag \\
    &
    + \lambda \left(\norm{ \papa{\bm{A}}{x}}_F^2 + \norm{ \papa{\bm{B}}{x}}_F^2 + \norm{ \papa{\bm{C}}{x}}_F^2\right)
    \label{eq:extension:LS-problem}
\end{align}
where $\|\cdot\|_F$ is the Frobenius norm and $\widehat{\bm A}, \widehat{\bm B}, \widehat{\bm C}$ are the approximated model coefficient functions. Since pointwise fitting is often ill-conditioned and prone to overfitting, we introduce a Tikhonov regularization term with parameter $\lambda$ to ensure smoothness of the coefficient functions in $x$. The optimization is subject to the mass-conservation constraint in Eq.~\eqref{eq:extension:LS-constraint}.

In practice, the equality-constrained problem is solved through the associated KKT system~\cite{lawson1995solving}. All derivatives are computed with spectral accuracy.

\subsubsection{\label{sec:extension-inhomogeneous-results}Results and discussion}

For both inhomogeneous flow cases, Eqs.~\eqref{eq:extension:inhomogeneous-flow-I} and~\eqref{eq:extension:inhomogeneous-flow-II}, the eGFM training data are generated on a coarse grid with $N_x=32$, $N_y=32$, $\Delta t = 8\times 10^{-3}$, $T_{\max}=8$, and zero initial condition $\bar c_0 \equiv 0$. Compared to the example in Sec.~\ref{sec:parallel}, more temporal snapshots are taken to improve the stability of the pointwise fit.

For modeling, we consider spectral-based and velocity-based reduced models with $n=1,2,\dots,5$. The eGFM datasets are generated with up to $n_{x\mathrm{mode}}=20$ and $n_{y\mathrm{mode}}=5$ in each case, giving $(20\times5)-1=99$ nontrivial forcing cases per variable type after the uniform case $(p,q)=(0,0)$ is omitted. In the coefficient-fitting procedure, we use $\lambda = 10^{-8}$, which is sufficient to stabilize the eGFM-based models. For comparison, we also report results from the direct-fit approach, for which a larger regularization parameter $\lambda = 10^{-4}$ is used, since the resulting models are less stable.

The test solution is obtained by solving Eq.~\eqref{eq:extension:ADE-HDM-inhom} on a fine grid with $N_x=192$, $N_y=48$, $\Delta t = 2\times 10^{-3}$, $T_{\max}=2$, and initial condition $\bar c_0(x)=\exp(-40x^2)$. The corresponding dispersion processes are shown in Figs.~\ref{fig:extension:flow-I-dispersion} and~\ref{fig:extension:flow-II-dispersion}.

\paragraph{Example 1: Inhomogeneous flow I.}

\begin{figure}[tbp]
    \centering
    \includegraphics{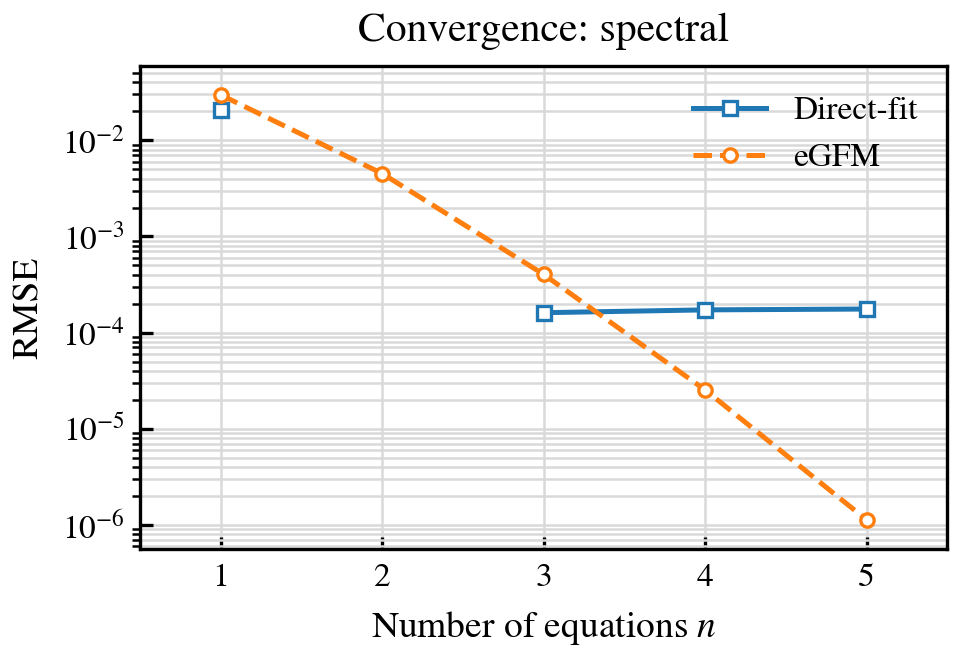}
    \caption{
        Comparison of the RMSE of \(\bar c\) for spectral-based reduced models in inhomogeneous flow I. For the eGFM-based models, we use \(n_{x\mathrm{mode}}=20\) and \(n_{y\mathrm{mode}}=n\). Only successful models are shown.
    }
    \label{fig:extension:flow-I-summary-spectral}
\end{figure}

For the spectral-based variables, the model accuracy is shown in Fig.~\ref{fig:extension:flow-I-summary-spectral}, which plots the RMSE of \(\bar c\) versus the number of equations \(n\). The eGFM-based models converge as \(n\) increases. By contrast, the direct-fit approach fails for $n=2$ and converges poorly for larger $n$.

\begin{figure}[tbp]
    \centering
    \includegraphics{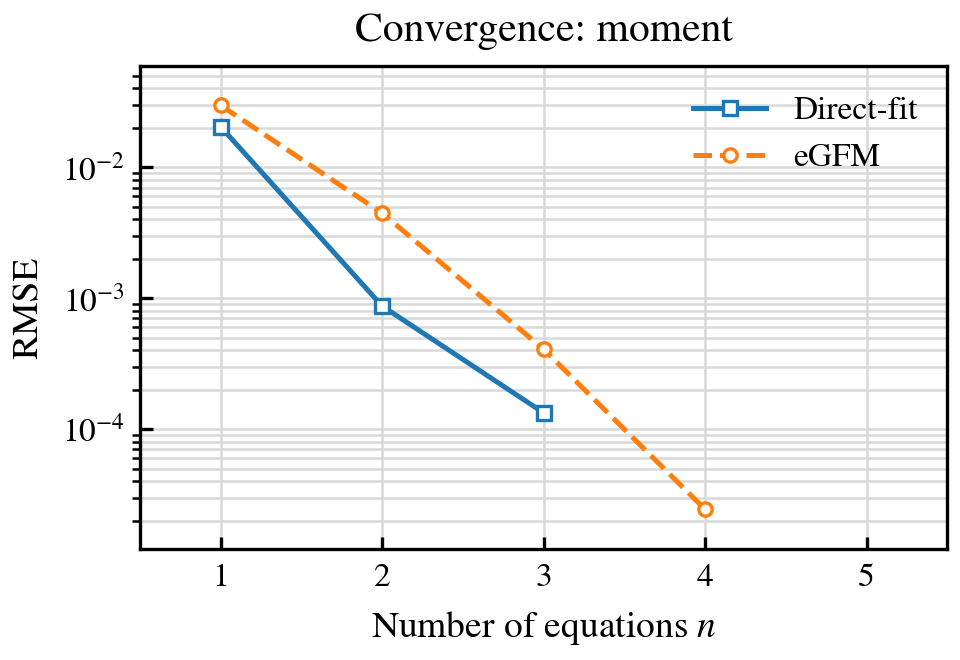}
    \caption{
        Comparison of the RMSE of \(\bar c\) for velocity-based reduced models in inhomogeneous flow I. For the eGFM-based models, we use \(n_{x\mathrm{mode}}=20\) and \(n_{y\mathrm{mode}}=n\). Only successful direct-fit models are shown.
    }
    \label{fig:extension:flow-I-summary-velocity}
\end{figure}

For the velocity-based variables, the corresponding comparison is shown in Fig.~\ref{fig:extension:flow-I-summary-velocity}. Both approaches converge over the range of \(n\) for which a stable model can be obtained: the direct-fit approach is slightly more accurate where it succeeds (\(n\le 3\)), while the eGFM-based models remain stable up to \(n=4\). For \(n=5\), neither approach yields a successful model, indicating an intrinsic limitation of this choice of reduced variables.

\begin{figure}[tbp]
    \centering
        \centering
        \includegraphics{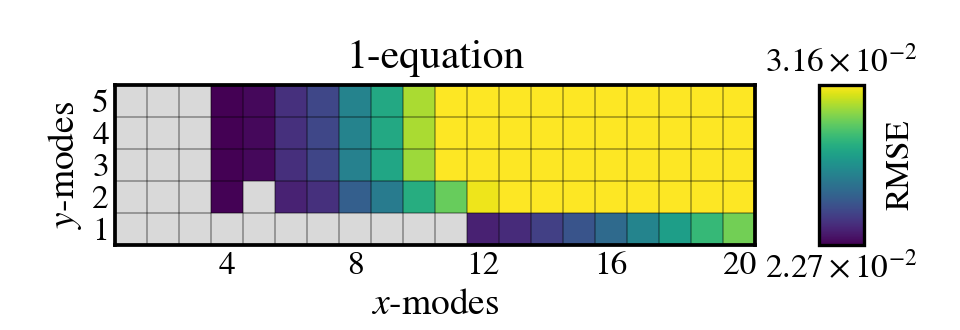}
    \hfill
        \centering
        \includegraphics{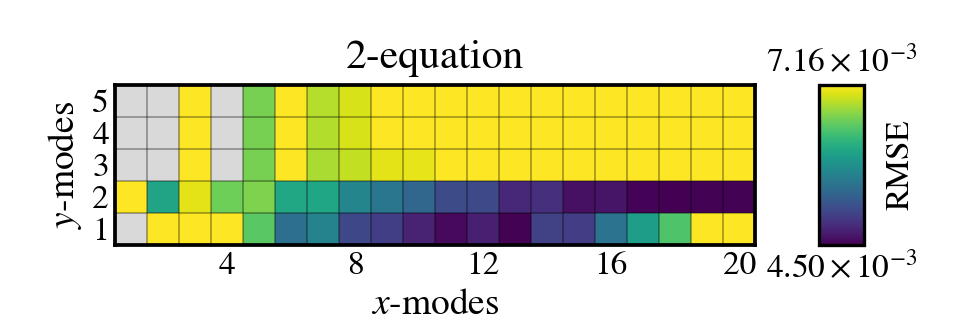}

        \centering
        \includegraphics{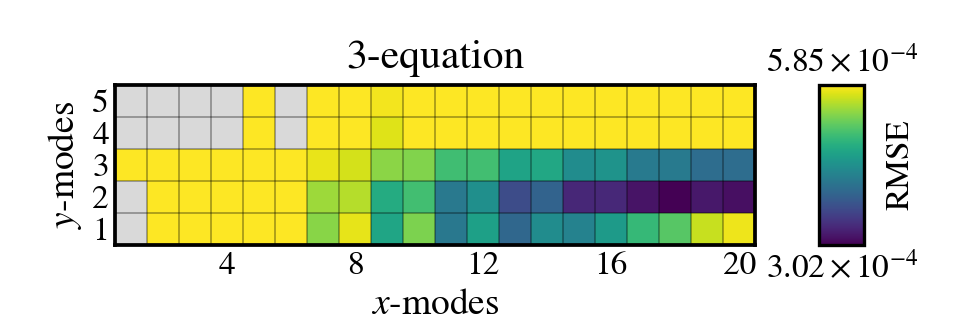}
    \hfill
        \centering
        \includegraphics{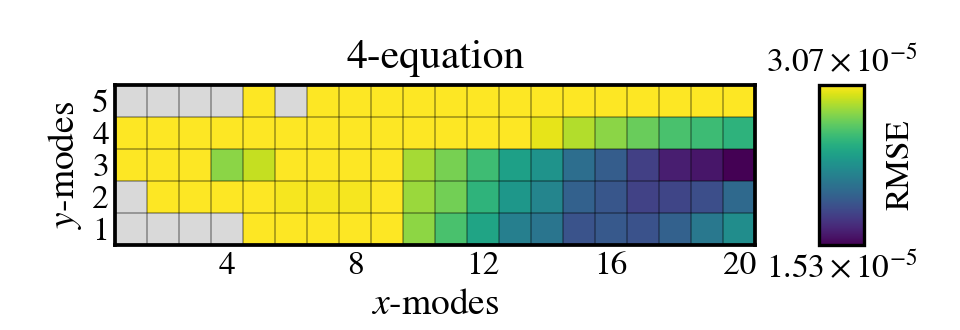}

        \centering
        \includegraphics{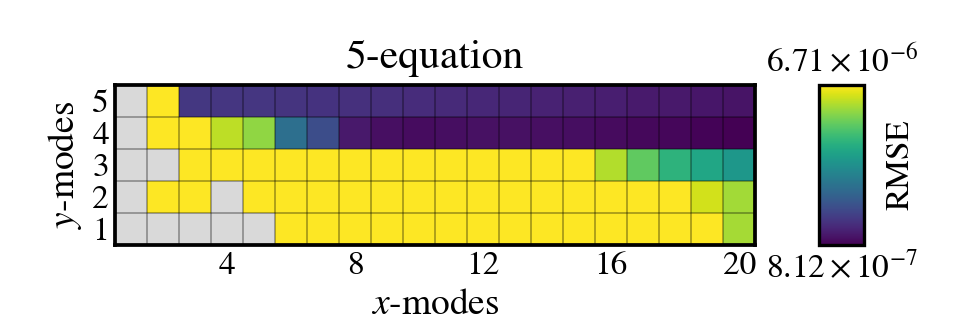}

    \caption{
        RMSE of eGFM spectral-based reduced models over the space of forcing subsets for \(n=1,\ldots,5\) in inhomogeneous flow I. Gray blocks indicate an empty forcing subset, fitting failure, or runtime blowup.
    }
    \label{fig:extension:flow-I-spectral-dataset}
\end{figure}

We next examine the effect of the training dataset on the eGFM-based models. For the spectral-based reduced models, Fig.~\ref{fig:extension:flow-I-spectral-dataset} shows the RMSE over the space of forcing subsets parameterized by \((n_{x\mathrm{mode}}, n_{y\mathrm{mode}})\) for each \(n\).  Four to five $x$-forcing modes are usually sufficient for a stable and accurate fit, and the lowest error is typically achieved near \(n_{y\mathrm{mode}}=n\). For \(n_{y\mathrm{mode}}\ge n+1\), increasing $n_{x\mathrm{mode}}$ tends to increase the error. These observations are consistent with the GFM principle and with the trends reported in Sec.~\ref{sec:parallel-results}.

\begin{figure}[tbp]
    \centering
        \centering
        \includegraphics{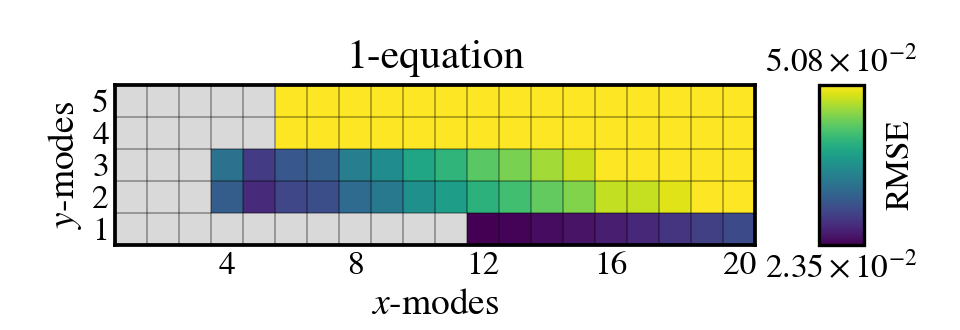}
    \hfill
        \centering
        \includegraphics{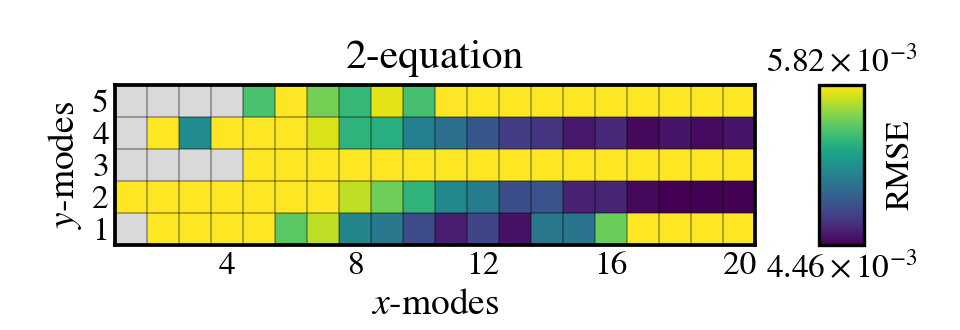}

        \centering
        \includegraphics{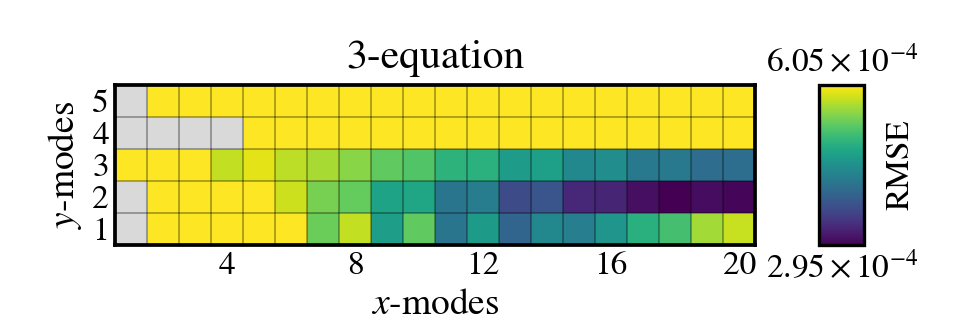}
    \hfill
        \centering
        \includegraphics{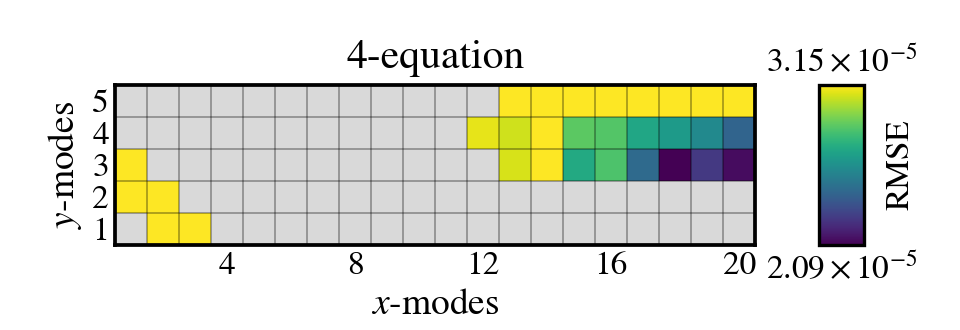}

    \caption{
        RMSE of eGFM velocity-based reduced models over the space of forcing subsets for \(n=1,\ldots,4\) in inhomogeneous flow I. Gray blocks indicate an empty forcing subset, fitting failure, or runtime blowup. The \(n=5\) case is omitted because no successful model is obtained.
    }
    \label{fig:extension:flow-I-velocity-dataset}
\end{figure}

For the velocity-based reduced models, the corresponding training-dataset effect is shown in Fig.~\ref{fig:extension:flow-I-velocity-dataset}, where the \(n=5\) case is omitted because no valid model is obtained. The dependence of accuracy on the forcing subset is qualitatively similar to the spectral-based case for \(n\le 3\). For \(n=4\), however, only a narrow combination of large \(n_{x\mathrm{mode}}\) and \(n_{y\mathrm{mode}}\ge 3\) yields a successful fit. This degradation reflects the limitation of the velocity-moment closure.

\paragraph{Example 2: Inhomogeneous flow II.}

\begin{figure}[tbp]
    \centering
    \includegraphics{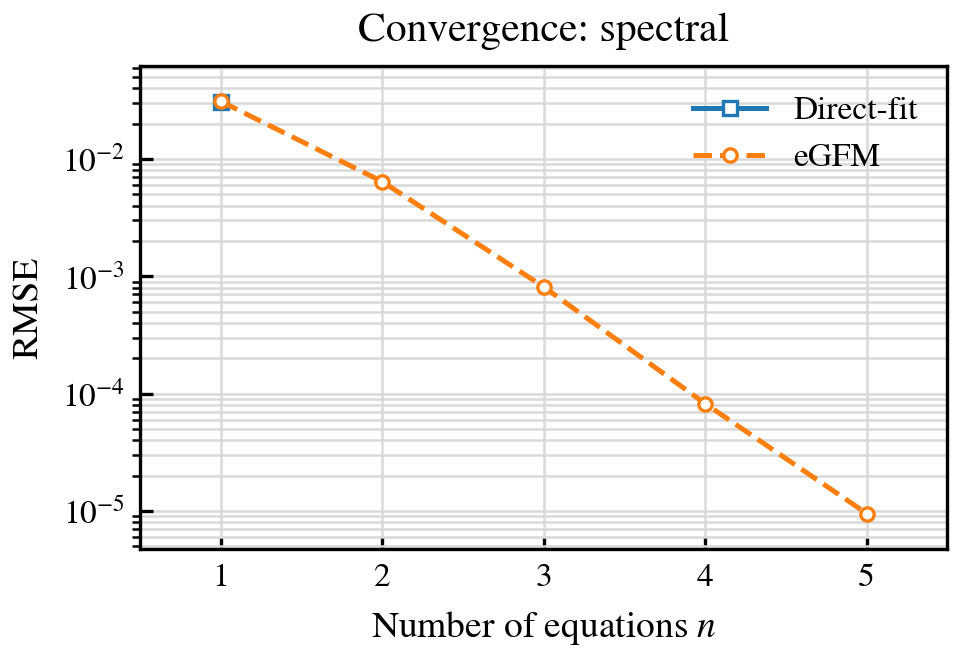}
    \caption{
        Comparison of the RMSE of \(\bar c\) for spectral-based reduced models in inhomogeneous flow II. For the eGFM-based models, we use \(n_{x\mathrm{mode}}=20\) and \(n_{y\mathrm{mode}}=n\). Only successful direct-fit models are shown.
    }
    \label{fig:extension:flow-II-summary-spectral}
\end{figure}

For the spectral-based variables, the model accuracy is presented in Fig.~\ref{fig:extension:flow-II-summary-spectral}. The direct-fit approach fails for all $n\ge 2$, whereas the eGFM-based models continue to converge as \(n\) increases.

\begin{figure}[tbp]
    \centering
    \includegraphics{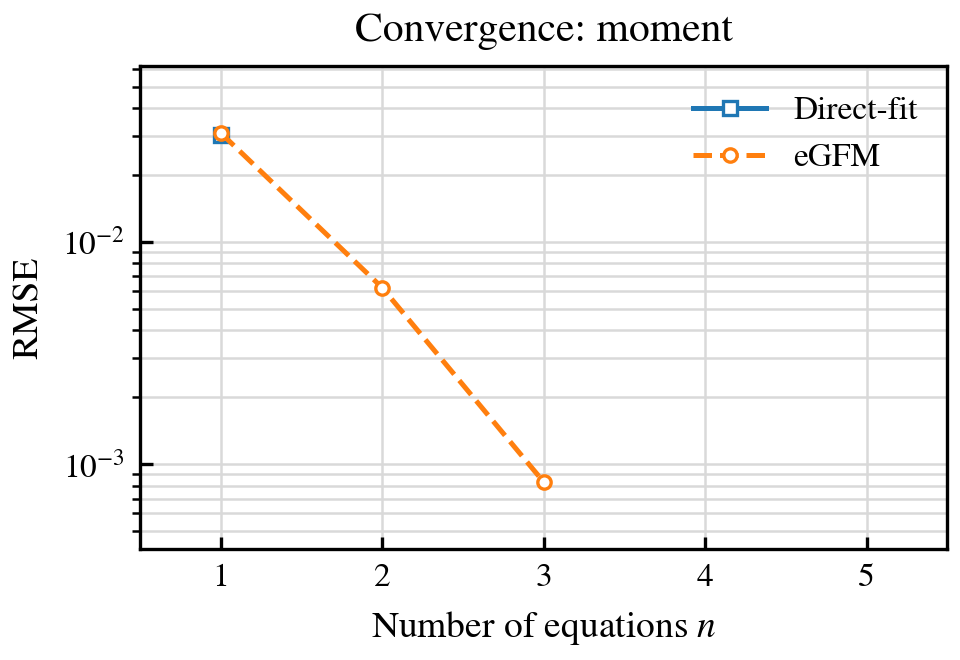}
    \caption{
        Comparison of the RMSE of \(\bar c\) for velocity-based reduced models in inhomogeneous flow II. For the eGFM-based models, we use \(n_{x\mathrm{mode}}=20\) and \(n_{y\mathrm{mode}}=n\).
        Only successful direct-fit models are shown.
    }
    \label{fig:extension:flow-II-summary-velocity}
\end{figure}

For the velocity-based variables, the corresponding comparison is shown in Fig.~\ref{fig:extension:flow-II-summary-velocity}. The eGFM-based models converge for \(n\le 3\), but no successful model is obtained for \(n\ge 4\); the direct-fit approach fails for all $n\ge 2$. This again reflects the limitation of the velocity-moment closure.

\begin{figure}[tbp]
    \centering
        \centering
        \includegraphics{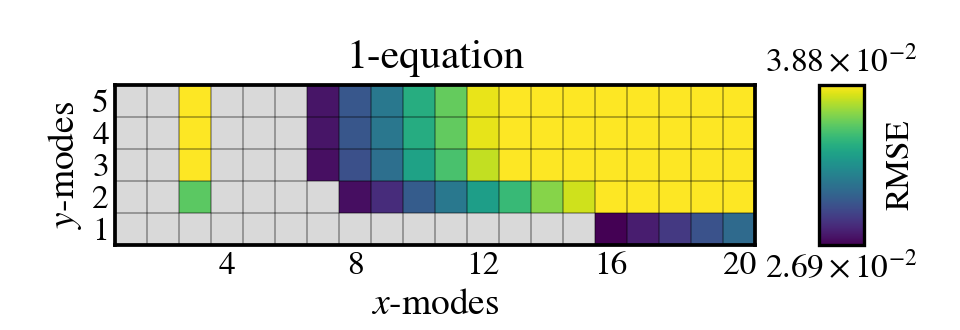}
    \hfill
        \centering
        \includegraphics{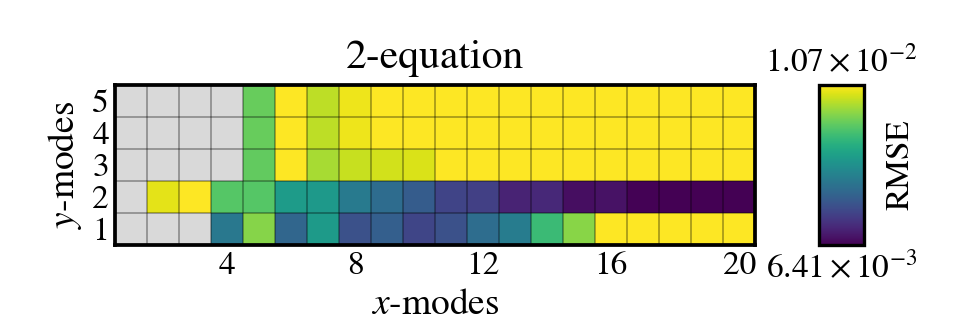}

        \centering
        \includegraphics{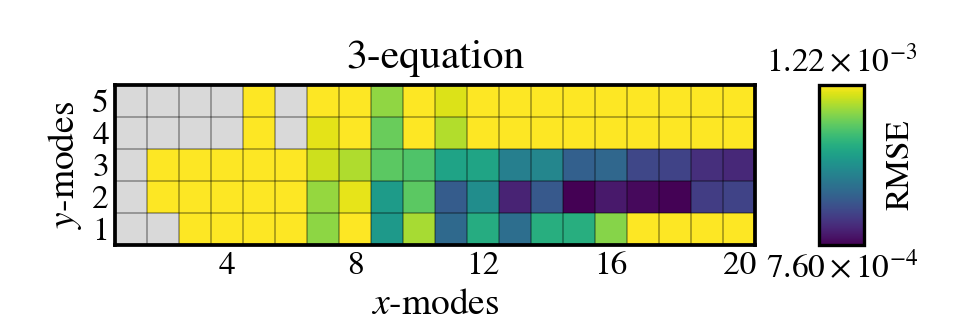}
    \hfill
        \centering
        \includegraphics{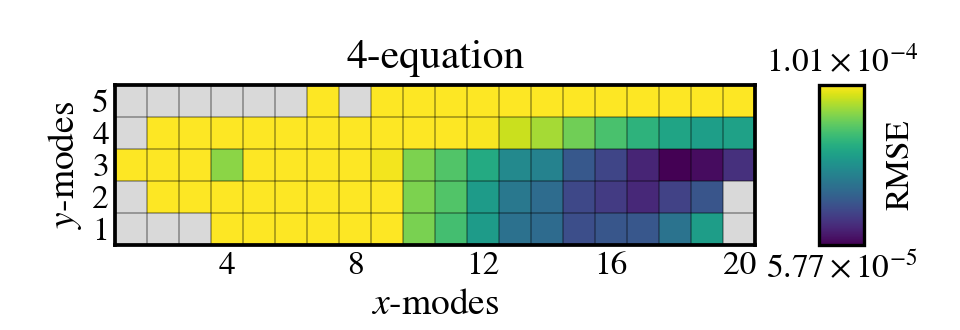}

        \centering
        \includegraphics{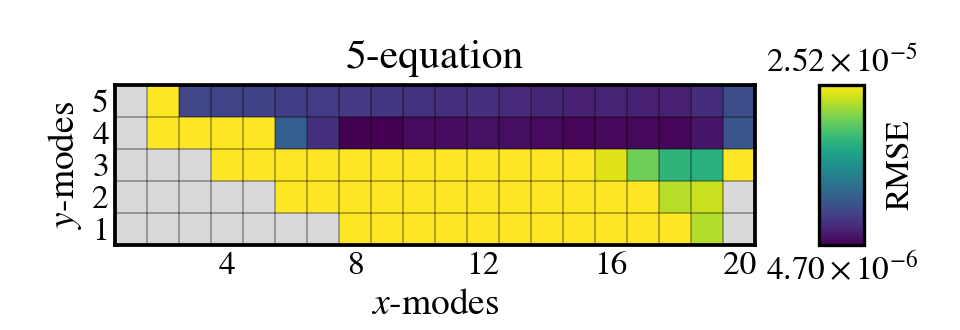}

    \caption{
        RMSE of eGFM spectral-based reduced models over the space of forcing subsets for \(n=1,\ldots,5\) in inhomogeneous flow II. Gray blocks indicate an empty forcing subset, fitting failure, or runtime blowup.
    }
    \label{fig:extension:flow-II-spectral-dataset}
\end{figure}

The effect of the training dataset on the eGFM-based models is shown in Fig.~\ref{fig:extension:flow-II-spectral-dataset} for the spectral-based case. The behavior is qualitatively similar to that of flow I.

\begin{figure}[tbp]
    \centering
        \centering
        \includegraphics{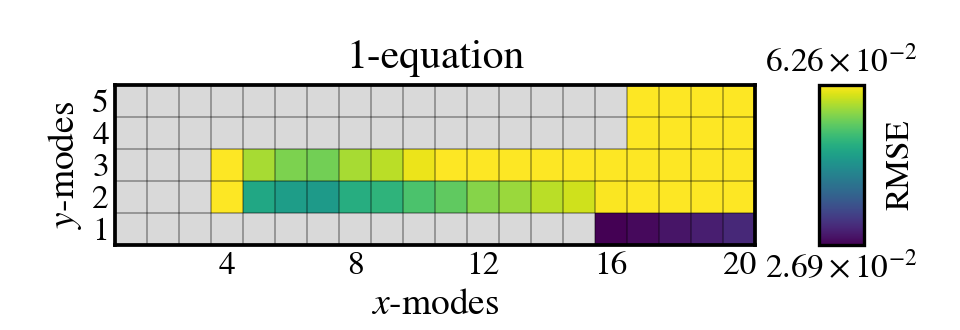}
    \hfill
        \centering
        \includegraphics{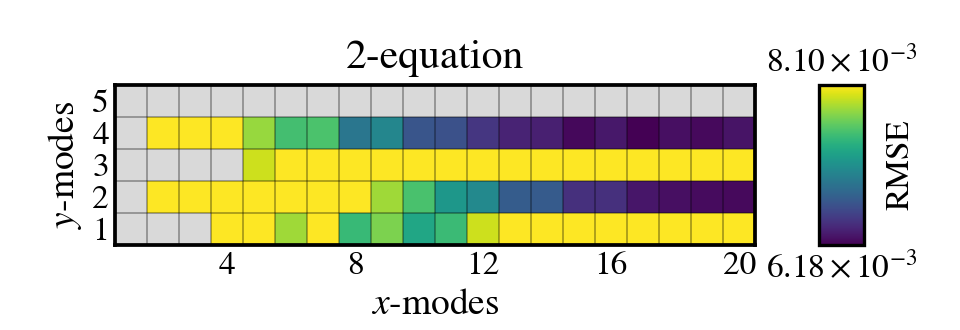}

        \centering
        \includegraphics{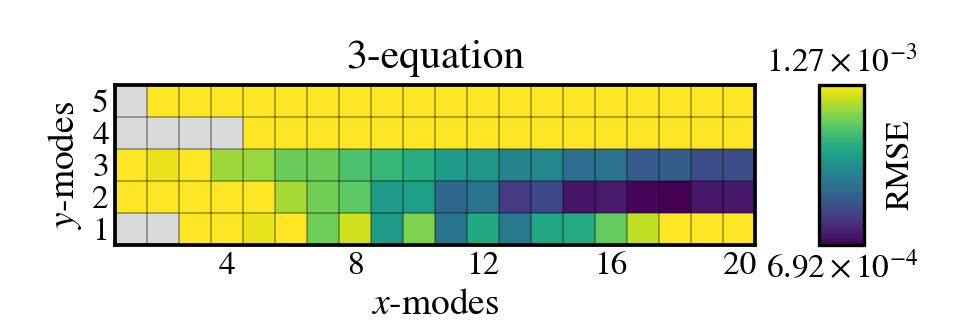}

    \caption{
        RMSE of eGFM velocity-based reduced models over the space of forcing subsets for \(n=1,2,3\) in inhomogeneous flow II. Gray blocks indicate an empty forcing subset, fitting failure, or runtime blowup. The \(n\ge 4\) cases are omitted because essentially no successful model is obtained.
    }
    \label{fig:extension:flow-II-velocity-dataset}
\end{figure}

For the velocity-based reduced models, the corresponding training-dataset effect is shown in Fig.~\ref{fig:extension:flow-II-velocity-dataset}, where the \(n\ge 4\) cases are omitted because no successful model is obtained, reflecting the same closure limitation noted above.

In summary, for both flows, eGFM shows comparable accuracy to the direct-fit approach where direct-fit succeeds, while remaining stable for a wider range of \(n\).
In the next part, we consider a further extension of the prototype problem in Eq.~\eqref{eq:parallel:ADE-HDM}, in which the flow field contains random temporal fluctuations.

\subsection{\label{sec:extension-random}Homogeneous shear flows with random coefficients}

\subsubsection{\label{sec:extension-random-problem}Problem statement}

We consider a modified version of Eq.~\eqref{eq:parallel:ADE-HDM} where the velocity carries a random temporal fluctuation:
\begin{equation}
    \frac{\partial c}{\partial t} + u(y,t) \frac{\partial c}{\partial x} = \frac{\partial^2 c}{\partial y^2}, \quad \Omega =\left(-\infty, \infty \right) \times \left[-\pi, \pi \right],\label{eq:extension:ADE-HDM-random}
\end{equation}
with periodic boundary conditions in $y$ and initial condition $c(x,y,0)=\bar c_0(x)$. The velocity is given by
\begin{equation}
    u(y,t) = \cos(y) \bigl( 1 + 0.5\,\beta(t) \bigr),
\end{equation}
where $\beta(t)$ is a zero-mean Gaussian process with the squared-exponential autocorrelation
\begin{equation}
    \mathbb{E}[\beta(t) \beta(t')] = \exp\left(-\frac{(t-t')^2}{2\ell^2}\right).
\end{equation}

To simplify notation, throughout this subsection the overbar $\bar{\bullet}$ denotes the combined cross-sectional and ensemble average.

\begin{figure}[tbp]
    \centering
        \centering
        \includegraphics{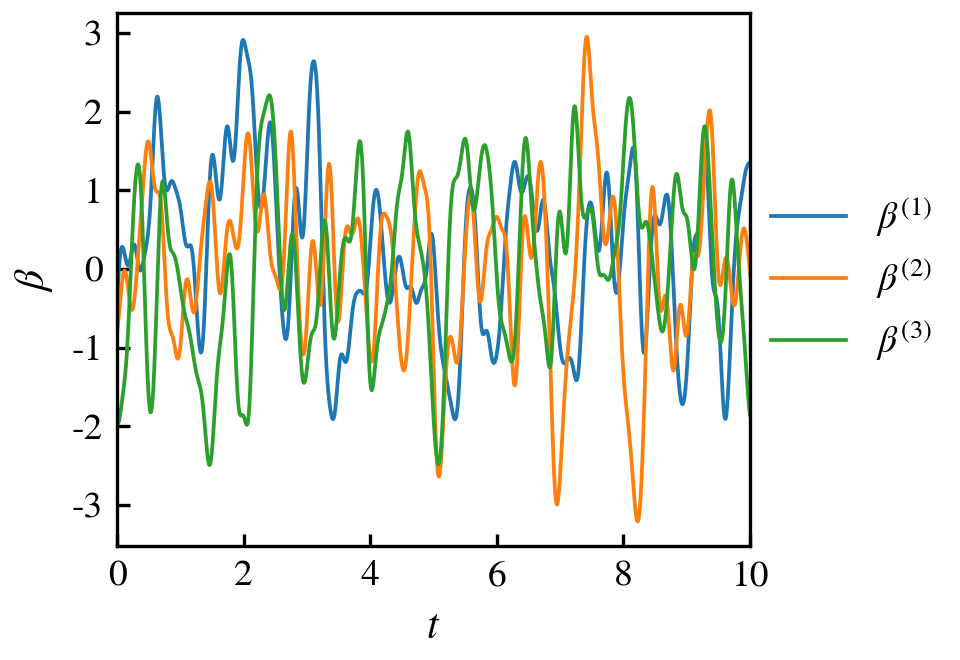}
    \hfill
        \centering
        \includegraphics{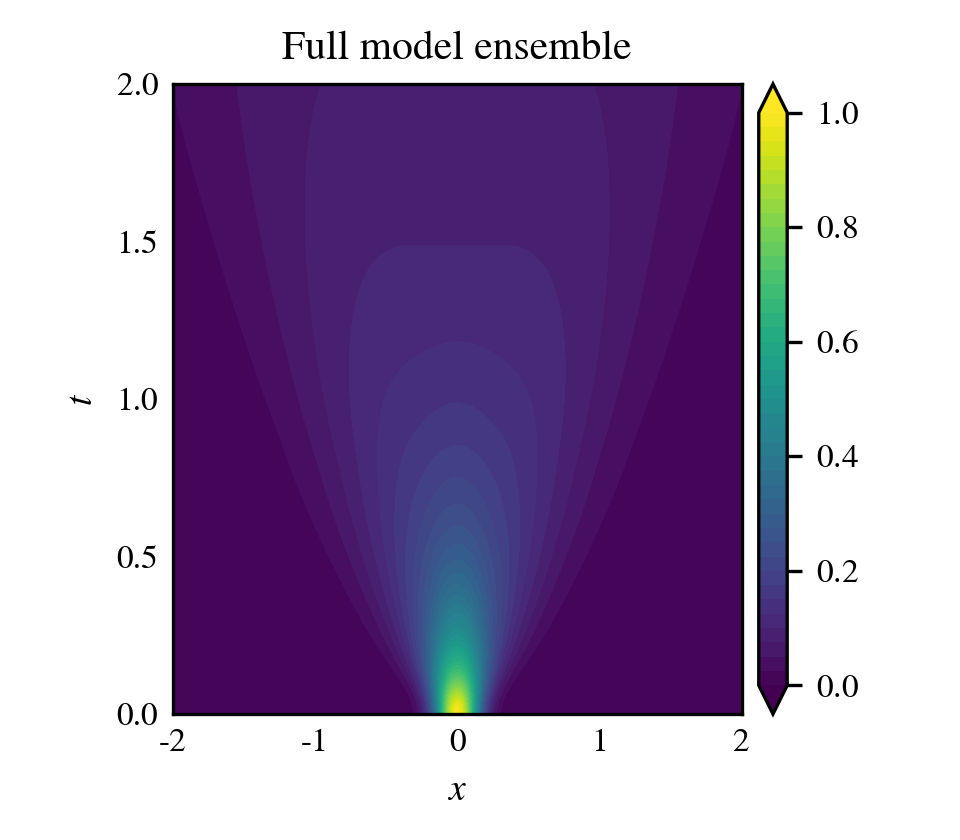}
    \caption{Passive scalar dispersion for the problem in Eq.~\eqref{eq:extension:ADE-HDM-random} with initial condition $\bar{c}_0(x) = \exp(-40x^2)$, ensemble-averaged over $10\,000$ realizations.}
    \label{fig:extension:random-dispersion}
\end{figure}

We set $\ell = 0.1$, corresponding to a finite temporal correlation. Visualizations of $\beta(t)$ and the resulting dispersion of $\bar{c}$ are shown in Fig.~\ref{fig:extension:random-dispersion}.

\subsubsection{\label{sec:extension-random-numerical}Numerical discovery of the reduced model}

\paragraph{Reduced variables and GFM forcing construction.}

As in Sec.~\ref{sec:parallel-problem}, we consider both spectral-based and velocity-based reduced variables. Since $u(y,t)$ is symmetric in $y$, the spectral-based variables follow the cosine-mode definition in Eq.~\eqref{eq:parallel:spectral-var-cos}. The velocity-based variables follow Eq.~\eqref{eq:parallel:velocity-vars}. As discussed in Sec.~\ref{sec:parallel-problem}, here we use a heuristic closure definition. The eGFM forcings for both choices follow Sec.~\ref{sec:parallel-gfm-forcing} without modification.

\paragraph{Model form.}

Because the problem is statistically homogeneous in $x$, the reduced model retains the constant-coefficient ansatz in Eq.~\eqref{eq:parallel:ADE-RM} of Sec.~\ref{sec:parallel}. Analytically, the finite temporal correlation of $\beta(t)$ implies that the ensemble mean $\bar{c}$ does not in general satisfy a local-in-time (Markovian) PDE, except in the limit $\ell\to 0$~\cite{Kubo1963,mori1965transport}. However, memory effects can be approximated by introducing additional equations~\cite{liu2023systematic}. We therefore adopt a Markovian ansatz, as is standard in engineering closure models, and assess its accuracy through the test error.

\paragraph{Simulation settings.}

The full and reduced models use the numerical schemes of Sec.~\ref{sec:parallel-simulation-settings}. The Gaussian process $\beta(t)$ is reconstructed by spectral interpolation, detailed in Sec.~\ref{sec:extension-random-results}.

\paragraph{Optimization problem.}

The coefficient fitting follows the least-squares procedure of Sec.~\ref{sec:intro-forcing}.
In post-processing, the closure terms are evaluated by averaging over $y$ and realizations.

\subsubsection{\label{sec:extension-random-results}Results and discussion}

The full and reduced models follow the numerical setup of Sec.~\ref{sec:parallel-simulation-settings}. We summarize below the discretization parameters specific to this problem and the procedure used to resolve $\beta(t)$.

The eGFM training data are generated on a coarse grid with $L_x=2\pi$, $N_x=32$, $N_y=32$, $T_{\max}=2$, and zero initial condition $\bar c_0\equiv 0$, averaged with 100 realizations. The time step is $\Delta t = 4\times 10^{-3}$ for the spectral-based fit and $\Delta t = 2\times 10^{-3}$ for the velocity-based fit, where the finer step is required for stability in the velocity-based case. We consider spectral-based reduced models with $n=1,\ldots,10$ and velocity-based reduced models with $n=1,\ldots,5$, since the velocity-based models fail to yield valid models for $n>5$ in this setting.

The test solution is obtained by solving Eq.~\eqref{eq:extension:ADE-HDM-random} on a fine grid with $L_x=4\pi$, $N_x=384$, $N_y=48$, $\Delta t = 4\times 10^{-3}$, $T_{\max}=2$, and initial condition $\bar c_0(x)=\exp(-40x^2)$, averaged with 10\,000 realizations. The direct-fit approach is applied to both cases for comparison, with coefficients fitted directly from the test solutions.

The Gaussian process $\beta(t)$ is sampled and stored on a uniform grid with spacing $\Delta t_\beta = 5\times 10^{-3}$ over the interval $[-20\ell,\,10+20\ell]$. The padding of length $20\ell$ on each side ensures that the reconstructed signal in $[0,10]$ is unaffected by truncation. In every simulation, $\beta(t)$ is reconstructed at the integration times required by the DOP853 scheme through spectral interpolation of the stored samples. These time steps are chosen to fully resolve the correlation time scale $\ell=0.1$. We have verified numerically that, for a fixed realization of $\beta(t)$, the discretization error is of order $\mathcal{O}(10^{-9})$ in $c$.

\paragraph{Spectral-based reduced models.}

\begin{figure}[tbp]
    \centering
    \includegraphics[width=\textwidth]{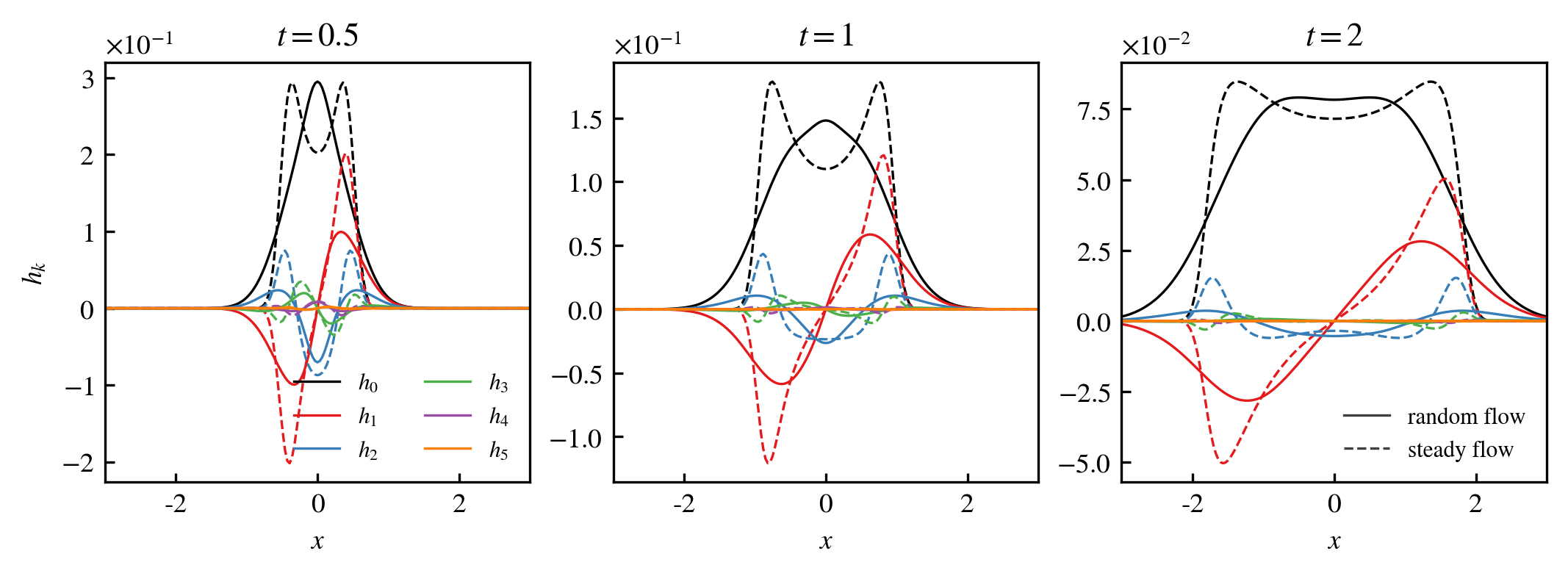}
    \caption{Profiles of the first six spectral modes, $h_0,\ldots,h_5$, for Eqs.~\eqref{eq:extension:ADE-HDM-random} and~\eqref{eq:intro:ADE-HDM-simple} at $t=0.5$, $1$, and $2$.}
    \label{fig:extension:random-spectral-profiles}
\end{figure}

Figure~\ref{fig:extension:random-spectral-profiles} compares the profiles of the first six spectral modes for the random-flow problem with their steady-flow counterparts in Fig.~\ref{fig:intro:spectral-profiles}. In both cases, the amplitudes decrease rapidly with the mode index. The random temporal fluctuations broaden the mean profile $h_0$ and weaken the higher modes, producing a different dispersion pattern from the steady case.

\begin{figure}[tbp]
    \centering
    \includegraphics{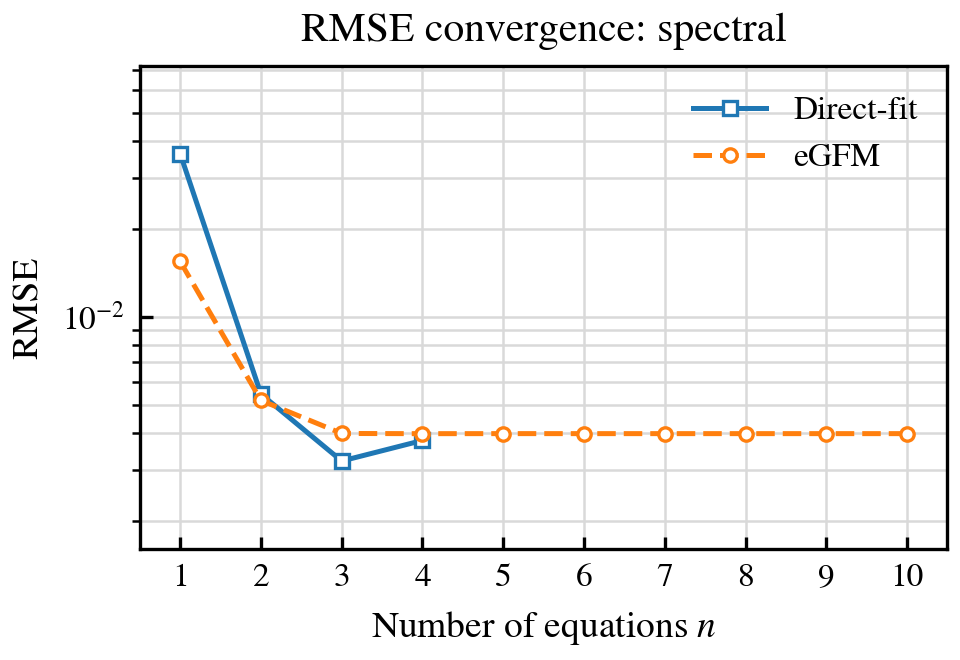}
    \caption{
        Comparison of the RMSE of \(\bar c\) for spectral-based reduced models in the random-flow case. For the eGFM-based models, we use \(n_{x\mathrm{mode}}=5\) and \(n_{y\mathrm{mode}}=n\).
    }
    \label{fig:extension:random-summary-spectral}
\end{figure}

The model accuracy as a function of $n$ is shown in Fig.~\ref{fig:extension:random-summary-spectral}. For $n\le 3$, the eGFM-based and direct-fit models attain comparable RMSE. For $n\ge 4$, the direct-fit approach fails, while the eGFM-based models remain stable but show no further improvement; the error reaches a plateau beyond $n=3$.

\begin{figure}[tbp]
    \centering
        \centering
        \includegraphics{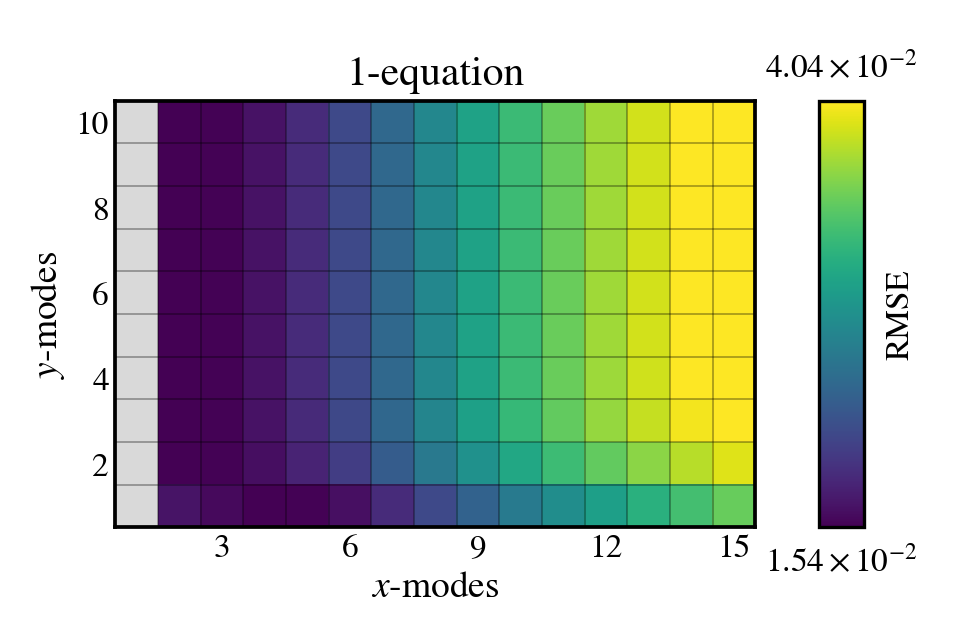}
    \hfill
        \centering
        \includegraphics{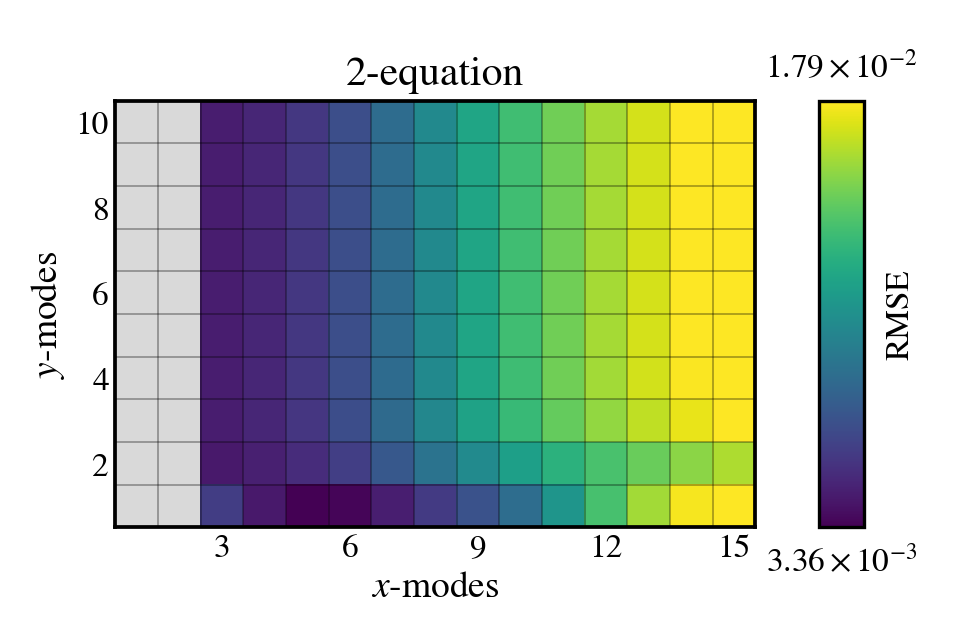}

        \centering
        \includegraphics{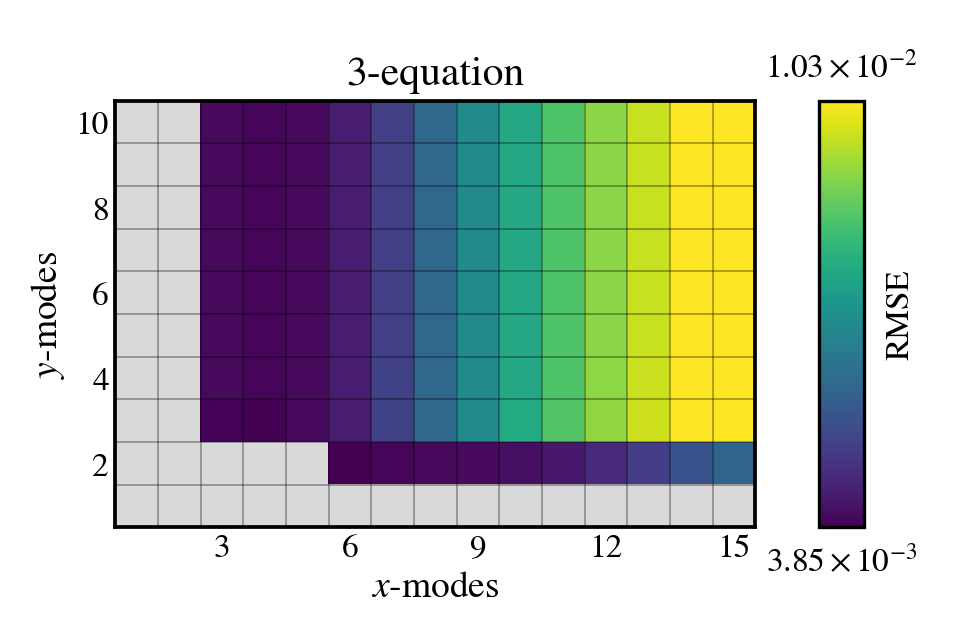}
    \hfill
        \centering
        \includegraphics{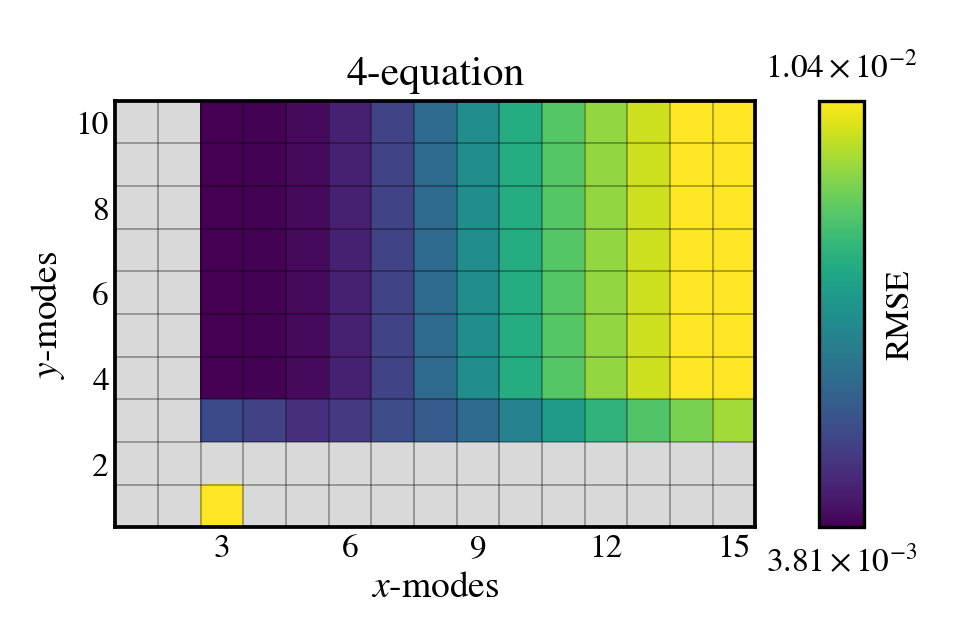}

        \centering
        \includegraphics{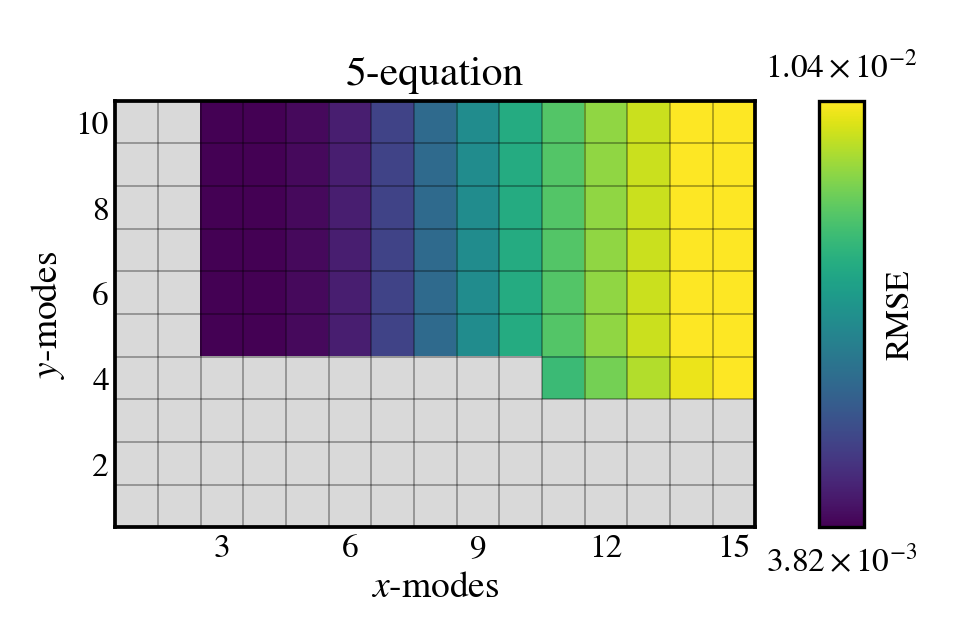}
    \hfill
        \centering
        \includegraphics{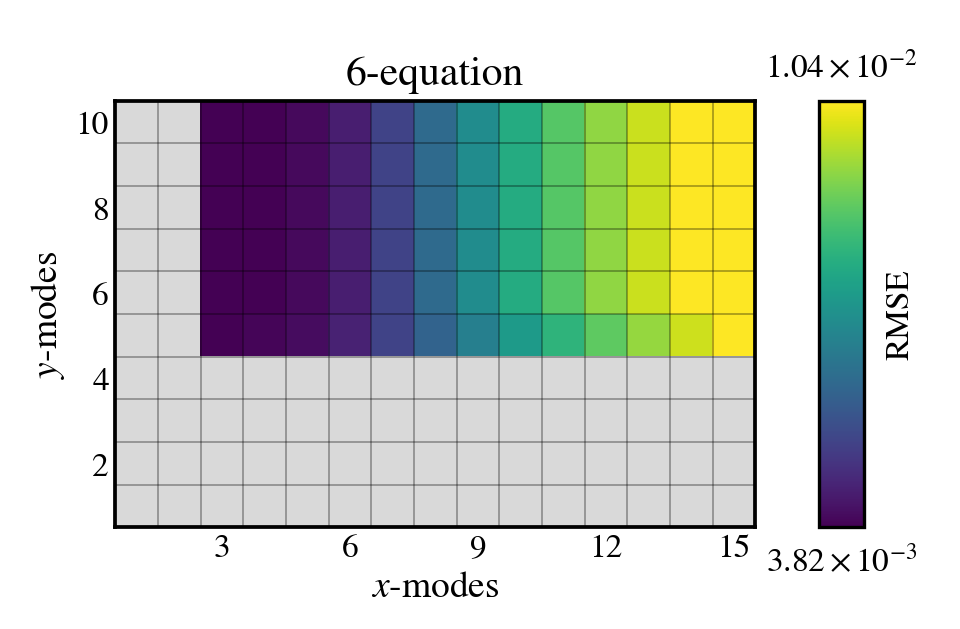}

    \caption{
        RMSE of eGFM spectral-based reduced models over the space of forcing subsets for \(n=1,\ldots,6\) in the random-flow case. Gray blocks indicate an empty forcing subset, fitting failure, or runtime blowup.
    }
    \label{fig:extension:random-spectral-dataset-1-6}
\end{figure}

\begin{figure}[tbp]
    \centering
        \centering
        \includegraphics{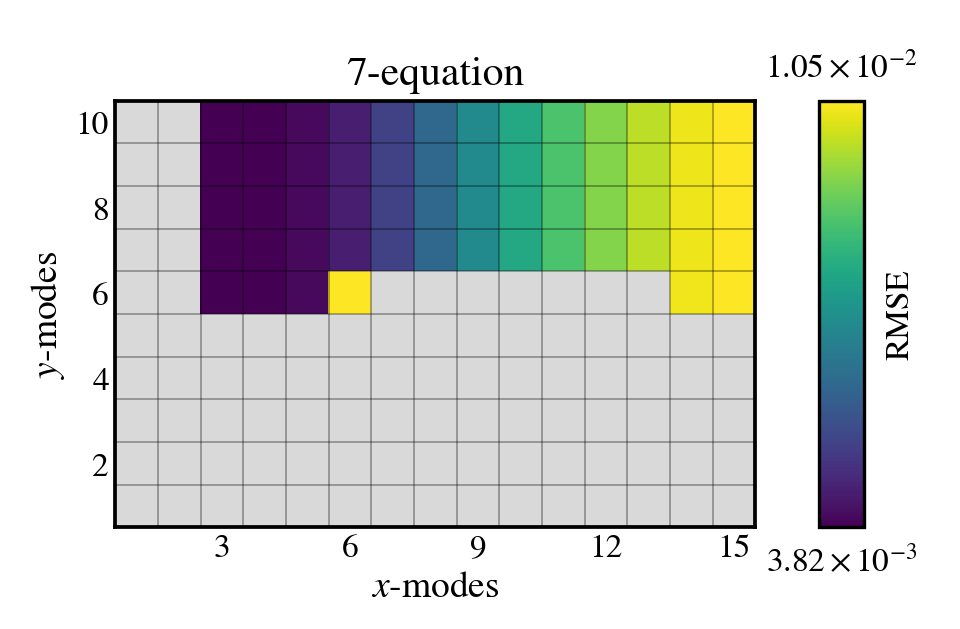}
    \hfill
        \centering
        \includegraphics{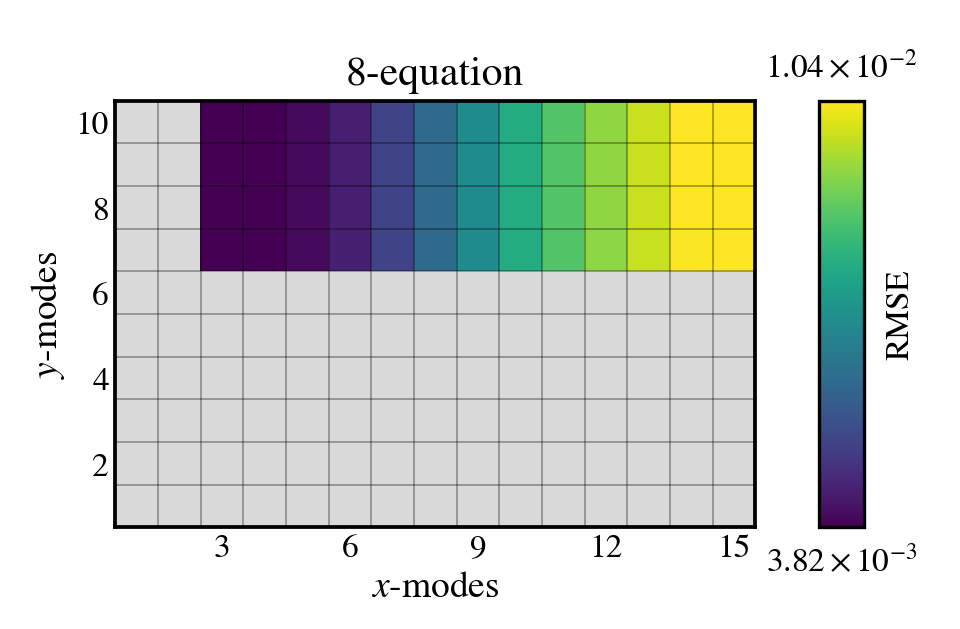}

        \centering
        \includegraphics{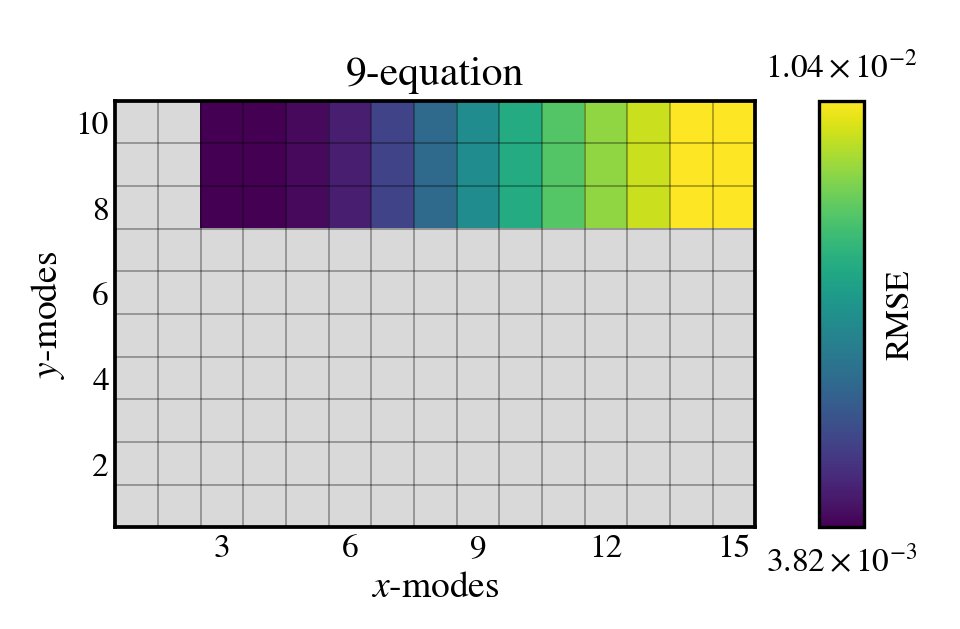}
    \hfill
        \centering
        \includegraphics{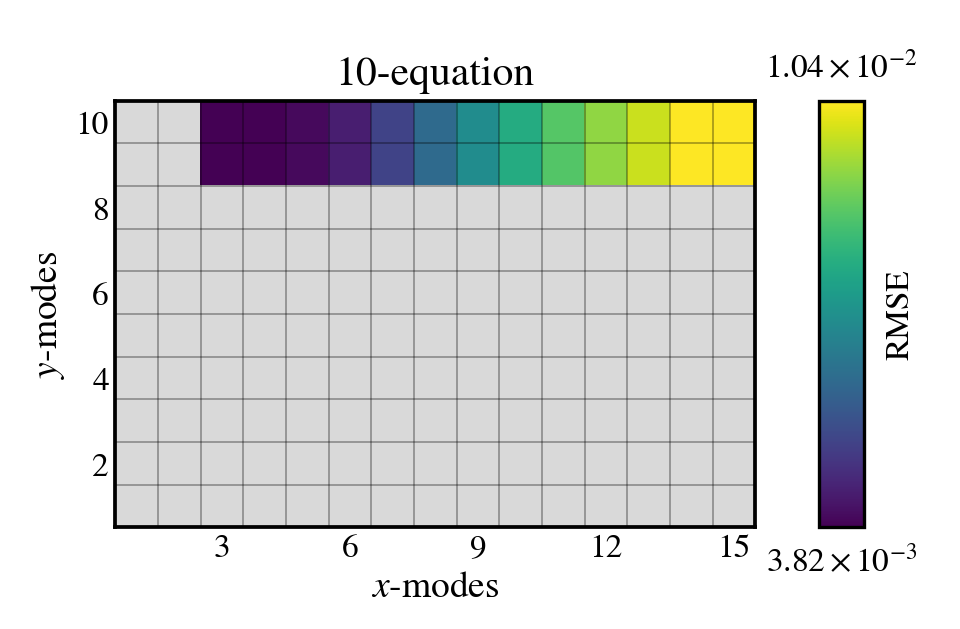}

    \caption{
        RMSE of eGFM spectral-based reduced models over the space of forcing subsets for \(n=7,\ldots,10\) in the random-flow case. Gray blocks indicate an empty forcing subset, fitting failure, or runtime blowup.
    }
    \label{fig:extension:random-spectral-dataset-7-10}
\end{figure}

The dependence on the training dataset is shown in Figs.~\ref{fig:extension:random-spectral-dataset-1-6} and~\ref{fig:extension:random-spectral-dataset-7-10}. In contrast to the previous examples, the present case exhibits qualitatively different trends: increasing $n_{x\mathrm{mode}}$ reduces the predictive accuracy, whereas the dependence on $n_{y\mathrm{mode}}$ remains relatively weak.

We interpret this behavior as evidence that the current spectral-based variables with the model ansatz in Eq.~\eqref{eq:parallel:ADE-RM} are not ideal for the random-flow problem. For $\ell=0.1$, the temporally correlated velocity fluctuations can introduce strong memory effects in the resolved dynamics. Such memory effects can in principle be modeled by adding suitable auxiliary variables~\cite{liu2023systematic}; however, our numerical results show that the spectral variables do not improve the performance for $n > 3$. Further improvement therefore requires using different reduced variables or model forms, rather than simply increasing $n$ or enlarging the admissible forcing subspace.

\paragraph{Velocity-based reduced models.}

\begin{figure}[tbp]
    \centering
    \includegraphics{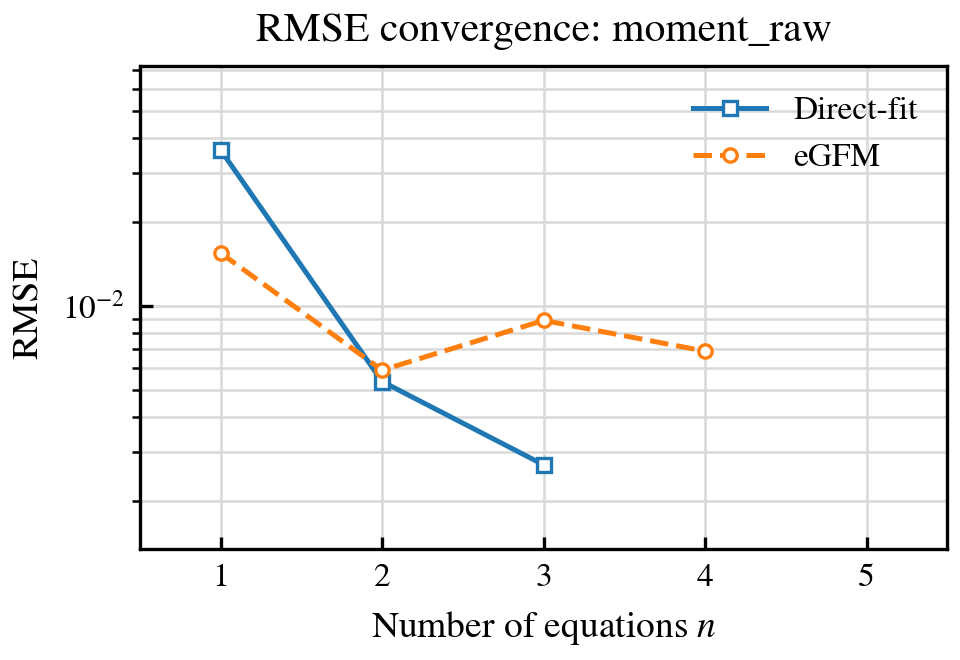}
    \caption{
        Comparison of the RMSE of \(\bar c\) for velocity-based reduced models in the random-flow case. For the eGFM-based models, we use \(n_{x\mathrm{mode}}=5\) and \(n_{y\mathrm{mode}}=n\).
    }
    \label{fig:extension:random-summary-velocity}
\end{figure}

The model accuracy as a function of $n$ is shown in Fig.~\ref{fig:extension:random-summary-velocity}. The direct-fit error decreases slowly through $n=3$ and then fails, while the eGFM-based models remain stable through $n=4$. Neither approach improves substantially beyond $n=2$.

\begin{figure}[tbp]
    \centering
        \centering
        \includegraphics{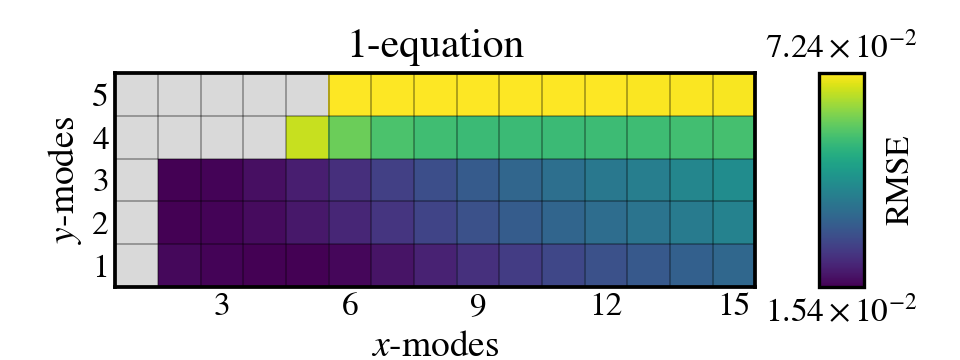}
    \hfill
        \centering
        \includegraphics{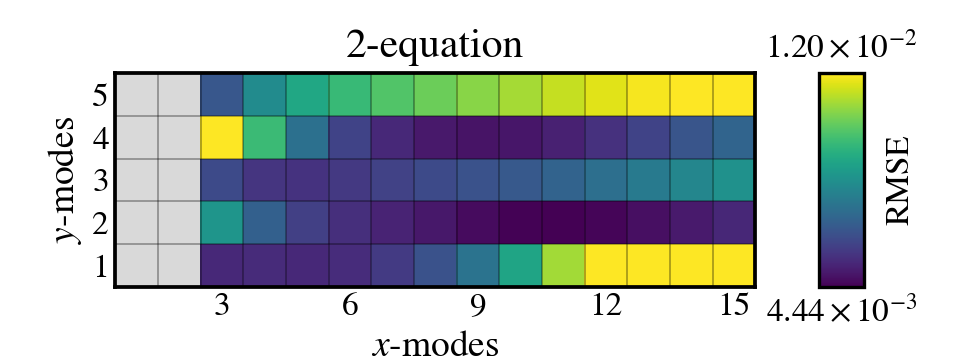}

        \centering
        \includegraphics{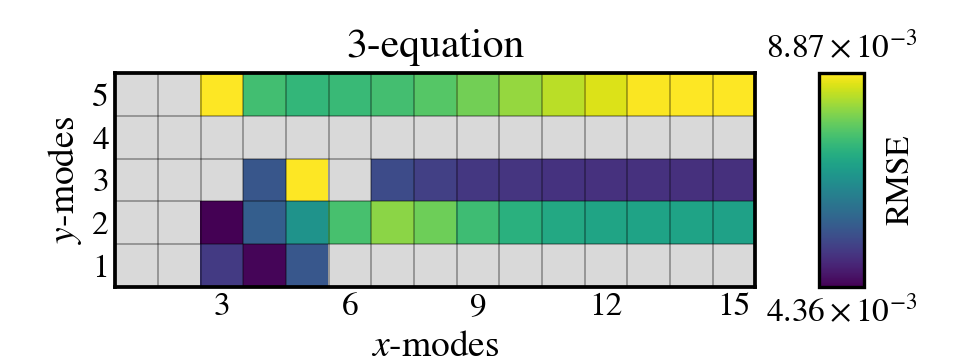}
    \hfill
        \centering
        \includegraphics{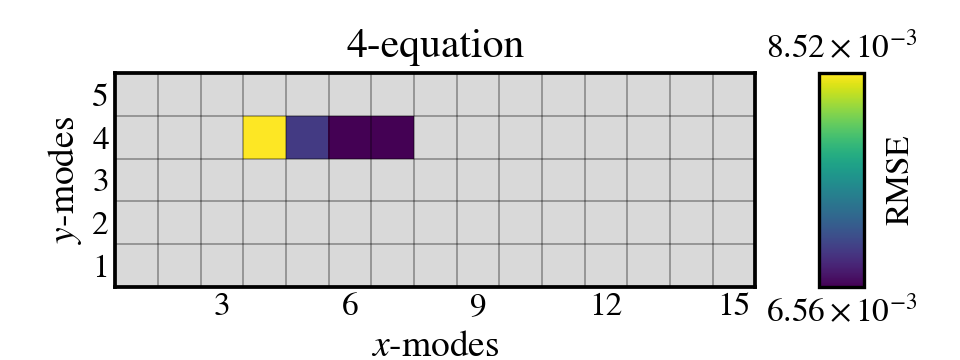}

    \caption{
        RMSE of eGFM velocity-based reduced models over the space of forcing subsets for \(n=1,\ldots,4\) in the random-flow case. Gray blocks indicate an empty forcing subset, fitting failure, or runtime blowup. The \(n=5\) case is omitted because no successful model is obtained.
    }
    \label{fig:extension:random-velocity-dataset}
\end{figure}

The dependence on the training dataset is shown in Fig.~\ref{fig:extension:random-velocity-dataset}. At $n=1$, the trends follow the spectral-based case. At $n=2$ and $n=3$, the trend partially follows the GFM principle: datasets with $n_{y\mathrm{mode}}=n$ and large $n_{x\mathrm{mode}}$ produce low error. For $n\ge 4$, however, the model fails for almost every forcing subset.
This breakdown again reflects mismatches in the variable choice or the model form.

\subsection{Summary}

In this section, two different extensions are studied in detail. For spatially inhomogeneous flows, a spatially dependent closure model ansatz is introduced, and the eGFM-based models remain accurate and stable using a constrained regularized least-squares procedure.

Meanwhile, the random-flow example exposes a limitation of the present choice of reduced variables and model form. For the random flow with finite temporal correlation, they are not capable of representing the resolved dynamics. Thus, increasing the number of variables within the same variable family or enlarging the forcing space does not necessarily improve accuracy. These results indicate that further studies on selecting appropriate reduced variables or model forms are necessary for this type of flow.

%% file: sections/5.turbulence.tex
\section{\label{sec:turbulence}Passive-scalar transport in homogeneous isotropic turbulence}

The previous examples use prescribed canonical velocity profiles. We next consider passive-scalar transport by homogeneous isotropic turbulence (HIT).

\subsection{\label{sec:turbulence-configuration}Problem statement}

The flow field is obtained from a modified large-eddy simulation (LES) of incompressible HIT. Following the linear-forcing approach~\cite{rosales2005linear,bassenne2016constant,homan2024reynolds}, we consider the following governing equation in a triply periodic domain \([-\pi,\pi]^3\):
\begin{equation}
    \papa{\bm u}{t}+\bm u\cdot\nabla\bm u=-\nabla p+\nabla\cdot(\nu_t\nabla\bm u)+\bm \Gamma\widetilde{\bm u}, \qquad \nabla\cdot\bm u=0,
    \label{eq:turbulence:NS}
\end{equation}
where the Smagorinsky--Lilly eddy viscosity is \(\nu_t=(C_s\Delta_u)^2\sqrt{2\,\bm S:\bm S}\)~\cite{smagorinsky1963general}. Here, \(\bm S=[\nabla\bm u+(\nabla\bm u)^T]/2\) is the strain-rate tensor, \(\Delta_u\) is the filter width on the velocity grid, and \(C_s=0.14\) is the Smagorinsky coefficient. We consider the infinite-Reynolds-number limit and therefore set molecular viscosity to zero. Following~\cite{homan2024reynolds}, the velocity used in the linear forcing is high-pass filtered in Fourier space to reduce dependence on the domain size:
\begin{equation}
    \widehat{\widetilde{\bm u}}(\bm k)=G(|\bm k|)\widehat{\bm u}(\bm k), \qquad G(k)=\begin{cases}0, & k\leq2,\\ \frac{1}{2}\left[1-\cos\!\left(\pi(k-2)\right)\right], & 2<k<3,\\ 1, & k\geq3.\end{cases}
    \label{eq:turbulence:forcing-filter}
\end{equation}
The Reynolds-stress tensor is defined by \(R_{ij}(t)=\overline{u_i'u_j'}\), where \(i,j\in\{1,2,3\}\), \(u_i'=u_i-\overline{u_i}\), and the overbar denotes the volume average. This convention of overbar is used only for the HIT flow and is distinct from the averaging used for scalars introduced below. The forcing matrix is diagonal with \(\bm \Gamma(t)=\operatorname{diag}(\gamma_1,\gamma_2,\gamma_3)\). Following~\cite{bassenne2016constant,homan2024reynolds}, its three coefficients are computed by a feedback controller with gain $0.45/\Delta t$ updated every time step to keep \(R_{ii}\) close to the HIT target value \(R_{ii}^{\mathrm{target}}=1\) (no summation in $i$).

\begin{figure}[tbp]
    \centering
    \includegraphics[width=\textwidth]{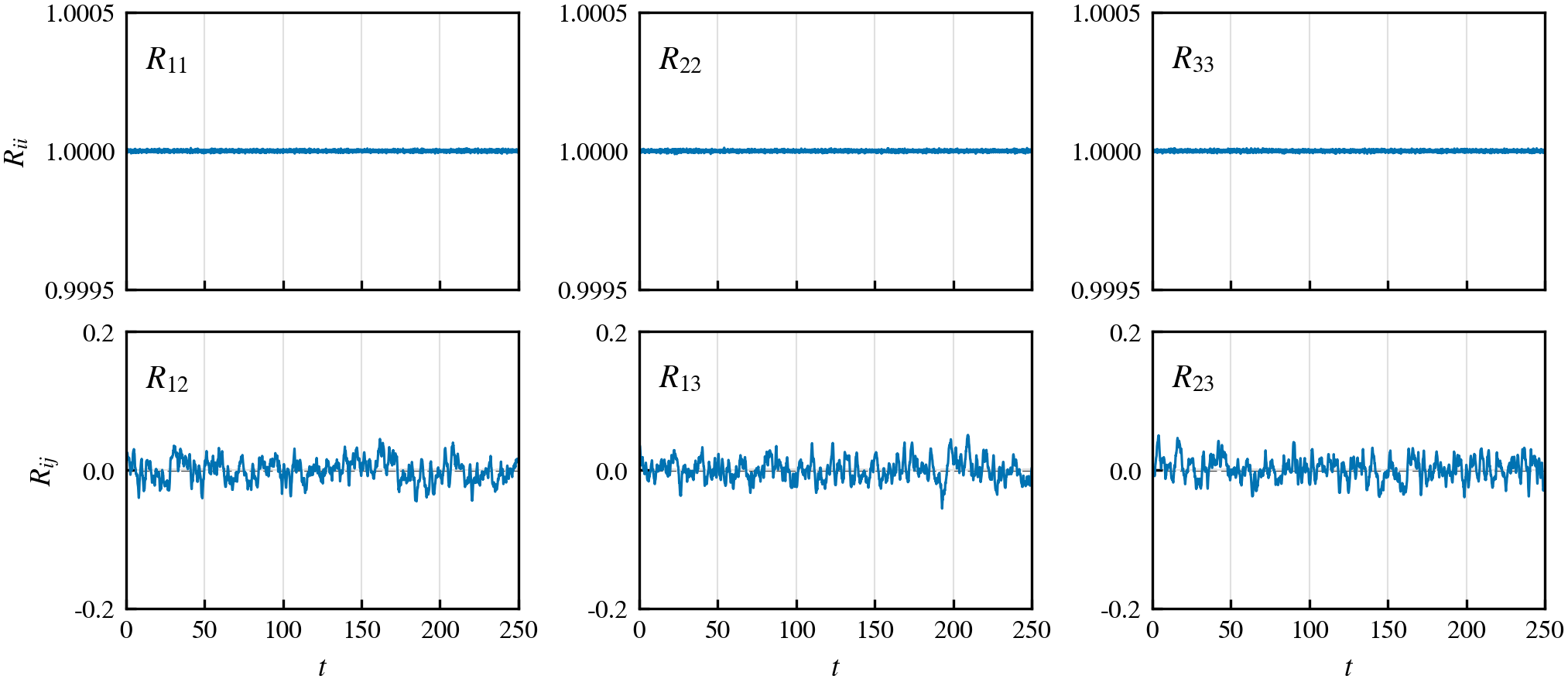}
    \caption{Time histories of the six Reynolds-stress components \(R_{ij}\) in a representative \(192^3\) statistically stationary HIT simulation.}
    \label{fig:turbulence:reynolds-stress}
\end{figure}

\begin{figure}[tbp]
    \centering
    \includegraphics[width=0.65\textwidth]{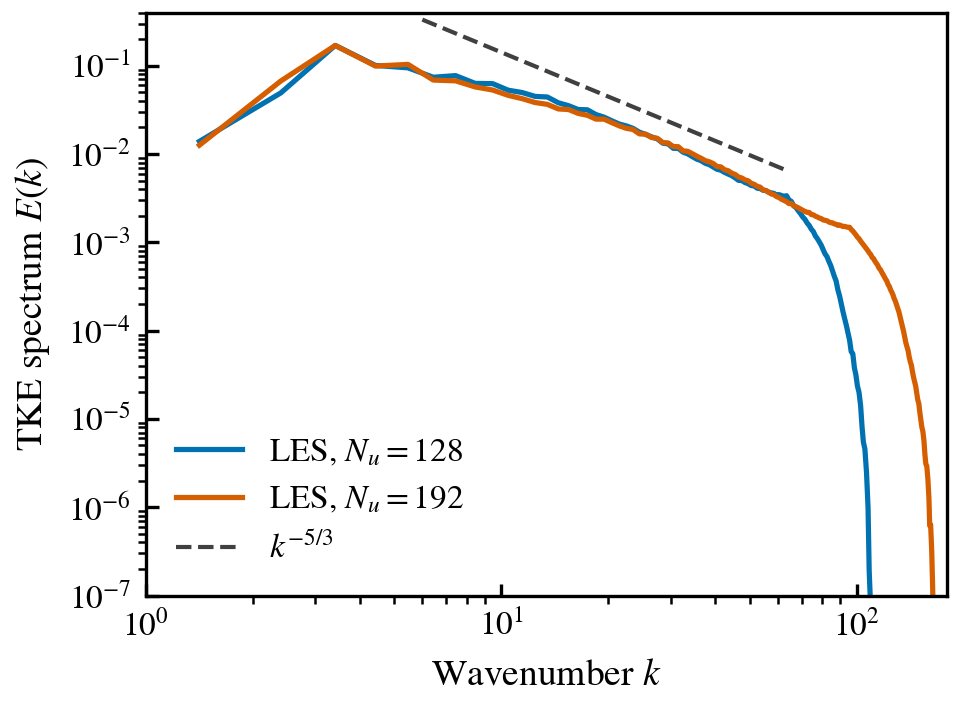}
    \caption{TKE spectra computed from stationary HIT velocity fields on the \(128^3\) and \(192^3\) grids.}
    \label{fig:turbulence:tke-spectrum}
\end{figure}

Figure~\ref{fig:turbulence:reynolds-stress} shows the time histories of all six Reynolds-stress components in a statistically stationary HIT simulation on a \(192^3\) velocity grid. Figure~\ref{fig:turbulence:tke-spectrum} compares the turbulent kinetic energy (TKE) spectra on the \(128^3\) and \(192^3\) velocity grids used below for training and testing, respectively. The two spectra agree over the low and intermediate wavenumbers and exhibit a long interval consistent with \(k^{-5/3}\) scaling.

Given the turbulent velocity field \(\bm u\), we consider a passive scalar \(c(\bm x,t)\) governed by
\begin{equation}
    \papa{c}{t}+\nabla\cdot(\bm u c)=D\nabla^2c, \qquad c(\bm x,0)=\exp(-40x^2),
    \label{eq:turbulence:scalar}
\end{equation}
in the same triply periodic domain, where \(D=0.002\) is the scalar diffusivity. No scalar sub-grid scale model is used. Although the flow is statistically homogeneous, the initial condition breaks the homogeneity in \(x\). We denote the average of \(c\) over the homogeneous directions $y$ and $z$ and over realizations by \(\bar c(x,t)\). Our goal is to construct a reduced transport model of the form introduced in Eq.~\eqref{eq:parallel:ADE-RM}, with \(\bar c\) as its primary resolved variable.

\subsection{\label{sec:turbulence-model}Reduced modeling}

\subsubsection{Reduced variables}

The turbulent velocity field contains many Fourier modes, so the spectral variables used in Sec.~\ref{sec:parallel-problem} would require a large number of equations. We instead use the velocity-moment variables introduced in the same section and define
\begin{equation}
    \phi_0=1, \qquad \phi_1=u_1 - \mathbb{E} [u_1] = u_1, \qquad \phi_2=u_1^2- \mathbb{E}[u_1^2]=u_1^2-1,
    \label{eq:turbulence:basis}
\end{equation}
and the corresponding resolved variables
\begin{equation}
    f_i(x,t)= \mathbb{E} \left\langle\phi_i c\right\rangle_{yz}, \qquad i=0,1,2.
    \label{eq:turbulence:resolved-variables}
\end{equation}
Here, \(\mathbb E\) denotes ensemble averaging, and $\left\langle \bullet \right\rangle_{yz}$ denotes the spatial averaging over the homogeneous directions $y$ and $z$.

As observed in the random-flow example of Sec.~\ref{sec:extension-random-numerical}, adding higher velocity moments gives little improvement. We therefore consider models with up to three equations.

\subsubsection{\label{sec:turbulence-egfm}eGFM forcing and model discovery}

Following Sec.~\ref{sec:parallel-gfm-forcing}, the eGFM training simulations solve Eq.~\eqref{eq:turbulence:scalar} with zero initial condition and additional body forces
\begin{equation}
    s_{p,q}(\bm x,t)=\cos\!\left(\frac{2\pi p x}{L_x}\right)\phi_q(\bm x,t), \qquad p=0,\ldots,n_{x\mathrm{mode}}-1, \quad q=0,\ldots,n_{y\mathrm{mode}}-1,
    \label{eq:turbulence:forcing}
\end{equation}
where the constant case \((p,q)=(0,0)\) is omitted. The full training set uses \(n_{x\mathrm{mode}}=15\) and \(n_{y\mathrm{mode}}=3\), giving 44 nontrivial forcing cases, which are advanced simultaneously with the same velocity field.

We fit the reduced model in Eq.~\eqref{eq:parallel:ADE-RM} using the procedure for random flows in Sec.~\ref{sec:extension-random-numerical}. For the main results, an \(n\)-equation model uses \(n_{x\mathrm{mode}}=15\) and \(n_{y\mathrm{mode}}=n\).

\subsubsection{Numerical settings}

We modify the in-house Fourier pseudospectral solver of Ref.~\cite{homan2024reynolds} to solve the Navier--Stokes and scalar equations. We use the standard \(3/2\)-dealiasing procedure~\cite{canuto2006spectral} and an RK4 time-integration scheme. We denote the velocity and scalar resolutions by \(N_u^3\) and \(N_c^3\), respectively. For reference data, we use a finer scalar grid \(N_c>N_u\) to resolve the small-scale structures, and the velocity is spectrally interpolated onto the scalar grid by zero-padding its Fourier coefficients.

The higher-resolution test data are generated with \(N_u=192\), \(N_c=512\), and \(\Delta t=10^{-3}\). We collect 25 velocity-field snapshots from the stationary state at intervals of 10 time units. These snapshots initialize 25 ensemble dispersion simulations of Eq.~\eqref{eq:turbulence:scalar}. These simulations are run until $T_{\max}=2$ and are then ensemble-averaged to form the test data.

The eGFM training data are generated with \(N_u=N_c=128\) and \(\Delta t=4\times10^{-3}\). We collect 10 velocity-field snapshots from the stationary state at intervals of 20 time units. These snapshots initialize 10 training simulations, each run for 20 time units. The resulting scalar statistics are ensemble-averaged to form the training data.

For comparison, we also apply the direct-fit approach from Sec.~\ref{sec:parallel-results}, fitting Eq.~\eqref{eq:parallel:ADE-RM} directly to the test data. Both the eGFM and direct-fit models are evaluated on the 512-point scalar \(x\)-grid using the reduced-model solver in Sec.~\ref{sec:parallel-simulation-settings}.

\subsection{\label{sec:turbulence-results}Results and discussion}

\subsubsection{Dispersion process}

\begin{figure}[tbp]
    \centering
    \includegraphics{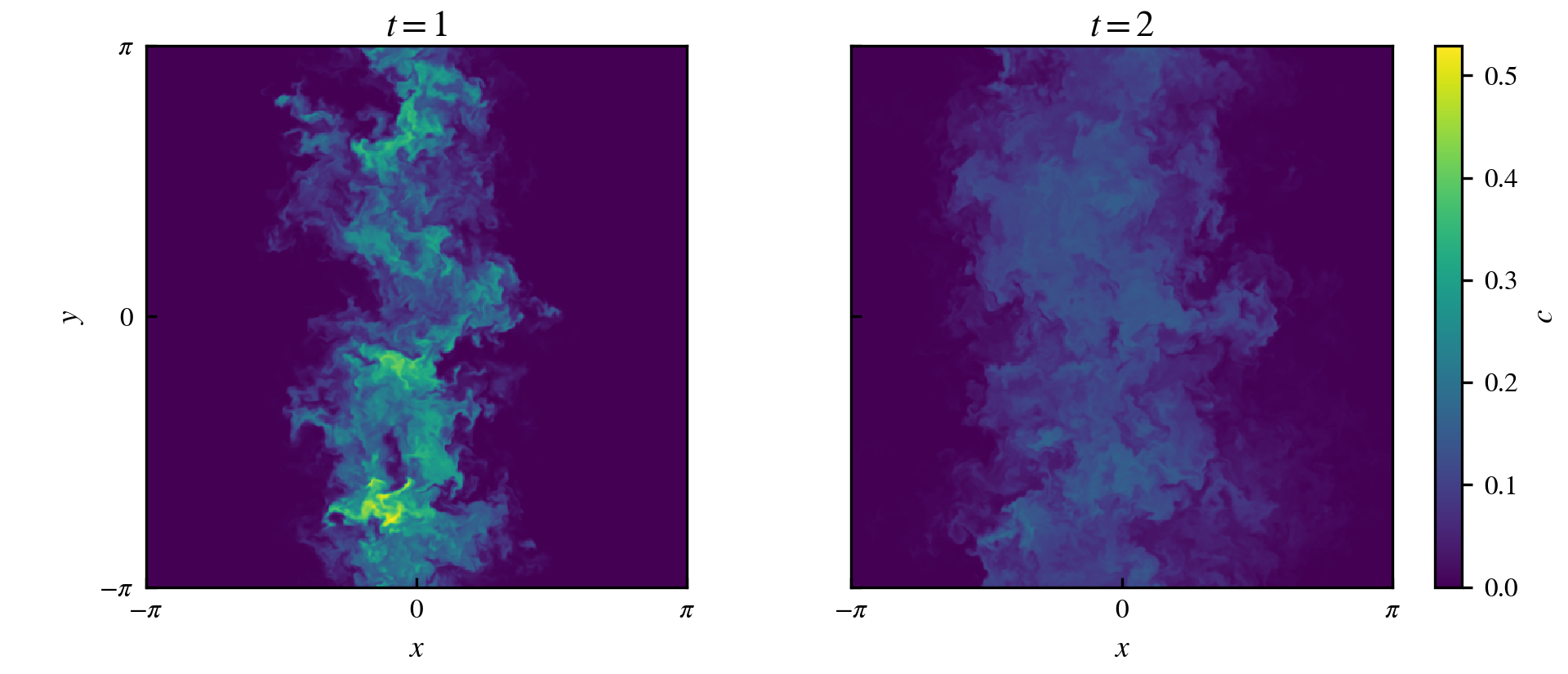}
    \caption{Instantaneous scalar fields \(c(x,y,z=0,t)\) from one test realization at \(t=1\) and \(2\).}
    \label{fig:turbulence:instantaneous-slices}
\end{figure}

We first visualize the process of scalar dispersion of the test runs. Figure~\ref{fig:turbulence:instantaneous-slices} shows instantaneous scalar fields from one test realization. The scalar develops small-scale structures as it is transported by the turbulent velocity field.

\begin{figure}[tbp]
    \centering
    \includegraphics{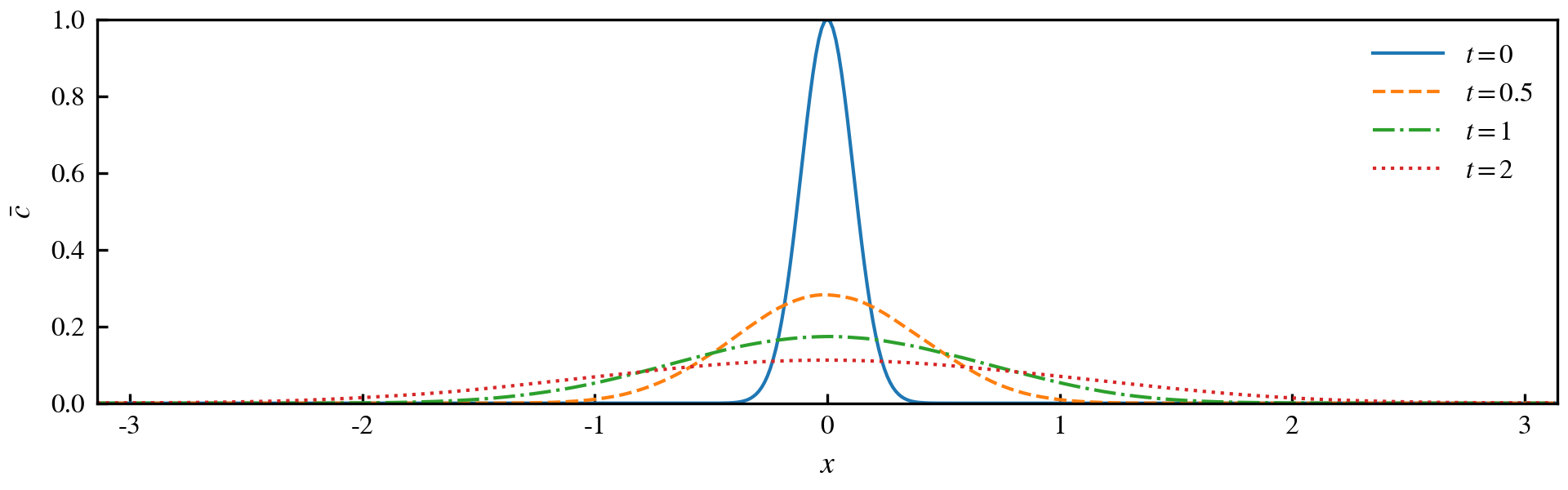}
    \caption{Ensemble-averaged scalar profiles \(\bar c(x,t)\) from the test case at \(t=0,0.5,1,\) and \(2\).}
    \label{fig:turbulence:dispersion-history}
\end{figure}

After averaging over the homogeneous directions and realizations, Fig.~\ref{fig:turbulence:dispersion-history} shows the scalar profiles from the test data. Compared with the parallel flow case in Fig.~\ref{fig:intro:contour}, a much smoother and broader dispersion process is observed.

\subsubsection{Identified models}

The fitted one-, two-, and three-equation eGFM models are reported below. We omit coefficients with magnitude below \(0.001\) and the coefficients are rounded accordingly. This changes the test RMSE by less than \(0.1\%\). Although stability was not imposed during fitting, all three identified models are linearly stable.
\begin{equation}
\papa{f_0}{t}=0.0673\papat{f_0}{x}.
\label{eq:turbulence:model-n1}
\end{equation}
\begin{align}
\papa{f_0}{t}={}&-\papa{f_1}{x}+0.002\papat{f_0}{x},\nonumber\\
\papa{f_1}{t}={}&-3.5279f_1-0.8307\papa{f_0}{x}+0.059\papat{f_1}{x}.
\label{eq:turbulence:model-n2}
\end{align}
\begin{align}
\papa{f_0}{t}={}&-\papa{f_1}{x}+0.002\papat{f_0}{x},\nonumber\\
\papa{f_1}{t}={}&-3.3334f_1-0.0067f_2-0.8963\papa{f_0}{x}-0.7546\papa{f_2}{x}+0.0107\papat{f_1}{x},\nonumber\\
\papa{f_2}{t}={}&-0.1825f_0+0.0151f_1-7.6978f_2-0.0013\papa{f_0}{x}-1.4497\papa{f_1}{x}\nonumber\\
&+0.0015\papa{f_2}{x}-0.032\papat{f_0}{x}+0.021\papat{f_2}{x}.
\label{eq:turbulence:model-n3}
\end{align}

\subsubsection{Model testing}

\begin{figure}[tbp]
    \centering
    \includegraphics{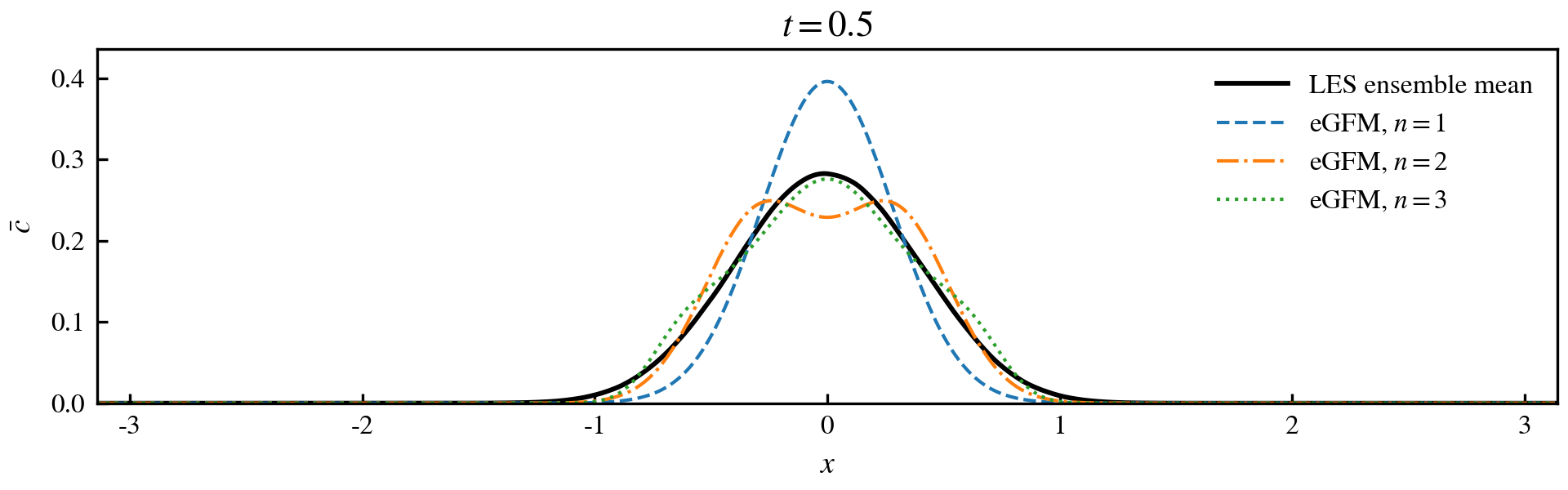}
    \includegraphics{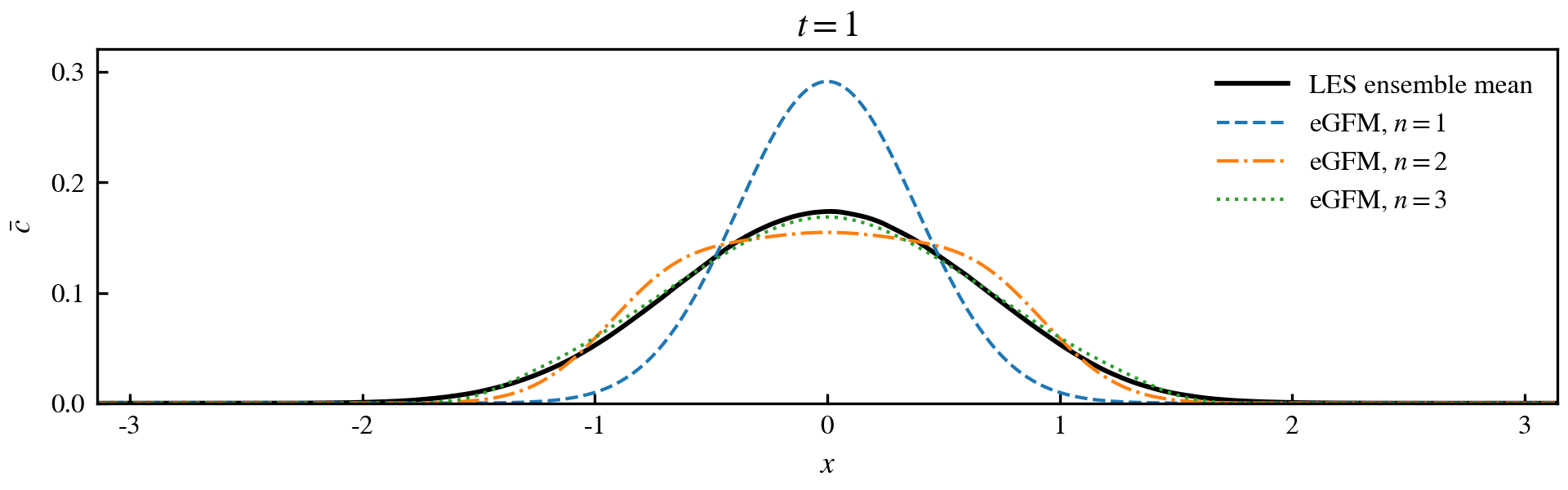}
    \includegraphics{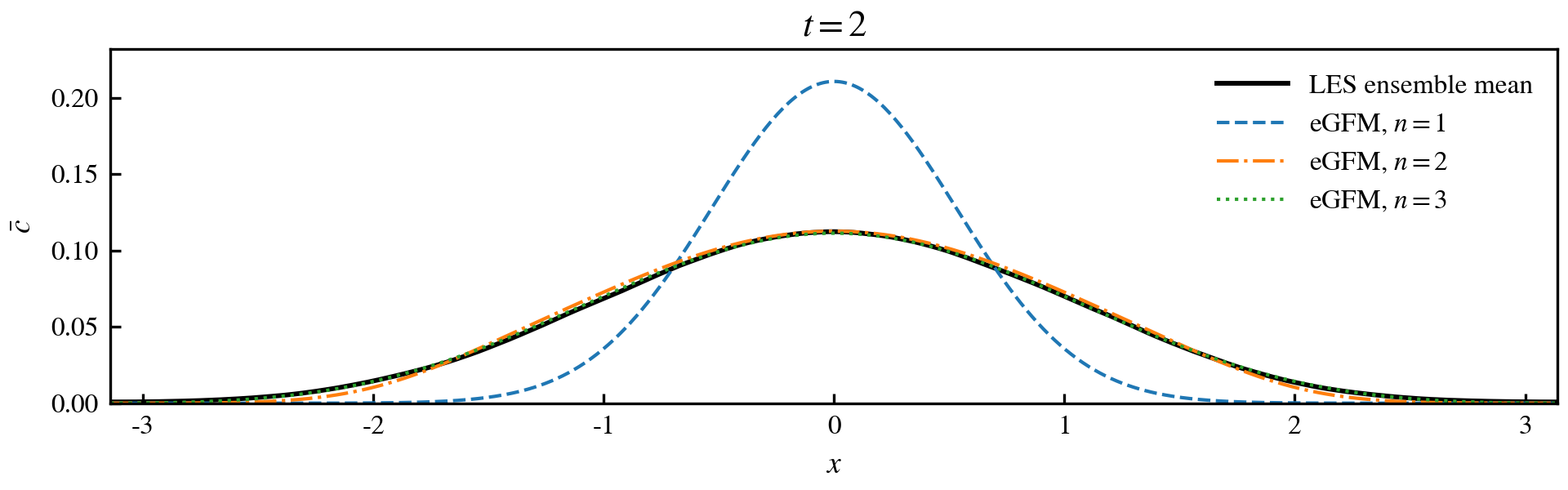}
    \caption{Comparison of \(\bar c\) from the test data and the one-, two-, and three-equation eGFM models at \(t=0.5,1,\) and \(2\). The eGFM models are trained with \(n_{x\mathrm{mode}}=15\) and \(n_{y\mathrm{mode}}=n\).}
    \label{fig:turbulence:dispersion-slices}
\end{figure}

Figure~\ref{fig:turbulence:dispersion-slices} compares the one-, two-, and three-equation eGFM predictions with the test data at \(t=0.5,1,\) and \(2\). The one-equation model disperses the scalar too slowly, while adding the scalar-flux equation captures most of the long-time dispersion behavior. The third equation further improves the transient prediction.

\begin{figure}[tbp]
    \centering
    \includegraphics{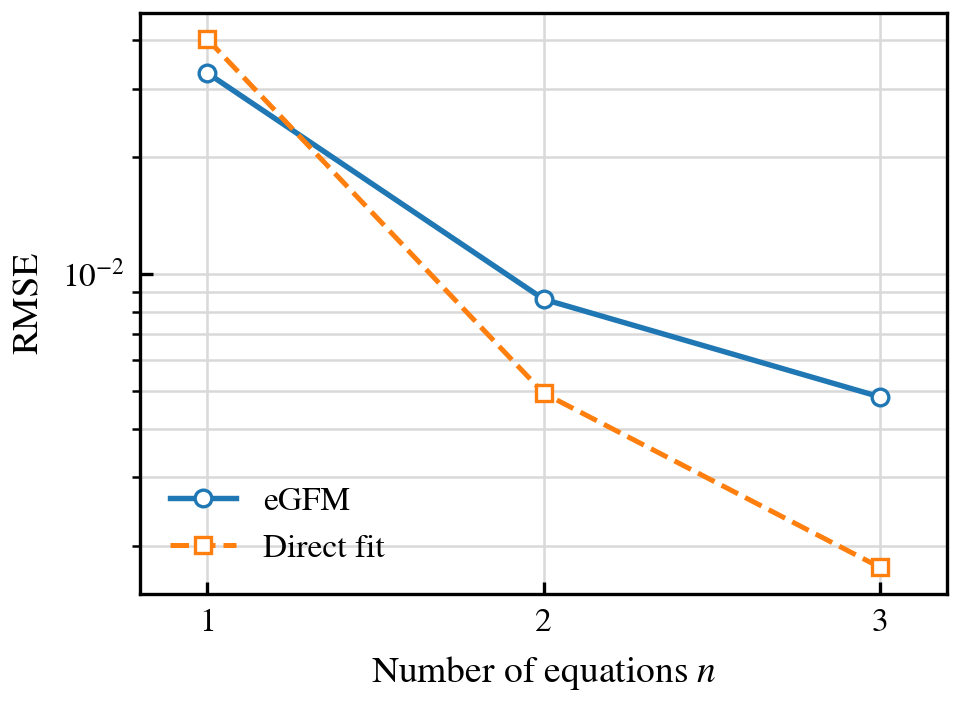}
    \caption{RMSE of \(\bar c\) for the eGFM-based and direct-fit models. The eGFM models use \(n_{x\mathrm{mode}}=15\) and \(n_{y\mathrm{mode}}=n\).}
    \label{fig:turbulence:rmse}
\end{figure}

Figure~\ref{fig:turbulence:rmse} compares the test errors of the eGFM-based and direct-fit models. The direct-fit models are more accurate for \(n\ge 2\). This is expected since they are trained on the test data, while the eGFM-based models are trained on lower-resolution forced simulations without using the Gaussian initial condition. Over the full test dataset, the two- and three-equation eGFM models have relative \(L^2\) errors in \(\bar c\) of approximately 9.55\% and 5.35\%, respectively (computed as \(\|\bar c_{\mathrm{model}}-\bar c_{\mathrm{reference}}\|_2/\|\bar c_{\mathrm{reference}}\|_2\) over all sampled $x$ and $t$).

\subsubsection{Effect of the training dataset}
\begin{figure}[tbp]
    \centering
    \includegraphics{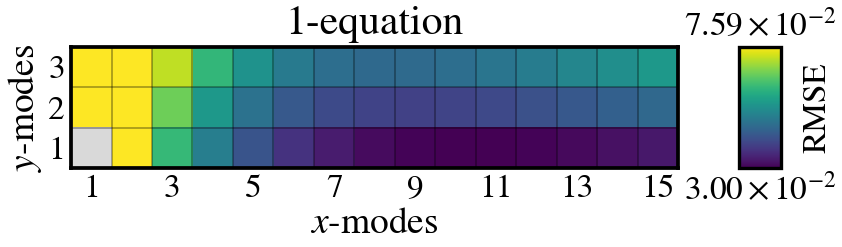}
    \hfill
    \includegraphics{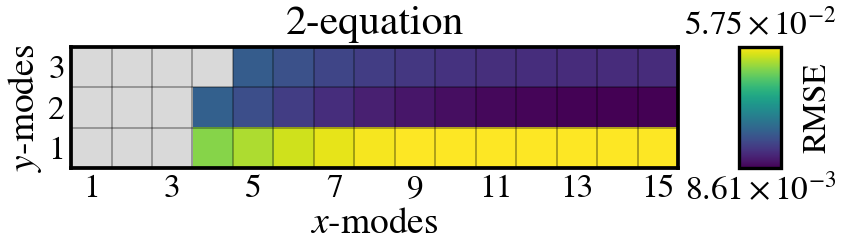}
    \includegraphics{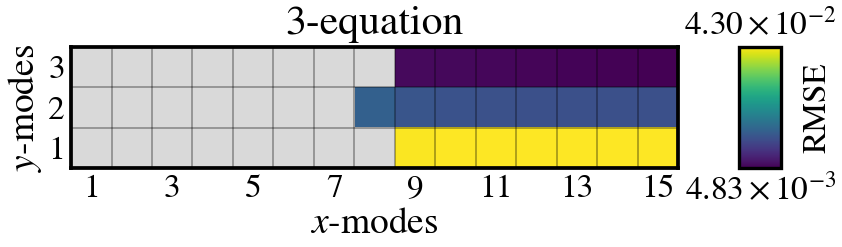}
    \caption{RMSE of eGFM-based reduced models over the space of forcing subsets for \(n=1,2,3\). Gray blocks indicate an empty forcing subset, fitting failure, or runtime blowup.}
    \label{fig:turbulence:dataset}
\end{figure}

Following the training-dataset study in Sec.~\ref{sec:parallel-results}, Fig.~\ref{fig:turbulence:dataset} shows the test errors as \(n_{x\mathrm{mode}}\) and \(n_{y\mathrm{mode}}\) are varied. Consistent with the GFM forcing principle, the lowest RMSE often occurs when \(n_{y\mathrm{mode}}=n\) and $n_{x\mathrm{mode}}$ is sufficiently large. The higher-order models require a richer training forcing set in $x$.

\subsection{Summary}

In this section, we apply eGFM to passive-scalar transport in homogeneous isotropic turbulence. The two-equation model captures most of the long-time dispersion behavior, while the three-equation model further improves the transient prediction. These results extend the eGFM framework from prescribed velocity fields to turbulent scalar transport.

%% file: sections/6.conclusion.tex
\section{\label{sec:conclusion}Conclusion}

This paper developed the Generalized Forcing Method (GFM) as a systematic framework for generating training data for closure models of linear transport PDEs.
The central motivation is that data-driven closure modeling requires TAAD datasets: tailored, accurate, affordable, and diverse. For dissipative multiscale systems, trajectories generated only by sampling initial conditions can rapidly collapse onto a low-dimensional set of dynamics, leading to ill-conditioned model fitting and inefficient use of high-fidelity simulations. To address this difficulty, GFM designs forcings that excite the resolved dynamics while preserving the target closure relation.
In this sense, GFM can be interpreted as a generalization of the Macroscopic Forcing Method (MFM) by extending the admissible forcing beyond the macroscopic field.

The formulation of the GFM principle is based on a resolved--unresolved decomposition of the full state space, induced by the chosen reduced variables. By analyzing the forced linear system, we showed that the target closure relation can be affected by both unresolved forcing and the initial condition. Therefore, the GFM principle uses zero initial conditions and restricts the forcing to the resolved subspace. Based on this principle, we introduced implicit GFM (iGFM), which prescribes trajectories of the resolved variables, and explicit GFM (eGFM), which constructs a basis of admissible forcings. iGFM is useful for probing closure structure, while eGFM provides a systematic procedure for training data generation.

We revisited the scalar dispersion problem in parallel shear flows, where an analytical closure family was derived from projecting the full state onto low Fourier modes and applying quasi-steady approximations. This analytical construction motivates the choice of spectral-based resolved variables and a linear parabolic PDE model ansatz, and provides reference models for assessing data-generated models.
Using this analytical closure family as a reference, we tested eGFM on parallel-flow dispersion problems with two different flow profiles. In both cases, eGFM recovered reduced models whose accuracy was comparable to the analytical models and the models fitted directly from the test data. From comparing the performance across training datasets, it is further confirmed that the forcing set must sufficiently span the resolved subspace, while forcing outside the admissible resolved subspace can degrade the accuracy of the trained model.

We also extended the framework to spatially inhomogeneous flows. In this setting, the reduced model requires spatially dependent coefficient matrices, and the fitting problem becomes more ill-conditioned. To address this, we proposed a constrained regularized fitting procedure that incorporates smoothness regularization and physical constraints such as mass conservation. Across the two inhomogeneous test cases, the eGFM-based models show better stability and accuracy than the corresponding direct-fit models.

Meanwhile, the random-flow example reveals a limitation associated with the choice of reduced variables. For flows with finite temporal correlation, the spectral-based and velocity-based variables considered here are not capable of representing the memory effects in a Markovian model. In this setting, increasing the number of resolved variables within the same variable family or enlarging the admissible forcing space does not systematically improve prediction accuracy.
This behavior indicates the need for reduced variables that are consistent with the structure of the target closure relation.
Another possible approach is to extend the model form beyond the parabolic PDE ansatz. Since the GFM forcing design is independent of the specific model form, the same data-generation strategy can also be used to train neural-network-based closure models.

We further applied eGFM to passive-scalar transport in homogeneous isotropic turbulence. The two-equation model captures most of the long-time dispersion behavior, while the three-equation model improves the transient prediction. Models fitted using the lower-resolution turbulence data are predictive when evaluated against the higher-resolution test data, demonstrating the applicability of eGFM to turbulent scalar transport.

Finally, we identify several directions for future work. First, the present study was restricted to linear passive scalar transport. Extensions to non-dissipative or coupled systems will require revisiting the forcing principles. For such systems, stability may also need to be imposed explicitly during model fitting. Second, the choice of resolved variables was prescribed. A problem- and target-dependent procedure for basis selection is needed, particularly for cases with strong nonlocality such as random flows. Third, statistical tools for sampling, data weighting, and uncertainty quantification should be developed in combination with GFM for reliability-based modeling. Fourth, extending GFM to nonlinear systems is substantially more challenging. In such cases, a unified principle may not be available, and a problem-dependent forcing design will likely be necessary.